\documentclass[12pt]{article}
\usepackage{amssymb}
\usepackage{amsfonts}
\usepackage{amsmath}
\usepackage{eurosym}
\usepackage{makeidx}
\usepackage{amssymb}
\usepackage{times}
\usepackage{anysize}

\marginsize{2cm}{2cm}{1cm}{1cm}
\newtheorem{theorem}{Theorem}[section]

\newtheorem{condition}[theorem]{Condition}

\newtheorem{corollary}[theorem]{Corollary}

\newtheorem{definition}[theorem]{Definition}

\newtheorem{lemma}[theorem]{Lemma}
\newtheorem{notation}[theorem]{Notation}

\newtheorem{proposition}[theorem]{Proposition}
\newtheorem{remark}[theorem]{Remark}

\newenvironment{proof}[1][Proof]{\noindent\textbf{#1.} }{\ \rule{0.5em}{0.5em}}
\input{tcilatex}
\begin{document}

\title{Large Deviations in Weakly Interacting Fermions -- Generating
Functions as Gaussian Berezin Integrals and Bounds on Large Pfaffians}
\author{N. J. B. Aza \and J.-B. Bru \and W. de Siqueira Pedra \and L. C. P.
A. M. M\"{u}ssnich}
\date{\today }
\maketitle

\begin{abstract}
We prove that the G\"{a}rtner--Ellis generating function of probability
distributions associated with KMS states of weakly interacting fermions on
the lattice can be written as the limit of logarithms of Gaussian Berezin
integrals. The covariances of the Gaussian integrals are shown to have a
uniform Pfaffian bound and to be summable in general cases of interest,
including systems that are \emph{not} translation invariant. The Berezin
integral representation can thus be used to obtain convergent expansions of
the generating function in terms of powers of its parameter. The derivation
and analysis of the expansions of logarithms of Berezin integrals are the
subject of the second part of the present work. Such technical results are
also useful, for instance, in the context of quantum information theory, in
the computation of relative entropy densities associated with fermionic
Gibbs states, and in the theory of quantum normal fluctuations for weakly
interacting fermion systems.\bigskip

\noindent \textbf{Keywords:} Berezin integral, constructive methods,
interacting fermions, determinant bound, large deviations. \bigskip

\noindent \textbf{AMS Subject Classification:} 81T08, 82C22, 46L53
\end{abstract}

%TCIMACRO{\TeXButton{\tableofcontents }{\tableofcontents}}%
%BeginExpansion
\tableofcontents%
%EndExpansion

\section{Introduction\label{Section intro LD}}

This paper is the first step towards large deviations (LD) analyses of
fermions that, at equilibrium, weakly interact with each other on lattices.
Various fields of theoretical physics and mathematics are concerned in this
study and, to make our results understandable to a wider audience, we
describe their mathematical framework in detail in Sections \ref{Section
Historical Overview}-\ref{Spin or CAR Unital Algebras copy(1)}, including a
historical overview (Section \ref{LD histiry}). Indeed, LD methods have been
used many times to study quantum systems from at least the 1980s.

To begin, we review recent LD results related to the present work. In 2000,
Lebowitz, Lenci and Spohn proved \cite{LLS00} that the particle density of
an ideal $d$-dimensional continuum quantum (Fermi or Bose) gas at thermal
equilibrium obeys a so-called large deviation principle (LDP), cf. \cite[%
Theorem II.1]{LLS00} and Section \ref{LD Formalism within Algebraic Quantum
Mechanics}. This is proven for arbitrary inverse temperatures $\beta \in 
\mathbb{R}^{+}$ and chemical potentials $\mu $. There is \emph{no}
interparticle interaction and the system is supposed to be in a single
phase, which is guaranteed by imposing differentiability of the pressure 
\cite[p. 1226]{LLS00}. For convergence, a clustering condition is assumed.

Two years later, Gallavotti, Lebowitz and Mastropietro \cite{GLM02}
contributed another proof that the particle density of (rarefied) quantum
gases fulfills an LDP at thermal equilibrium. The authors consider systems
of particles satisfying either Boltzmann, Bose, or Fermi statistics, but we
focus here on results for lattice Fermi systems. In this case, the proof is
done for \emph{weakly interacting} fermions via \emph{Berezin integrals}%
\footnote{%
Also called Grassmann or Grassmann-Berezin integrals.} and tree expansions
for \emph{correlation functions} of the equilibrium state (cf. \cite[%
Equation (7.11)]{GLM02}), from which the (limiting)\emph{\ }logarithmic
moment generating function (Section \ref{LD Formalism within Algebraic
Quantum Mechanics}) is obtained. This approach is strongly connected to the
one we develop presently. The complete proof of the LDP in \cite{GLM02} may
be achieved by getting bounds for large determinants (or Pfaffians)
associated with the full covariance (cf. \cite[Expression (7.23)]{GLM02}),
which is, nonetheless, \emph{not} done by the authors. Instead, they perform
a multiscale decomposition of the covariance to bound the resulting series
of determinants afterwards. See \cite[Item (7) on page 855]{GLM02}. Avoiding
the multiscale analysis is one of our main motivations here. We also show
that it is not necessary to use \emph{correlation functions} to construct
the (limiting)\emph{\ }logarithmic moment generating function $\mathrm{J}$
via functional integrals:

\begin{itemize}
\item We directly write $\mathrm{J}$ as a Gaussian Berezin integral with a
covariance that\emph{\ differs} from the one for correlation functions. See
below Equations (\ref{non-auto operateor})-(\ref{discrete time covariance}).

\item Then, we prove that this covariance is summable and satisfies a
Pfaffian bound in the desired way.
\end{itemize}

\noindent In contrast to \cite{GLM02}, the method presented here, which will
be heavily exploited in \cite{LD2}, is \emph{not} restricted to particle
density observables $\{\mathrm{D}_{l}\}_{l\in \mathbb{R}^{+}}$ (\ref{density
observables}). It can be applied to any thermodynamic sequences $\{\mathrm{E}%
_{l}^{\Psi }\}_{l\in \mathbb{R}^{+}}$ (\ref{average K}) associated with
densities of any other physical quantity that can be encoded in a
translation-invariant and finite-range interaction $\Psi $ (see Section \ref%
{Thermodynamic Sequences of Observables}). Before explaining in more detail
our technical approach in relation to the one of \cite{GLM02}, we proceed
with the review of recent LD results in quantum statistical mechanics.

In 2004, Neto\v{c}n\'{y} and Redig \cite{netovcny2004large} studied quantum
spin systems at thermal equilibrium. They demonstrated the existence of an
LDP and a central limit theorem at sufficiently high temperatures for
thermodynamic sequences $\{\mathrm{E}_{l}^{\Psi }\}_{l\in \mathbb{R}^{+}}$
of density observables encoded in \emph{zero-range}\footnote{%
A zero-range interaction $\Psi $ yields a sequence of the form $\mathrm{E}%
_{l}^{\Psi }=\frac{1}{|\Lambda _{l}|}\sum_{x\in \Lambda _{l}}\Psi _{\{x\}}$,
which is a particular case of the sequences defined by Equation (\ref%
{average K}). See Section \ref{Thermodynamic Sequences of Observables}.}
interactions $\Psi $. This result thus holds true for more general
thermodynamic sequences than just the particle density case used in previous
studies. Note that the central limit theorem in \cite{netovcny2004large} is
a consequence of the Bryc theorem (Theorem \ref{Coro estimate super
importante}), which will also be used in \cite{LD2}.

The following year, Lenci and Rey-Bellet \cite{lenci2005large} took into
consideration empirical means (space averages) $\{\mathrm{M}_{l}^{A}\}_{l\in 
\mathbb{R}^{+}}$ of a finite-volume element $A$ in either spin or CAR $%
C^{\ast }$-algebra, as defined below by (\ref{emperical mean}). This
quantity refers to thermodynamic sequences $\{\mathrm{E}_{l}^{\Psi }\}_{l\in 
\mathbb{R}^{+}}$ of density observables for more general
translation-invariant, finite-range interactions $\Psi $, as defined below
by (\ref{average K}). See also Equation (\ref{equation mean}). In contrast
to \cite{netovcny2004large}, they did not exploit cluster expansions, but
the inequality 
\begin{equation}
\left\vert \ln \mathrm{tr}\left( A\mathrm{e}^{H+K}\right) -\ln \mathrm{tr}%
\left( A\mathrm{e}^{H}\right) \right\vert \leq \sup_{0\leq t\leq 1}\sup_{-%
\frac{1}{2}\leq w\leq \frac{1}{2}}\left\Vert \mathrm{e}^{-w\left(
H+tK\right) }K\mathrm{e}^{w\left( H+tK\right) }\right\Vert _{\mathcal{B}%
\left( \mathbb{C}^{n}\right) }\ ,  \label{bound moniteur salsa}
\end{equation}%
for any matrices $A,H,K$ so that $A>0$, $H=H^{\ast }$ and $K=K^{\ast }$,
where $\mathrm{tr}$ denotes the normalized trace. See \cite[Lemma 3.6]%
{lenci2005large}. $\mathcal{B}(\mathcal{X})$ denotes the space of linear
operators acting on a vector space $\mathcal{X}$. The uniqueness of the
thermal equilibrium state, i.e., the Kubo-Martin-Schwinger (KMS) state, is
used as a key argument, in contrast to \cite{netovcny2004large} (cf. Remark %
\ref{LDP and KMS states} and \cite[Remark 2.14]{netovcny2004large}). They
obtain an LDP for the empirical average of finite-volume observables either
in dimension one at any temperature, or for any finite dimension at
sufficiently high temperatures. Their proof works for both spin and
fermionic lattice systems. See \cite[Theorem 3.2]{lenci2005large}.

Note that the bound (\ref{bound moniteur salsa}) is only useful if its
right-hand side behaves as $\left\Vert K\right\Vert _{\mathcal{B}\left( 
\mathbb{C}^{n}\right) }$, uniformly with respect to the choices of the
self-adjoint matrices $H$ and $K$ in $\mathcal{B}(\mathbb{C}^{n})$. This is
related to the following open problem: to find sufficient conditions on an
interaction $\Phi $, as defined in Section \ref{Thermodynamic Sequences of
Observables}, such that, for fixed $A$ in (finite-volume) spin or CAR $%
C^{\ast }$-algebra, the norm of $\mathrm{e}^{H_{L}^{\Phi }}A\mathrm{e}%
^{-H_{L}^{\Phi }}$ does not diverge\ in the thermodynamic limit $%
L\rightarrow \infty $. Here, as defined below by Equation (\ref{Hamiltonians}%
), $H_{L}^{\Phi }$ is the Hamiltonian associated with the interaction $\Phi $
in some cubic box $\Lambda _{L}$ of side length $L\in \mathbb{R}^{+}$. Araki 
\cite{A69} proved this non-divergence for all translation-invariant,
finite-range interactions in the one-dimensional case. Nevertheless, the
assertion was recently proven to be \emph{false} in dimension two \cite%
{Bouch}, at least for sufficiently large $\left\vert \Phi \right\vert $. One
of the advantages of constructing the logarithmic moment generating function
via Gaussian Berezin integrals, together with the corresponding tree
expansion we will use in \cite{LD2}, is that such uniform bounds are \emph{%
not} needed, at least not explicitly.

In 2011, Ogata and Rey-Bellet \cite{OR11} made use of the Ruelle-Lanford
function \cite{Ruelle1965,Lanford1973} to technically simplify and unify the
proof of previous LD\ results for quantum spin systems with a slight
extension: Their proofs are based on the notion of \textquotedblleft
asymptotic decoupled states\textquotedblright\ as defined in the classical
case in \cite[Definitions 3.2 and 3.3]{P02}, i.e., states $\rho $ on spin $%
C^{\ast }$-algebra for which there is a function $c:\mathbb{N}\rightarrow
\lbrack 0,\infty )$ such that, for any $L\in \mathbb{N}$, and any
non-negative observables $A,B$ with corresponding supports $\Lambda
^{(A)}\subset \Lambda _{L}$ and $\Lambda ^{(B)}\subset \mathbb{Z}%
^{d}\backslash \Lambda _{L}$, 
\begin{equation*}
\lim\limits_{L\rightarrow \infty }\frac{c(L)}{\left\vert \Lambda
_{L}\right\vert }=0\qquad \text{and}\qquad \mathrm{e}^{-c(L)}\rho (A)\rho
(B)\leq \rho (AB)\leq \mathrm{e}^{c(L)}\rho (A)\rho (B)\ .
\end{equation*}%
There are few quantum systems for which KMS states are known to satisfy this
condition, for instance at dimension one \cite{A69} and at high temperatures 
\cite{A74}, like the cases considered in \cite%
{netovcny2004large,lenci2005large}.

More recently, De Roeck, Maes, Neto\v{c}n\'{y} and Sch\"{u}tz studied in 
\cite{RoMaNeSc} (identically distributed (i.d.)) empirical means of
operators associated with zero-range interactions for quantum spin systems
at thermal equilibrium, via the corresponding distributions\ as defined
below by (\ref{fluctuation measure}). They focus on Gibbsianness or
quasi-locality of the (limiting) distribution, which is a stronger feature
than the property of asymptotic decoupling used in \cite{OR11}, which in
turn is stronger than an LD property. See, e.g., \cite[Theorem 4.4]{RoMaNeSc}%
. Like in \cite{netovcny2004large,lenci2005large,OR11} they study the high
temperature regime, but also the \emph{low }temperature situation for a
class of \emph{gapped} quantum spin systems. In all studied regimes, there
is a \emph{unique} KMS state. They use a polymer model together with either
a high-temperature cluster expansion or an expansion around the ground state
that is reminiscent of the quantum Pirogov-Sinai theory.

In the case of \emph{weakly} interacting fermions on lattices, we aim to
show both an LDP and a central limit theorem for the translation-invariant
thermal equilibrium states at \emph{any} temperature and \emph{any}
dimension for \emph{any} thermodynamic sequences $\{\mathrm{E}_{l}^{\Psi
}\}_{l\in \mathbb{R}^{+}}$ (\ref{average K}) that is encoded in a
translation-invariant and finite-range interaction $\Psi $. The first
technical step towards this result is performed in the current paper.

We study fermion systems in their algebraic formulation via \emph{self-dual}
CAR algebra \cite[Chap. 6]{EK98}, a concept first introduced by Araki in 
\cite{A68}. Araki's view point does not seem to have strongly percolated
into the general mathematical physics community\footnote{%
For instance, the term \textquotedblleft self--dual CAR
algebra\textquotedblright\ does not even appear in the celebrated textbook 
\cite{BratteliRobinson}. It is only implicitly there in the discussion of
Bogoliubov transformations at the end of \cite[Section 5.2.2.1]%
{BratteliRobinson}.}, even though it can be used to elegantly treat \emph{%
non-gauge invariant} quadratic models. Therefore, the preliminary
discussions of the results are restricted to the well-known CAR algebra
setting. Below, $\mathrm{CAR}(\mathfrak{h})$ denotes the CAR $C^{\ast }$%
-algebra associated with any finite-dimensional Hilbert space $\mathfrak{h}$%
, like for fermion systems on a finite lattice.

Using the tracial state $\mathrm{tr}$ on $\mathrm{CAR}(\mathfrak{h})$, i.e.,
the normalized trace on the finite-dimensional fermionic Fock space
representation (Section \ref{Fock}, in particular Remark \ref{bound norm a
CAR copy(2)}), we explain in Section \ref{Feynman--Kac--like Formula} how
the quantity%
\begin{equation}
\mathrm{tr}(\mathrm{e}^{-\beta \mathbf{H}}\mathrm{e}^{sK})\ ,\qquad \beta
\in \mathbb{R}^{+},\ s\in \mathbb{R}\quad \text{and}\quad \mathbf{H}=\mathbf{%
H}^{\ast },K=K^{\ast }\in \mathrm{CAR}(\mathfrak{h})\ ,  \label{trace}
\end{equation}%
can be written as a limit of \emph{Berezin integrals}. This is related to a
version of the Feynman-Kac formula proven here. See Theorem \ref{coro
equation cool2} and Remark \ref{coro equation cool2 copy(1)}.

We consider fermionic Hamiltonians of the form%
\begin{equation}
\mathbf{H}=\mathrm{d}\Gamma (h)+\mathrm{d}\Upsilon (g)+W\ ,\qquad W=W^{\ast
}\in \mathrm{CAR}(\mathfrak{h})\ ,  \label{form}
\end{equation}%
where, for any $h=h^{\ast }\in \mathcal{B}(\mathfrak{h})$ and \emph{%
antilinear} operator $g=-g^{\ast }$ on $\mathfrak{h}$, 
\begin{eqnarray}
\mathrm{d}\Gamma (h) &\doteq &\sum\limits_{i,j\in J}\left\langle \psi
_{i},h\psi _{j}\right\rangle _{\mathfrak{h}}a\left( \psi _{i}\right) ^{\ast
}a\left( \psi _{j}\right) \ ,  \label{form1} \\
\mathrm{d}\Upsilon (g) &\doteq &\frac{1}{2}\sum\limits_{i,j\in J}\left(
\left\langle \psi _{i},g\psi _{j}\right\rangle _{\mathfrak{h}}a\left( \psi
_{i}\right) ^{\ast }a\left( \psi _{j}\right) ^{\ast }+\overline{\left\langle
\psi _{i},g\psi _{j}\right\rangle _{\mathfrak{h}}}a\left( \psi _{j}\right)
a\left( \psi _{i}\right) \right) \ .  \label{form2}
\end{eqnarray}%
Here, $\{\psi _{j}\}_{j\in J}$ is any orthonormal basis\footnote{$\mathrm{d}%
\Gamma (h),\mathrm{d}\Upsilon (g)$\ are self-adjoint elements which do not
depend on the choice of the orthonormal basis, but on the choice of
generators $\{a(\varphi )\}_{\varphi \in \mathfrak{h}}$ of $\mathrm{CAR}(%
\mathfrak{h})$.} of $\mathfrak{h}$, and $a\left( \varphi \right) ,a\left(
\varphi \right) ^{\ast }\in \mathrm{CAR}\left( \mathfrak{h}\right) $ are the
usual fermionic annihilation/creation operators of a fermion in the state $%
\varphi \in \mathfrak{h}$. As is usual, $h^{\ast }$ and $g^{\ast }$ are the
adjoints\footnote{%
The adjoint of an antilinear map $g$ is defined by the condition $%
\left\langle \varphi _{1},g^{\ast }\varphi _{2}\right\rangle _{\mathfrak{h}%
}=\left\langle \varphi _{2},g\varphi _{1}\right\rangle _{\mathfrak{h}}$, $%
\varphi _{1},\varphi _{2}\in \mathfrak{h}$.} of $h$ and $g$, respectively.

The self-adjoint, even element $\mathrm{d}\Gamma (h)\in \mathrm{CAR}\left( 
\mathfrak{h}\right) $ is the second quantization\ of the one-particle
Hamiltonian $h\in \mathcal{B}(\mathfrak{h})$. It represents a gauge
invariant model of free fermions. The non-gauge invariant quadratic part of $%
\mathbf{H}$ is represented by the self-adjoint and even element $\mathrm{d}%
\Upsilon (g)\in \mathrm{CAR}\left( \mathfrak{h}\right) $. Such
\textquotedblleft off-diagonal\textquotedblright\ terms appear, for
instance, in the celebrated BCS theory of superconductivity. Note that it is
well-known \cite{Berezin,A68} that quadratic fermionic Hamiltonians like $%
\mathrm{d}\Gamma (h)+\mathrm{d}\Upsilon (g)$ can always be transformed via
some (Bogoliubov) unitary transformation into $\mathrm{d}\Gamma (\tilde{h}%
)+\lambda $ for some constant $\lambda \in \mathbb{R}$ and matrix $\tilde{h}%
\in \mathcal{B}(\mathfrak{h})$. Such a transformation is, in general, not
explicitly given and also changes the element $W$, which encodes in (\ref%
{form}) the interparticle interaction of the fermion system. It is, however,
technically advantageous to avoid such a transformation: Instead of
modifying the model, we adapt our Grassmann-algebra construction. In this
context, the self-dual CAR framework turns out to be much more natural than
the usual CAR setting.

If the element $W$ is taken to be an \emph{even} element of $\mathrm{CAR}(%
\mathfrak{h})$, like for any many-fermion model in physics, then we show
that (\ref{trace}) can be written in terms of a limit of \emph{Gaussian}
Berezin integrals in the following sense:%
\begin{equation}
\frac{\mathrm{tr}(\mathrm{e}^{-\beta \left( \mathrm{d}\Gamma \left( h\right)
+\mathrm{d}\Upsilon (g)+W\right) }\mathrm{e}^{sK})}{\mathrm{tr}\left( 
\mathrm{e}^{-\beta \left( \mathrm{d}\Gamma \left( h\right) +\mathrm{d}%
\Upsilon (g)\right) }\right) }=\lim_{n\rightarrow \infty }\int \mathrm{d\mu }%
_{C_{h,g}^{(n)}}(\mathfrak{H}^{(n)})\mathrm{e}^{\mathcal{W}_{W,sK}^{(n)}}\ .
\label{trace2}
\end{equation}%
%
%
%
%
%
%
%
%
%
%
%
%
%
%
%
%
%
%
%
%
%
%
%
%
%
%
%
%
%
%
%
%
%
%
%
%
%
%
%
%
%
%
%
%
%
%\begin{equation*}
%[++L \frac{\mathrm{tr}(\mathrm{e}^{-\beta \left( \mathrm{d}\Gamma \left( h\right)
%+\mathrm{d}\Upsilon (g)+W\right) }\mathrm{e}^{sK})}{\mathrm{tr}\left(
%\mathrm{e}^{-\beta \left( \mathrm{d}\Gamma \left( h\right) +\mathrm{d}%
%\Upsilon (g) + W \right) }\right) }=\lim_{n\rightarrow \infty }\int \mathrm{d\mu }%
%_{C_{h,g}^{(n)}}(\mathfrak{H}^{(n)})\mathrm{e}^{\mathcal{W}_{W,sK}^{(n)}}\ .
%\label{trace2}
% L++]
%\end{equation*}%
Inside this limit, 
\begin{equation*}
\int \mathrm{d\mu }_{C_{h,g}^{(n)}}(\mathfrak{H}^{(n)})
\end{equation*}%
stands for a Gaussian Berezin integral on some Grassmann algebra over $%
\mathfrak{H}^{(n)}$ associated with an explicit covariance $C_{h,g}^{(n)}$
only dependent on $h$, $g$, $\beta $ and $n\in \mathbb{N}$. $\mathcal{W}%
_{W,sK}^{(n)}$ is the result of a canonical mapping from $\mathrm{CAR}(%
\mathfrak{h})$ to the Grassmann algebra over $\mathfrak{H}^{(n)}$ and only
depends on $W$, $sK$, $\beta $ and $n\in \mathbb{N}$. For more details, see
Section \ref{Feynman--Kac--like Formula}, in particular Corollary \ref%
{satz.spurformel copy(1)}.

In the future application to LDP for \emph{lattice} fermion systems at
equilibrium, $\beta $ is the inverse temperature and $s\in \mathbb{R}$ will
be the parameter of the\emph{\ }logarithmic moment generating function. For $%
L_{\mathrm{f}}\geq L_{\mathrm{i}}\geq l\geq 0$, we take $\mathfrak{h}=\ell
^{2}\left( \Lambda _{L_{\mathrm{f}}}\right) $ where $\Lambda _{L_{\mathrm{f}%
}}\subset \mathbb{Z}^{d}$ ($d\in \mathbb{N}$) is a centered cubic box of
side length $\mathcal{O}\left( L_{\mathrm{f}}\right) $, see Section \ref%
{Spin or CAR Unital Algebras} and Equation (\ref{eq:boxesl1}). The elements $%
W=|\Lambda _{L_{\mathrm{i}}}|\mathrm{E}_{L_{\mathrm{i}}}^{\Phi _{\mathrm{i}%
}} $ and $K=\left\vert \Lambda _{l}\right\vert \mathrm{E}_{l}^{\Psi }$ will
be associated with two arbitrary translation-invariant, finite-range
interactions $\Phi _{\mathrm{i}}$ and $\Psi $, respectively, as defined
below by (\ref{average K}). In this case, Equation (\ref{trace2}) allows us
to write the logarithmic moment generating function%
\begin{equation}
\mathrm{J}_{L_{\mathrm{f}},L_{\mathrm{i}},l}^{\Psi }\left( s\right) \doteq 
\frac{1}{\left\vert \Lambda _{l}\right\vert }\ln \frac{\mathrm{tr}(\mathrm{e}%
^{-\beta (\mathrm{d}\Gamma \left( h\right) +\mathrm{d}\Upsilon (g)+|\Lambda
_{L_{\mathrm{i}}}|\mathrm{E}_{L_{\mathrm{i}}}^{\Phi _{\mathrm{i}}})}\mathrm{e%
}^{s|\Lambda _{l}|\mathrm{E}_{l}^{\Psi }})}{\mathrm{tr}\left( \mathrm{e}%
^{-\beta (\mathrm{d}\Gamma \left( h\right) +\mathrm{d}\Upsilon (g)+|\Lambda
_{L_{\mathrm{i}}}|\mathrm{E}_{L_{\mathrm{i}}}^{\Phi _{\mathrm{i}}})}\right) }%
\ ,  \label{equation generating funct}
\end{equation}%
associated with $\{\mathrm{E}_{l}^{\Psi }\}_{l\in \mathbb{R}^{+}}$ in the
Gibbs state, as the limit of Gaussian Berezin integrals for any inverse
temperature\footnote{%
The case $\beta =0$ corresponds to use the trace state instead of the Gibbs
state and is not considered here. It can be studied by much simpler methods
since, in this case, the logarithmic moment generating function is a
pressure.}$\ \beta \in \mathbb{R}^{+}$,\ $s\in \mathbb{R}$ and $L_{\mathrm{f}%
}\geq L_{\mathrm{i}}\geq l\geq 0$.

We intend to use in \cite{LD2} the so-called \emph{Brydges-Kennedy tree
expansion} to show, at any dimension $d\geq 1$ and any fixed inverse
temperature\footnote{%
More precisely, $\beta $ must not be small, as in most previous results for $%
d>1$, but rather the \emph{interparticle} interaction $\Phi _{\mathrm{i}}$
has to be small enough, depending on $\beta $.} $\beta $, the existence of
an analytic continuation to a centered disk in the complex plane of the
logarithmic moment generating function%
\begin{equation}
\mathrm{J}^{\Psi }\left( s\right) \doteq \lim_{l\rightarrow \infty }\lim_{L_{%
\mathrm{i}}\rightarrow \infty }\lim_{L_{\mathrm{f}}\rightarrow \infty }%
\mathrm{J}_{L_{\mathrm{f}},L_{\mathrm{i}},l}^{\Psi }\left( s\right)
\label{limit}
\end{equation}%
associated with $\{\mathrm{E}_{l}^{\Psi }\}_{l\in \mathbb{R}^{+}}$ and any
weak$^{\ast }$ accumulation point\footnote{%
The function $\mathrm{J}^{\Psi }$ a priori depends on the weak$^{\ast }$
accumulation point of Gibbs states, fixed by the choice of a subsequence of $%
L_{\mathrm{f}}\geq L_{\mathrm{i}}\geq 0$.} of the sequence of Gibbs states
of weakly interacting fermions on lattices. Our method thus applies to
systems for which the uniqueness of the KMS state is not known. As explained
in Section \ref{LD Formalism within Algebraic Quantum Mechanics}, such a
result yields an LDP and a central limit theorem for the corresponding
observables.

\begin{remark}
\label{remark QH testing}\mbox{
}\newline
The method we set forth here (and later in \cite{LD2}) can be used to study
relative entropy densities \cite[Section 2.6]{JOPP12} and quantum normal
fluctuations \cite[Section 6]{V11} for fermion systems on lattices. It can
also be useful in the theory of \textquotedblleft quantum hypothesis
testing\textquotedblright\ via Chernoff and Hoefding bounds. See for
instance \cite[Section VI]{HiaiMosoOga}. For more explanations, see \cite%
{LD2}.
\end{remark}

The convergence of the Brydges-Kennedy tree expansion in non-relativistic
fermionic constructive quantum field theory is ensured at weak interaction,
provided that (i) Pfaffians arising in the expansion can be efficiently
bounded and (ii) the covariance is sum%
%TCIMACRO{\TeXButton{\-}{\-}}%
%BeginExpansion
\-%
%EndExpansion
mable:

\begin{itemize}
\item[(i)] In Section \ref{sect det bounds copy(1)} we provide (up to an
exponential term with small rate) \emph{sharp} bounds on Pfaffians of
fermionic covariances $C_{h,g}^{(n)}$ appearing in (\ref{trace2}). These
bounds are \emph{uniform} with respect to the choice of operators $h=h^{\ast
}$, $g=-g^{\ast }$ and the dimension of $\mathfrak{h}$. \emph{No}
translation invariance is required.

\item[(ii)] In Section \ref{sec Covariances as Correlations of Quasi--Free
States} we explain how one obtains the summability of the covariance $%
C_{h,g}^{(n)}$ from the decay of the two-point correlation functions. In
Section \ref{Section summability}, we prove the summability of the
covariance at \emph{any} inverse temperature $\beta \in \mathbb{R}^{+}$ for
general lattice fermion systems. In the gapped case, the covariance is
proven to be \emph{uniformly} summable with respect to $\beta \in \mathbb{R}%
^{+}$, as generally expected and already verified in various similar cases
considered in the literature. The proofs are based on the celebrated
Combes-Thomas bound \cite[Theorem 10.5]{AizenmanWarzel}. In particular, the
translation invariance is again \emph{not} required for the summability of
the covariance.
\end{itemize}

\noindent In contrast to (ii), observe that the Pfaffians associated with
fermionic covariances are not deduced from the corresponding Pfaffians or
determinants for two-point correlations. Indeed, the covariances $%
C_{h,g}^{(n)}$, $n\in \mathbb{N}$, appearing in (\ref{trace2}) are related
to a \emph{non-autonomous }difference equation: To simplify the discussion%
\footnote{%
The general case can be mapped to the\ special case $g=0$ by unitary
transformations.}, assume that $g=0$. Let $\ell _{\mathrm{ap}}^{2}(\mathbb{T}%
_{n};\mathfrak{h})$ be the space of antiperiodic $\mathfrak{h}$-valued
functions on the discrete torus%
\begin{equation*}
\mathbb{T}_{n}\doteq \left\{ \left( k-n_{\beta }+1\right) n^{-1}\beta :k\in
\left\{ 0,\ldots ,2n_{\beta }-1\right\} \right\} \subset \left( -\beta
-1,\beta +1\right]
\end{equation*}%
with $n_{\beta }\doteq n+\left\lfloor n/\beta \right\rfloor $, $\left\lfloor
x\right\rfloor $ being the largest natural number smaller than $x\in \mathbb{%
R}^{+}$. For any matrix $h=h^{\ast }\in \mathcal{B}(\mathfrak{h})$, define $%
\mathfrak{u}_{h}\in \mathcal{B}(\ell _{\mathrm{ap}}^{2}(\mathbb{T}_{n};%
\mathfrak{h}))$ by 
\begin{equation}
\lbrack \mathfrak{u}_{h}f]\left( \alpha \right) \doteq \left\{ 
\begin{array}{lll}
h\left( f\left( \alpha \right) \right) & \text{for} & \alpha \in \mathbb{T}%
_{n}\cap \left( 0,\beta \right] \ , \\ 
0 & \text{for} & \alpha \in \mathbb{T}_{n}\cap \left( \beta ,\beta +1\right]
\ ,%
\end{array}%
\right.  \label{non-auto operateor}
\end{equation}%
for any $f\in \ell _{\mathrm{ap}}^{2}(\mathbb{T}_{n};\mathfrak{h})$. Denote
by $\partial \in \mathcal{B}(\ell _{\mathrm{ap}}^{2}(\mathbb{T}_{n};%
\mathfrak{h}))$ the discrete time derivative operator. Then, the \emph{%
discrete time covariance} used to define the corresponding Gaussian Berezin
integrals for $g=0$ is equal to%
\begin{equation}
C_{h,0}^{(n)}=-2\left( \partial +\mathfrak{u}_{h}\right) ^{-1}\in \mathcal{B}%
(\ell _{\mathrm{ap}}^{2}(\mathbb{T}_{n};\mathfrak{h}))\ .
\label{discrete time covariance}
\end{equation}%
This representation of covariances is not used further in this paper. It is
nonetheless useful in the present discussion to highlight the difference
between $C_{h,g}^{(n)}$ and well-known covariances associated with
correlation functions.

The covariances appearing in a similar construction for the generating
functions of correlations are related, rather, to an \emph{autonomous}
difference equation on the torus $\mathbb{\tilde{T}}_{n}\doteq \mathbb{T}%
_{n}\cap \left( -\beta ,\beta \right] $, which refers to the case $\alpha
\in \mathbb{T}_{n}\cap \left( 0,\beta \right] $ in the right-hand side of (%
\ref{non-auto operateor}). This is related to the case $k<n$ in Corollary %
\ref{satz.spurformel copy(1)}. In this situation, as is done for instance in 
\cite[Section 3]{BGPS}, \cite[Section 3.2]{GM}, \cite[Section 5.1.]{GMP},
large determinants (or Pfaffians) of the covariances can be dealt with by
using a combination of multiscale analyses for the so-called Matsubara UV
problem together with the usual Gram bound for determinants. It is precisely
what is done in \cite[Section 7]{GLM02} for weakly interacting fermions on a
lattice: They consider the sequence $\{\mathrm{E}_{l}^{\Psi }=\mathrm{D}%
_{l}\}_{l\in \mathbb{R}^{+}}$ of particle density observables (\ref{density
observables}) and expand the exponential $\mathrm{e}^{s\mathrm{E}_{l}^{\Psi
}}=\mathrm{e}^{s\mathrm{D}_{l}}$ in (\ref{equation generating funct}) to
compute the resulting series of correlation functions via constructive
methods. The choice $\mathrm{E}_{l}^{\Psi }=\mathrm{D}_{l}$ strongly
simplifies the combinatorics because $\mathrm{D}_{l}$, $l\in \mathbb{R}^{+}$%
, are \emph{projection operators}. A generalization of this approach to any
thermodynamic sequence $\{\mathrm{E}_{l}^{\Psi }\}_{l\in \mathbb{R}^{+}}$ is
not clear.

Meanwhile, applying multiscale decompositions in the Matsubara frequency
together with the usual Gram bound for covariances like $C_{h,0}^{(n)}$ 
\emph{would not directly work}, because the operators $\mathfrak{u}_{h}$ and 
$\partial $ in (\ref{discrete time covariance}) do \emph{not} commute, in
contrast to the autonomous case. Last but not least, these studies using
multiscale methods require Bogoliubov transformations to recover the gauge
invariant case ($g=0$).

However, \cite{dSPS} showed that such multiscale analyses to tackle the
Matsubara UV problem are \emph{not} necessary, at least in the autonomous
case, by proving a new bound for determinants that generalizes the original
Gram bound, see \cite[Theorem 1.3]{dSPS}. In the same spirit, we recently
derived \cite{universal} bounds that also do not need the UV regularization
of the Matsubara frequency, with the technical advantage that the given
covariance does not need to be decomposed as in \cite[Eq. (8)]{dSPS} to
obtain determinant bounds. Moreover, the estimate in \cite{universal} is 
\emph{sharp} (or optimal) and holds true for \emph{all} (even unbounded
cases, not being limited to semibounded) one-particle Hamiltonians. It is
important to stress that our approach to Pfaffian (or determinant) bounds
does not preclude the need of multiscale analysis to handle IR singularities
of covariances, as it is done, for instance, in \cite{GM,GMP}. Note however
that a positive temperature or the presence of a gap (at zero energy) in the
dispersion relation of free particles regularizes such IR\ singularities and
our method makes any multiscale analysis superfluous, at least for a small
enough interaction strength, in these two cases.

As remarked in \cite[Section 1.4]{universal}, using multiscale decomposition
to address the Matsubara UV problem can render the analysis less transparent
and avoiding this kind of procedure brings various technical benefits: In
the finite-dimensional case, we extend the proof of \cite[Corollary 1.4]%
{universal} to bound large Pfaffians of fermionic covariances $C_{h,g}^{(n)}$
in the non-autonomous cases, allowing us in a futur work \cite{LD2} to
directly compute, at weak coupling, the logarithmic moment generating
function (\ref{limit}) via constructive methods. In fact, if the free
Hamiltonian is generic, then our Pfaffians bounds yield the convergence of
usual expansions (in particular the Brydges-Kennedy expansion) for the
generating function, whenever the interaction strength is small as compared
to $\beta ^{-(d+1)}$. If the spectrum of the one-particle free Hamiltonian
has positive distance $\delta $ to zero (i.e., the Hamiltonian has a
spectral gap of size $\delta >0$) then the convergence radius is of order $%
\delta ^{d+1}$, uniformly with respect to the temperature. In particular,
the limit $\beta \rightarrow \infty $ can be taken and large deviation
results can be proven for ground states, in this case. Note that we
represent the logarithmic moment generating function of large deviation
theory in a way that is formally the same as the usual one for the pressure
of fermion systems, while keeping good Pfaffian (or determinant) bounds.
Thus, very likely, also the gapless case can be handled at zero temperature,
by combining our results with (a suitable version of) usual multiscale
decompositions to tame the IR singularity of covariances at small energies.

In \cite{universal}, constructions involving quasi-free states in suitably
chosen CAR algebras are used, whereas in the present case Grassmann-algebra
computations turn out to be more efficient by allowing us to represent
determinants as\emph{\ traces}. Like in \cite[Section 1.6]{universal}, H\"{o}%
lder inequalities for non-commutative $L^{p}$-spaces, here for Schatten
norms (see (\ref{Holder jean sans bras0})-(\ref{Holder jean sans bras})),
are pivotal tools and replace the celebrated Gram bounds on which \emph{all}
previous results, like \cite{BGPS,dSPS,GLM02,GM,GMP}, are explicitly based.

\begin{remark}
\label{translatino invariance}\mbox{
}\newline
Since our methods do not need translation invariance, our results are also
useful for handling usual correlation functions as well as logarithmic
moment generating functions, for fermionic systems with random, but ergodic,
interactions, as defined for instance in \cite{OhmVI}.
\end{remark}

To conclude, the main results are Theorem \ref{coro equation cool2}, \ref%
{Therem summabilityI} and \ref{Therem summabilityII} as well as Corollaries %
\ref{satz.spurformel copy(1)}, \ref{Pfaffian bounds I} and \ref{Pfaffian
bounds II}. Additionally, the mathematical methods used here are developed
in a systematic and general way. Some of them are, to some extent,
well-known, like for instance the trace formula (Theorem \ref%
{satz.spurformel}). Other ingredients are less standard, like the definition
of the \textquotedblleft circle\textquotedblright\ product (Definition \ref%
{definition star}) and the Chernoff product formula (Definition \ref{def XA}%
, Theorem \ref{lemma exp}), both introduced in the Ph.D. thesis \cite{Walter}%
. However, the extension of this technology to general \emph{self-dual} CAR
algebra is new. In fact, this paper will serve as a reference for further
research works.\medskip

\noindent The paper is organized as follows:

\begin{itemize}
\item Section \ref{Section Gen unc as Grassmann int} presents self-dual CAR
algebra as well as the corresponding quasi-free dynamics and states. It sets
up the primary mathematical framework used in the present study.

\item Section \ref{Section Grassman} establishes a relation between
self-dual CAR algebra and Grassmann (or exterior) algebra. By using the
Berezin integral, we define in particular a family of \textquotedblleft
circle\textquotedblright\ products, indexed by so-called basis projections,
converting any Grassmann algebra into a family of self-dual CAR\ algebras.

\item In Section \ref{Fermionic Path Integral}, we derive a Feynman-Kac-like
formula for logarithmic moment generating functions by using Gaussian
Berezin integrals and an appropriately chosen basis projection.

\item Section \ref{sect det bounds} studies so-called Pfaffian bounds of the
fermionic covariances of the Gaussian Berezin integrals that represent
logarithmic moment generating functions as well as their expression as
functions of two-point correlation functions. The summability of the
covariance is also proven here for very general fermion systems on the
lattice.

\item CAR algebra is more widely known than its self-dual counterpart. We
thus provide in Section \ref{notation hilbert copy(1)} a pedagogical
introduction to CAR algebra in relation to self-dual CAR algebra. For the
reader's convenience, we also add a concise explanation of the Fock
representation of CAR algebra.

\item Finally, we describe in Sections \ref{Section Historical Overview} and %
\ref{Spin or CAR Unital Algebras copy(1)} the overall conceptual framework
beyond the concrete technical problems tackled here in order to highlight
our results. This also makes the discussions in the introduction more
precise. In particular, in Section \ref{Section Historical Overview} we
mathematically define the large deviations (LD) formalism loosely discussed
above, while Section \ref{Spin or CAR Unital Algebras copy(1)} explains its
applications to quantum spin systems or fermions on lattices.
\end{itemize}

\begin{notation}
\label{remark constant}\mbox{
}\newline
\emph{(i)} A norm on a generic vector space $\mathcal{X}$ is denoted by $%
\Vert \cdot \Vert _{\mathcal{X}}$ and the identity map of $\mathcal{X}$ by $%
\mathbf{1}_{\mathcal{X}}$. The space of all bounded linear operators on $(%
\mathcal{X},\Vert \cdot \Vert _{\mathcal{X}}\mathcal{)}$ is denoted by $%
\mathcal{B}(\mathcal{X})$. The unit element of any algebra $\mathcal{X}$ is
denoted by $\mathfrak{1}$, provided it exists. The scalar product of any
Hilbert space $\mathcal{X}$ is denoted by $\langle \cdot ,\cdot \rangle _{%
\mathcal{X}}$ and $\mathrm{Tr}_{\mathcal{X}}$ represents the usual trace on $%
\mathcal{B}(\mathcal{X})$. \newline
\emph{(ii)} For each $k\in \mathbb{N}_{0}$, denote by $\mathcal{X}^{(k)}$ a
copy of some vector space $\mathcal{X}$. The corresponding copy of $\xi \in 
\mathcal{X}$ is denoted by $\xi ^{(k)}$.\newline
\emph{(iii)} We denote by $D$ any positive and finite generic constant.
These constants do not need to be the same from one statement to another.
\end{notation}

\section{Self-dual CAR Algebra and Bilinear Hamiltonians\label{Section Gen
unc as Grassmann int}}

\subsection{Self-dual CAR Algebra\label{Self--dual CAR Algebras}}

If not otherwise stated, $\mathcal{\mathcal{H}}$ always stands for a
finite-dimensional (complex) Hilbert space with even dimension $\mathrm{dim}%
\mathcal{\mathcal{H}}\in 2\mathbb{N}$. Let $\mathfrak{A}$ be an antiunitary
involution on $\mathcal{\mathcal{H}}$, i.e., an antilinear map from $%
\mathcal{\mathcal{H}}$ to $\mathcal{\mathcal{H}}$ such that $\mathfrak{A}%
^{2}=\mathbf{1}_{\mathcal{\mathcal{H}}}$ and 
\begin{equation*}
\left\langle \mathfrak{A}\varphi _{1},\mathfrak{A}\varphi _{2}\right\rangle
_{\mathcal{H}}=\left\langle \varphi _{2},\varphi _{1}\right\rangle _{%
\mathcal{H}}\ ,\qquad \varphi _{1},\varphi _{2}\in \mathcal{\mathcal{H}}\ .
\end{equation*}%
The space $\mathcal{\mathcal{H}}$ endowed with the involution $\mathfrak{A}$
is named a \emph{self-dual Hilbert space} and yields \emph{self-dual CAR%
\footnote{%
CAR refers to canonical anti--commutation relations.} algebra}, which is
defined as follows:

\begin{definition}[Self-dual CAR algebra]
\label{def Self--dual CAR Algebras}\mbox{
}\newline
A self-dual CAR algebra $\mathrm{sCAR}(\mathcal{H},\mathfrak{A})\equiv (%
\mathrm{sCAR}(\mathcal{H},\mathfrak{A}),+,\cdot ,\ast )$ is a $C^{\ast }$%
-algebra generated by a unit $\mathfrak{1}$ and a family $\{\mathrm{B}%
(\varphi )\}_{\varphi \in \mathcal{\mathcal{H}}}$ of elements satisfying
Conditions \emph{(a)-(c)}: \newline
\emph{(a)} The map $\varphi \mapsto \mathrm{B}\left( \varphi \right) ^{\ast
} $ is (complex) linear.\newline
\emph{(b)} $\mathrm{B}(\varphi )^{\ast }=\mathrm{B}(\mathfrak{A}(\varphi ))$
for any $\varphi \in \mathcal{H}$.\newline
\emph{(c)} The family $\{\mathrm{B}(\varphi )\}_{\varphi \in \mathcal{%
\mathcal{H}}}$ satisfies the CAR: For any $\varphi _{1},\varphi _{2}\in 
\mathcal{\mathcal{H}}$,%
\begin{equation}
\mathrm{B}(\varphi _{1})\mathrm{B}(\varphi _{2})^{\ast }+\mathrm{B}(\varphi
_{2})^{\ast }\mathrm{B}(\varphi _{1})=\left\langle \varphi _{1},\varphi
_{2}\right\rangle _{\mathcal{H}}\,\mathfrak{1}\ .  \label{CAR Grassmann III}
\end{equation}
\end{definition}

To the best of our knowledge, this notion was introduced by Araki in \cite%
{A68}. He changed his definition two years later in \cite{A70} by replacing
the linearity of the map $\varphi \mapsto \mathrm{B}\left( \varphi \right)
^{\ast }$ with its antilinearity. We maintain here the original definition
from \cite{A68}, and explicitly add the fact that $\mathrm{sCAR}(\mathcal{H},%
\mathfrak{A})$ is not only a $\ast $-algebra but also a $C^{\ast }$-algebra
(see, e.g., \cite[Lemma 4.5]{A68}).

\begin{remark}
\label{Remark bounded generator}\mbox{
}\newline
By the CAR\ (\ref{CAR Grassmann III}), the antilinear map $\varphi \mapsto 
\mathrm{B}\left( \varphi \right) $ is necessarily injective and contractive.
Therefore, $\mathcal{H}$ can be imbedded in $\mathrm{sCAR}(\mathcal{H},%
\mathfrak{A})$.
\end{remark}

\begin{remark}
\mbox{
}\newline
Self-dual CAR algebras are also well-defined even if the Hilbert space $%
\mathcal{H}$ has infinite or odd dimension. In the odd case, they are $\ast $%
-isomorphic to the direct product of a CAR\ algebra and a two-dimensional
abelian algebra, see \cite[Lemmata 3.3, 3.7]{A68}. This situation is,
however, not important in our study. The infinite-dimension case only shows
up in Section \ref{Section summability}, when we consider $\mathcal{H}%
_{\infty }$. See again \cite[Lemma 3.3]{A68}.
\end{remark}

Strictly speaking, Conditions (a)-(c) of Definition \ref{def Self--dual CAR
Algebras} only define self-dual CAR algebras up to Bogoliubov $\ast $%
-automorphisms\footnote{%
An analogous result for CAR algebra is, for instance, given by \cite[Theorem
5.2.5]{BratteliRobinson}.} (see (\ref{Bogoliubov automorphism})). In Section %
\ref{Section Grassman} we construct explicit $\ast $-isomorphic self-dual
CAR algebras from $\mathcal{H}$ and $\mathfrak{A}$. This is done via the
following concept of \emph{basis projection} \cite[Definition 3.5]{A68},
which also highlights the relationship between CAR algebras and their
self-dual counterparts.

\begin{definition}[Basis projections]
\label{def basis projection}\mbox{
}\newline
A basis projection associated with $(\mathcal{H},\mathfrak{A})$ is an
orthogonal projection $P\in \mathcal{B}(\mathcal{H})$\ satisfying $\mathfrak{%
A}P\mathfrak{A}=P^{\bot }\doteq \mathbf{1}_{\mathcal{H}}-P$. We denote by $%
\mathfrak{h}_{P}$ the range $\mathrm{ran}P$ of the basis projection $P$.
\end{definition}

\noindent Note that $\mathfrak{h}_{P}$ must satisfy the conditions 
\begin{equation}
\mathfrak{A}(\mathfrak{h}_{P})=\mathfrak{h}_{P}^{\bot }\text{\qquad
and\qquad }\mathfrak{A}(\mathfrak{h}_{P}^{\bot })=\mathfrak{h}_{P}
\label{frakA and perp}
\end{equation}%
for any basis projection $P$.

By \cite[Lemma 3.3]{A68}, an explicit basis projection $P\in \mathcal{B}(%
\mathcal{H})$ associated with $(\mathcal{H},\mathfrak{A})$ can always be
constructed because $\mathrm{dim}\mathcal{\mathcal{H}}\in 2\mathbb{N}$.
Moreover, $\varphi \mapsto (\mathfrak{A}\varphi )^{\ast }$ is a unitary map
from $\mathfrak{h}_{P}^{\bot }$ to the dual space $\mathfrak{h}_{P}^{\ast }$%
. In this case we can identify $\mathcal{H}$ with%
\begin{equation}
\mathcal{H}\equiv \mathfrak{h}_{P}\oplus \mathfrak{h}_{P}^{\ast }
\label{definition H bar}
\end{equation}%
and 
\begin{equation}
\mathrm{B}\left( \varphi \right) \equiv \mathrm{B}\left( P\varphi \right) +%
\mathrm{B}\left( \mathfrak{A}P^{\bot }\varphi \right) ^{\ast }.
\label{map iodiote}
\end{equation}%
Therefore, there is a natural isomorphism of $C^{\ast }$-algebras from $%
\mathrm{sCAR}(\mathcal{H},\mathfrak{A})$ to the CAR algebra $\mathrm{CAR}(%
\mathfrak{h}_{P})$ generated by the unit $\mathfrak{1}$ and $\{\mathrm{B}%
_{P}(\varphi )\}_{\varphi \in \mathfrak{h}_{P}}$. See Definition \ref{def
Self--dual CAR Algebras copy(1)} and also \cite[Lemma 3.3]{A68} for more
details. In other words, a basis projection $P$ can be used to \emph{fix}
so-called \emph{annihilation} and \emph{creations} operators.

Sections \ref{Section Gen func as Grassmann int copy(1)}-\ref{notation
hilbert} provide a pedagogical construction of self-dual CAR algebra
starting from a CAR algebra $\mathrm{CAR}(\mathfrak{h})$ associated with
some one-particle Hilbert space $\mathfrak{h}$, as explained in Definition %
\ref{def Self--dual CAR Algebras copy(1)}. CAR algebra is more widely known
than its self-dual analogue. However, the notion of self-dual CAR algebra is
more flexible because the one-particle Hilbert space is not fixed anymore.
In fact, for each basis projection $P$ associated with $(\mathcal{H},%
\mathfrak{A})$, by (\ref{definition H bar}), $\mathfrak{h}_{P}$ can be seen
as a one-particle Hilbert space. As shown in \cite{A68,A70}, this approach
to CAR naturally arises in the diagonalization of quadratic fermionic
Hamiltonians (Definition \ref{def trace state copy(1)}), via Bogoliubov
transformations defined as follows:

For any unitary operator $U\in \mathcal{B}(\mathcal{H})$ such that $U%
\mathfrak{A=A}U$, the family of elements $\mathrm{B}(U\varphi )_{\varphi \in 
\mathcal{\mathcal{H}}}$ satisfies Conditions (a)-(c) of Definition \ref{def
Self--dual CAR Algebras} and, together with the unit $\mathfrak{1}$,
generates $\mathrm{sCAR}(\mathcal{H},\mathfrak{A})$. Like in \cite[Section 2]%
{A70}, such a unitary operator $U\in \mathcal{B}(\mathcal{H})$ commuting
with the antiunitary map $\mathfrak{A}$ is named a \emph{Bogoliubov
transformation}, and the unique $\ast $-automorphism $\mathbf{\chi }_{U}$\
such that 
\begin{equation}
\mathbf{\chi }_{U}\left( \mathrm{B}(\varphi )\right) =\mathrm{B}(U\varphi )\
,\qquad \varphi \in \mathcal{\mathcal{H}}\ ,
\label{Bogoliubov  automorphism}
\end{equation}%
is called in this case a \emph{Bogoliubov }$\ast $\emph{-automorphism}. Note
that a Bogoliubov transformation $U\in \mathcal{B}(\mathcal{H})$ always
satisfies 
\begin{equation}
\mathrm{det}\left( U\right) =\mathrm{det}\left( \mathfrak{A}U\mathfrak{A}%
\right) =\overline{\mathrm{det}\left( U\right) }=\pm 1\ .
\label{orientation}
\end{equation}%
(Recall that, by definition, a Bogoliubov transformation is unitary.) If $%
\mathrm{det}\left( U\right) =1$, we say that $U$ has \emph{positive}
orientation. See Lemma \ref{Lemma Berezin orientation}. Otherwise $U$ is
said to have \emph{negative} orientation. These properties are also called
even and odd \cite[Chap. 6]{EK98}.

Clearly, if $P$\ is a basis projection associated with $(\mathcal{H},%
\mathfrak{A})$\ and $U\in \mathcal{B}(\mathcal{H})$ a Bogoliubov
transformation, then $P_{U}\doteq UPU^{\ast }$ is another basis projection.
Conversely, for any pair of basis projections $P_{1},P_{2}\in \mathcal{B}(%
\mathcal{H})$ there is a (generally not unique) Bogoliubov transformation $U$
such that $P_{2}=UP_{1}U^{\ast }$. See \cite[Lemma 3.6]{A68}. In particular,
Bogoliubov transformations map one-particle Hilbert spaces onto one another.

In Section \ref{Section Grassman}, we show that self-dual CAR algebra
naturally arises in the framework of Grassmann algebra. Self-dual CAR
algebra also allows us to elegantly treat non-gauge invariant models. That
is why we focus on $\mathrm{sCAR}(\mathcal{H},\mathfrak{A})$ and employ the
terminology $\mathrm{CAR}(\mathfrak{h})$ of (common) CAR algebra only to
pedagogical discussions.

\subsection{Bilinear Elements of Self-dual CAR Algebra\label{Sect Bilinear
Elements}}

An element $A\in \mathrm{sCAR}(\mathcal{H},\mathfrak{A})$, satisfying $%
\mathbf{\chi }_{-\mathbf{1}_{\mathcal{H}}}(A)=A$ or $\mathbf{\chi }_{-%
\mathbf{1}_{\mathcal{H}}}(A)=-A$ (see (\ref{Bogoliubov automorphism})), is
called, respectively, \emph{even} or \emph{odd}. The subspace of even
elements is a sub-$C^{\ast }$-algebra of $\mathrm{sCAR}(\mathcal{H},%
\mathfrak{A})$.

An important class of even elements in quantum physics is provided by
quadratic fermionic Hamiltonians. These refer to the self-adjoint (even)
elements of the CAR algebra which are quadratic in the creation and
annihilation operators, like for instance the Bogoliubov approximation of
the celebrated (reduced) BCS model. In the context of self-dual CAR algebra,
those elements are called \emph{bilinear Hamiltonians} by Araki \cite[Lemma
4.4,\ Definition 4.7]{A68} and are self-adjoint bilinear elements:

\begin{definition}[Bilinear elements of self-dual CAR algebra]
\label{def trace state copy(1)}\mbox{
}\newline
Given an orthonormal basis $\{\psi _{i}\}_{i\in I}$ of $\mathcal{H}$, we
define the bilinear element associated with $H\in \mathcal{B}(\mathcal{H})$
to be 
\begin{equation*}
\langle \mathrm{B},H\mathrm{B}\rangle \doteq \sum\limits_{i,j\in
I}\left\langle \psi _{i},H\psi _{j}\right\rangle _{\mathcal{H}}\mathrm{B}%
\left( \psi _{j}\right) \mathrm{B}\left( \psi _{i}\right) ^{\ast }\ .
\end{equation*}
\end{definition}

\noindent Note that $\langle \mathrm{B},H\mathrm{B}\rangle $ \emph{does not
depend} on the particular choice of the orthonormal basis, but does depend
on the choice of generators $\{\mathrm{B}(\varphi )\}_{\varphi \in \mathcal{%
\mathcal{H}}}$ of the self-dual CAR algebra $\mathrm{sCAR}(\mathcal{H},%
\mathfrak{A})$. Moreover, 
\begin{eqnarray}
\langle \mathrm{B},H\mathrm{B}\rangle &=&\sum\limits_{i,j\in I}\left\langle 
\mathfrak{A}\psi _{i},H\psi _{j}\right\rangle _{\mathcal{H}}\mathrm{B}\left(
\psi _{j}\right) \mathrm{B}\left( \psi _{i}\right) =\sum\limits_{i,j\in
I}\left\langle \psi _{i},H\mathfrak{A}\psi _{j}\right\rangle _{\mathcal{H}}%
\mathrm{B}\left( \psi _{j}\right) ^{\ast }\mathrm{B}\left( \psi _{i}\right)
^{\ast }  \notag \\
&=&\sum\limits_{i,j\in I}\left\langle \psi _{j},\mathfrak{A}H^{\ast }%
\mathfrak{A}\psi _{i}\right\rangle _{\mathcal{H}}\mathrm{B}\left( \psi
_{j}\right) ^{\ast }\mathrm{B}\left( \psi _{i}\right)
\label{equation facilissimo}
\end{eqnarray}%
for all $H\in \mathcal{B}(\mathcal{H})$, and by (\ref{CAR Grassmann III}),
bilinear elements of $\mathrm{sCAR}(\mathcal{H},\mathfrak{A})$ have adjoints
equal to%
\begin{equation}
\langle \mathrm{B},H\mathrm{B}\rangle ^{\ast }=\langle \mathrm{B},H^{\ast }%
\mathrm{B}\rangle \ ,\qquad H\in \mathcal{B}(\mathcal{H})\ .
\label{selfadjoint bilinear}
\end{equation}%
See, e.g., \cite[Eqs. (7.5)--(7.6)]{A70}.

\emph{Bilinear Hamiltonians} are then defined as bilinear elements
associated with \emph{self-adjoint} operators $H=H^{\ast }\in \mathcal{B}(%
\mathcal{H})$. They include all second quantizations of one-particle
Hamiltonians (see Equations (\ref{form1}) and (\ref{second quant II})), but
also models that are \emph{not gauge invariant} (see Equations (\ref{form2})
and (\ref{second quant III})). Important models in condensed matter physics,
like in the BCS theory of superconductivity, are bilinear Hamiltonians that
are \emph{not} gauge invariant.

To study all bilinear elements of self-dual CAR algebra, one can reduce the
analysis of bilinear elements to a special class of operators $H$:

\begin{lemma}[The set of bilinear elements of self--dual CAR algebra]
\label{lemma bilinear CAR}\mbox{ }\newline
Given a finite-dimensional self-dual CAR algebra $\mathrm{sCAR}(\mathcal{H},%
\mathfrak{A})$,%
\begin{equation*}
\left\{ \langle \mathrm{B},H\mathrm{B}\rangle +\lambda \mathfrak{1}:\lambda
\in \mathbb{C},\ H\in \mathcal{B}(\mathcal{H})\right\} =\left\{ \langle 
\mathrm{B},H\mathrm{B}\rangle +\lambda \mathfrak{1}:\lambda \in \mathbb{C},\
H\in \mathcal{B}(\mathcal{H}),\ H^{\ast }=-\mathfrak{A}H\mathfrak{A}\right\}
\ .
\end{equation*}
\end{lemma}

\begin{proof}
Observe that, for any $H\in \mathcal{B}(\mathcal{\mathcal{H})}$,%
\begin{equation}
\langle \mathrm{B},H\mathrm{B}\rangle +\langle \mathrm{B},\mathfrak{A}%
H^{\ast }\mathfrak{A}\mathrm{B}\rangle =\mathrm{Tr}_{\mathcal{H}}\left(
H\right) \mathfrak{1}  \label{equation idiote 2}
\end{equation}%
(cf. \cite[Eq. (7.7)]{A70}). This implies that, for any $H\in \mathcal{B}(%
\mathcal{\mathcal{H})}$,%
\begin{equation}
\langle \mathrm{B},H\mathrm{B}\rangle =\langle \mathrm{B},\mathbf{K}\left(
H\right) \mathrm{B}\rangle +\frac{1}{2}\mathrm{Tr}_{\mathcal{H}}\left(
H\right) \mathfrak{1}  \label{equation idiote 3}
\end{equation}%
with $\mathbf{K}$ being the linear map from $\mathcal{B}(\mathcal{\mathcal{H}%
)}$ to itself defined by%
\begin{equation}
\mathbf{K}\left( H\right) \doteq \frac{1}{2}\left( H-\mathfrak{A}H^{\ast }%
\mathfrak{A}\right) \text{ },\qquad H\in \mathcal{B}(\mathcal{\mathcal{H})}\
.  \label{equation idiote 3bis}
\end{equation}%
(This map already appears in \cite[Eqs. (4.16), (5.16)]{A68}.) Since, for
any $H\in \mathcal{B}(\mathcal{\mathcal{H})}$, 
\begin{equation}
\mathbf{K}\left( H\right) ^{\ast }=\mathbf{K}\left( H^{\ast }\right) =-%
\mathfrak{A}\mathbf{K}\left( H\right) \mathfrak{A}\ ,
\label{equation idiote 4}
\end{equation}%
the second equality of the lemma results from (\ref{equation idiote 3}).
\end{proof}

Therefore, by Lemma \ref{lemma bilinear CAR}, all of our analysis of
bilinear elements can be restricted, \emph{without loss of generality}, to
operators $H\in \mathcal{B}(\mathcal{H})$ satisfying $H^{\ast }=-\mathfrak{A}%
H\mathfrak{A}$. We call such operators \emph{self-dual operators}:

\begin{definition}[Self-dual operators]
\label{def one particle hamiltinian}\mbox{
}\newline
A self-dual operator on $(\mathcal{H},\mathfrak{A})$ is an operator $H\in 
\mathcal{B}(\mathcal{H})$ satisfying the equality $H^{\ast }=-\mathfrak{A}H%
\mathfrak{A}$. If, additionally, $H$ is self-adjoint, then we say that it is
a self-dual Hamiltonian on $(\mathcal{H},\mathfrak{A})$.
\end{definition}

Note that a self-dual operator $H$ has zero trace, i.e., 
\begin{equation*}
\mathrm{Tr}_{\mathcal{H}}\left( H\right) =\langle \mathrm{B},H\mathrm{B}%
\rangle +\langle \mathrm{B},\mathfrak{A}H^{\ast }\mathfrak{A}\mathrm{B}%
\rangle =0\ ,
\end{equation*}%
by Equation (\ref{equation idiote 2}).

We say that the basis projection $P$ (Definition \ref{def basis projection})
(block-) \textquotedblleft diagonalizes\textquotedblright\ the self-dual
operator $H\in \mathcal{B}(\mathcal{H})$ whenever%
\begin{equation}
H=\frac{1}{2}\left( PH_{P}P-P^{\bot }\mathfrak{A}H_{P}^{\ast }\mathfrak{A}%
P^{\bot }\right) \text{ },\qquad \text{with}\qquad H_{P}\doteq 2PHP\in 
\mathcal{B}(\mathfrak{h}_{P})\text{ }.  \label{kappabisbiskappabisbis}
\end{equation}%
In this situation, we also say that the basis projection $P$\ diagonalizes $%
\langle \mathrm{B},H\mathrm{B}\rangle $, similarly to \cite[Definition 5.1]%
{A68}: By Definition \ref{def Self--dual CAR Algebras} and Equation (\ref%
{kappabisbiskappabisbis}), for any orthonormal basis $\{\psi _{j}\}_{j\in J}$
of $\mathfrak{h}_{P}$,%
\begin{equation}
\langle \mathrm{B},H\mathrm{B}\rangle =2\sum\limits_{i,j\in J}\left\langle
\psi _{i},H\psi _{j}\right\rangle _{\mathcal{H}}\mathrm{B}\left( \psi
_{j}\right) \mathrm{B}\left( \psi _{i}\right) ^{\ast }+\mathrm{Tr}_{\mathcal{%
H}}\left( P^{\bot }HP^{\bot }\right) \mathfrak{1}\ .
\label{second quantizzation00}
\end{equation}%
Compare this equation with (\ref{kappa})-(\ref{kappabis}) and (\ref{second
quant II}).

By the spectral theorem, for any self-dual \emph{Hamiltonian} $H$ on $(%
\mathcal{H},\mathfrak{A})$, there is always a basis projection $P$
diagonalizing $H$. In quantum physics, as discussed in Section \ref%
{Self--dual CAR Algebras}, $\mathfrak{h}_{P}$ is in this case the \emph{%
one-particle Hilbert space} and $H_{P}$ the \emph{one-particle Hamiltonian}.
In particular, the first term of the right-hand side of (\ref{second
quantizzation00}) is the so-called \emph{second quantization} of $H_{P}$.

\subsection{Quasi-Free Dynamics and States}

Bilinear Hamiltonians are used to define so-called \emph{quasi-free}
dynamics: For any $H=H^{\ast }\in \mathcal{B}(\mathcal{H})$, we define the
continuous group $\{\tau _{t}\}_{t\in {\mathbb{R}}}$ of $\ast $-auto%
%TCIMACRO{\TeXButton{\-}{\-}}%
%BeginExpansion
\-%
%EndExpansion
morphisms of $\mathrm{sCAR}(\mathcal{H},\mathfrak{A})$ by%
\begin{equation*}
\tau _{t}(A)\doteq \mathrm{e}^{-it\langle \mathrm{B},H\mathrm{B}\rangle }A%
\mathrm{e}^{it\langle \mathrm{B},H\mathrm{B}\rangle }\ ,\qquad A\in \mathrm{%
sCAR}(\mathcal{H},\mathfrak{A}),\ t\in {\mathbb{R}}
\end{equation*}%
(see\ Definition \ref{def trace state copy(1)} and Equation (\ref{equation
idiote 2}). Provided $H$ is a self-dual Hamiltonian on $(\mathcal{H},%
\mathfrak{A})$ (Definition \ref{def one particle hamiltinian}), this group
is a quasi-free dynamics, that is, a continuous group of Bogoliubov $\ast $%
-automorphisms, as defined in Equation (\ref{Bogoliubov automorphism}). This
can be seen from the following assertion:

\begin{lemma}[From bilinear Hamiltonians to quasi-free dynamics]
\label{Lemma quasi-free}\mbox{
}\newline
Take any self-dual operator $H$ on $(\mathcal{H},\mathfrak{A})$ (Definition %
\ref{def one particle hamiltinian}). Then, for all $z\in \mathbb{C}$ and $%
\varphi \in \mathcal{H}$,%
\begin{equation*}
\exp \left( -\frac{z}{2}\langle \mathrm{B},H\mathrm{B}\rangle \right) 
\mathrm{B}\left( \varphi \right) ^{\ast }\exp \left( \frac{z}{2}\langle 
\mathrm{B},H\mathrm{B}\rangle \right) =\mathrm{B}\left( \mathrm{e}%
^{zH}\varphi \right) ^{\ast }\ .
\end{equation*}
\end{lemma}

\begin{proof}
The proof is completely standard. We outline it for completeness. As usual, $%
[B_{1},B_{2}]\doteq B_{1}B_{2}-B_{2}B_{1}$ is the commutator of two elements 
$B_{1},B_{2}\in \mathrm{sCAR}(\mathcal{H},\mathfrak{A})$.

Fix $H\in \mathcal{B}(\mathcal{H})$ and $\varphi \in \mathcal{H}$.
Straightforward computations using Definitions \ref{def Self--dual CAR
Algebras} and \ref{def trace state copy(1)}, together with the properties of
the antiunitary involution $\mathfrak{A}$, lead to the equality 
\begin{equation}
\frac{1}{2}\left[ \langle \mathrm{B},H\mathrm{B}\rangle ,\mathrm{B}\left(
\varphi \right) ^{\ast }\right] =-\mathrm{B}\left( \mathbf{K}\left( H\right)
\varphi \right) ^{\ast }\ ,  \label{tototototot}
\end{equation}%
where $\mathbf{K}$ is the linear map on $\mathcal{B}(\mathcal{H})$ defined
by (\ref{equation idiote 3bis}). This statement also refers to \cite[Eqs.
(4.15)--(4.16)]{A68}. In particular, if $H^{\ast }=-\mathfrak{A}H\mathfrak{A}
$ (cf. Definition \ref{def one particle hamiltinian}), then $\mathbf{K}%
\left( H\right) =H$ and, by (\ref{tototototot}), it follows that, for any $%
z\in \mathbb{C}$,%
\begin{equation*}
\partial _{z}\left\{ \exp \left( \frac{z}{2}\langle \mathrm{B},H\mathrm{B}%
\rangle \right) \mathrm{B}\left( \mathrm{e}^{zH}\varphi \right) ^{\ast }\exp
\left( -\frac{z}{2}\langle \mathrm{B},H\mathrm{B}\rangle \right) \right\}
=0\ ,
\end{equation*}%
which proves the lemma.
\end{proof}

Bilinear Hamiltonians are also used to define \emph{quasi-free} states.
States are positive and normalized linear functionals $\rho \in \mathrm{sCAR}%
(\mathcal{H},\mathfrak{A})^{\ast }$, i.e., $\rho (\mathfrak{1})=1$ and $\rho
(A^{\ast }A)\geq 0$ for all $A\in \mathrm{sCAR}(\mathcal{H},\mathfrak{A})$.
They are said to be \emph{quasi-free} when, for all $N\in \mathbb{N}_{0}$
and $\varphi _{0},\ldots ,\varphi _{2N}\in \mathcal{H}$,%
\begin{equation}
\rho \left( \mathrm{B}\left( \varphi _{0}\right) \cdots \mathrm{B}\left(
\varphi _{2N}\right) \right) =0\text{ },  \label{ass O0-00}
\end{equation}%
while, for all $N\in \mathbb{N}$ and $\varphi _{1},\ldots ,\varphi _{2N}\in 
\mathcal{H}$,%
\begin{equation}
\rho \left( \mathrm{B}\left( \varphi _{1}\right) \cdots \mathrm{B}\left(
\varphi _{2N}\right) \right) =\mathrm{Pf}\left[ \rho \left( \mathbb{O}%
_{k,l}\left( \mathrm{B}(\varphi _{k}),\mathrm{B}(\varphi _{l})\right)
\right) \right] _{k,l=1}^{2N}\ ,  \label{ass O0-00bis}
\end{equation}%
where 
\begin{equation*}
\mathbb{O}_{k,l}\left( A_{1},A_{2}\right) \doteq \left\{ 
\begin{array}{ccc}
A_{1}A_{2} & \text{for} & k<l\ , \\ 
-A_{2}A_{1} & \text{for} & k>l\ , \\ 
0 & \text{for} & k=l\ .%
\end{array}%
\right.
\end{equation*}%
In Equation (\ref{ass O0-00bis}), $\mathrm{Pf}$ is the usual Pfaffian
defined by 
\begin{equation}
\mathrm{Pf}\left[ M_{k,l}\right] _{k,l=1}^{2N}\doteq \frac{1}{2^{N}N!}%
\sum_{\pi \in \mathcal{S}_{2N}}\left( -1\right) ^{\pi
}\prod\limits_{j=1}^{N}M_{\pi \left( 2j-1\right) ,\pi \left( 2j\right) }
\label{Pfaffian}
\end{equation}%
for any $2N\times 2N$ skew-symmetric matrix $M\in \mathrm{Mat}\left( 2N,%
\mathbb{C}\right) $. Note that (\ref{ass O0-00bis}) is equivalent to the
definition given either in \cite[Definition 3.1]{A70} or in \cite[Equation
(6.6.9)]{EK98}.

Quasi-free states are therefore particular states\ that are uniquely defined
by two-point correlation functions, via (\ref{ass O0-00})-(\ref{ass O0-00bis}%
). In fact, a quasi-free state $\rho $ is uniquely defined by its so-called 
\emph{symbol}, that is, a positive operator $S_{\rho }\in \mathcal{B}(%
\mathcal{H})$ such that%
\begin{equation}
0\leq S_{\rho }\leq \mathbf{1}_{\mathcal{H}}\qquad \text{and}\qquad S_{\rho
}+\mathfrak{A}S_{\rho }\mathfrak{A=}\mathbf{1}_{\mathcal{H}}\text{ },
\label{symbol}
\end{equation}%
through the conditions%
\begin{equation}
\left\langle \varphi _{1},S_{\rho }\varphi _{2}\right\rangle _{\mathcal{H}%
}=\rho \left( \mathrm{B}(\varphi _{1})\mathrm{B}(\mathfrak{A}\varphi
_{2})\right) \ ,\qquad \varphi _{1},\varphi _{2}\in \mathcal{H}\ .
\label{symbolbis}
\end{equation}%
For more details on symbols of quasi-free states, see \cite[Lemma 3.2]{A70}%
\footnote{%
As already remarked above, \cite[Lemma 3.2]{A70} defines the map $\varphi
\mapsto \mathrm{B}(\varphi )$ to be linear and the corresponding definition
of symbols changes accordingly.}. Conversely, any self-adjoint operator
satisfying (\ref{symbol}) uniquely defines a quasi-free state through
Equation (\ref{symbolbis}). See \cite[Lemma 3.3]{A70}. In physics, $S_{\rho
} $ is called the \emph{one-particle density matrix} of the system.

Quasi-free states obviously depend on the choice of generators of the
self-dual CAR algebra. An example of a quasi-free state is provided by the
tracial state:

\begin{definition}[Tracial state]
\label{def trace state}\mbox{
}\newline
The tracial state $\mathrm{tr}\in \mathrm{sCAR}(\mathcal{H},\mathfrak{A}%
)^{\ast }$ is the quasi-free state with symbol $S_{\mathrm{tr}}\doteq \frac{1%
}{2}\mathbf{1}_{\mathcal{H}}$.
\end{definition}

\noindent This is the usual tracial state known for CAR\ algebra, as one can
see from Remark \ref{bound norm a CAR copy(2)}. It can be used to highlight
the relationship between quasi-free states and bilinear Hamiltonians:

\begin{lemma}[From Bilinear Hamiltonians to quasi-free states]
\label{Lemma quasi free state}\mbox{
}\newline
Take $\beta \in (0,\infty )$\ and any self-dual Hamiltonian $H$ on $(%
\mathcal{H},\mathfrak{A})$ (Definition \ref{def one particle hamiltinian}).
Then, the positive operator $(1+\mathrm{e}^{-\beta H})^{-1}$ satisfies
Condition (\ref{symbol}) and is the symbol of a quasi-free state $\rho _{H}$
satisfying%
\begin{equation}
\rho _{H}(A)=\frac{\mathrm{tr}\left( A\exp \left( \frac{\beta }{2}\langle 
\mathrm{B},H\mathrm{B}\rangle \right) \right) }{\mathrm{tr}\left( \exp
\left( \frac{\beta }{2}\langle \mathrm{B},H\mathrm{B}\rangle \right) \right) 
}\ ,\qquad A\in \mathrm{sCAR}(\mathcal{H},\mathfrak{A})\ .
\label{Gibbs states}
\end{equation}
\end{lemma}

\begin{proof}
Let $H\in \mathcal{B}(\mathcal{H})$ be such that $H=H^{\ast }=-\mathfrak{A}H%
\mathfrak{A}$. Then, by using a Taylor series, one verifies that the map 
\begin{equation*}
\beta \mapsto f\left( \beta \right) \doteq \mathfrak{A}(1+\exp (\beta
H))^{-1}\mathfrak{A-}(1+\exp (\beta \mathfrak{A}H\mathfrak{A}))^{-1}
\end{equation*}%
vanishes for small enough $\beta \in \mathbb{R}$ and, since this function is
real analytic on the whole real line $\mathbb{R}$, it can only be the zero
function on $\mathbb{R}$. Using elementary computations, it follows that $%
S=(1+\mathrm{e}^{-\beta H})^{-1}$ satisfies (\ref{symbol}) for any $\beta
\in \mathbb{R}$.

Note that the denominator of (\ref{Gibbs states}) is strictly positive
because the tracial state $\mathrm{tr}$ is faithful. Equation (\ref{Gibbs
states}) follows by observing that the states of both sides of this equality
are the \emph{unique} KMS state of the same quasi-free dynamics (cf. Lemma %
\ref{Lemma quasi-free} and \cite[Theorem 3]{A70}). We omit the details and
refer to \cite[Corollary 6.3]{A70} for the complete proof.
\end{proof}

\noindent The state $\rho _{H}$ is named the \emph{Gibbs state}, or thermal
equilibrium state, associated with the self-dual (one-particle) Hamiltonian $%
H$ on $(\mathcal{H},\mathfrak{A})$ at fixed $\beta \in (0,\infty )$. The
parameter $\beta \in (0,\infty )$ is physically interpreted as being the 
\emph{inverse temperature} of the fermion system.

\begin{remark}
\mbox{
}\newline
By Definition \ref{def basis projection}, any basis projection associated
with $(\mathcal{H},\mathfrak{A})$ can be seen as a symbol of a quasi-free
state on $\mathrm{sCAR}(\mathcal{H},\mathfrak{A})$. Such state is pure and
called a \emph{Fock state} \cite[Lemma 4.3]{A70}. Araki shows in \cite[%
Lemmata 4.5--4.6]{A70} that any quasi-free state can be seen as the
restriction of a quasi-free state on $\mathrm{sCAR}(\mathcal{H}\oplus 
\mathcal{H},\mathfrak{A}\oplus (-\mathfrak{A}))$, the symbol of which is a
basis projection associated with $(\mathcal{H}\oplus \mathcal{H},\mathfrak{A}%
\oplus (-\mathfrak{A}))$. This procedure is called purification of the
quasi-free state.
\end{remark}

The important issue of the present study is to show that so-called
Brydges-Kennedy tree expansions allow a construction of the logarithmic
moment generating function (\ref{equation generating funct}) of Gibbs states
associated with Hamiltonians of the form (\ref{form}), that is, 
\begin{equation}
\mathbf{H}=-\frac{1}{2}\langle \mathrm{B},H\mathrm{B}\rangle +W\ ,\qquad
W=W^{\ast }\in \mathrm{sCAR}(\mathcal{H},\mathfrak{A})^{\ast }\ ,
\label{formbis}
\end{equation}%
with $H\in \mathcal{B}(\mathcal{H})$ such that $H=H^{\ast }=-\mathfrak{A}H%
\mathfrak{A}$, provided $W$ is sufficiently small in an appropriate sense.

\section{Realization of Self-dual CAR Algebra as a Grassmann Algebra\label%
{Section Grassman}}

\subsection{Grassmann Algebra\label{Section Grassmann}}

\emph{Grassmann algebra}, also called \emph{exterior algebra}, is a
well-known mathematical structure, see, e.g., \cite[Appendix]{D73}, or \cite%
{FKT02} for an exposition in the context of QFT. For completeness and to fix
notation, we outline its construction: \medskip

\noindent \underline{(i):} Let $\mathcal{X}$ be a topological vector space.
Then, $\mathcal{X}^{\ast }$ denotes, as is usual, the space of all
continuous linear functionals on $\mathcal{X}$. For every $n\in {\mathbb{N}}$
and $x_{1}^{\ast },\ldots ,x_{n}^{\ast }\in \mathcal{X}^{\ast }$, we define
the completely antisymmetric $n$-linear form $x_{1}^{\ast }\wedge \cdots
\wedge x_{n}^{\ast }$ from $\mathcal{X}^{n}$ to ${\mathbb{C}}$ by 
\begin{equation}
x_{1}^{\ast }\wedge \cdots \wedge x_{n}^{\ast }(y_{1},\ldots ,y_{n})\doteq 
\mathrm{det}\left( (x_{k}^{\ast }(y_{l}))_{k,l=1}^{n}\right) \ ,\qquad
y_{1},\ldots ,y_{n}\in \mathcal{X}\ .  \label{eq mulit linear}
\end{equation}%
In particular, for any permutation $\pi $ of $n\in {\mathbb{N}}$ elements
with sign $(-1)^{\pi }$, 
\begin{equation}
x_{1}^{\ast }\wedge \cdots \wedge x_{n}^{\ast }\doteq (-1)^{\pi }x_{\pi
(1)}^{\ast }\wedge \cdots \wedge x_{\pi (n)}^{\ast }\ ,\qquad x_{1}^{\ast
},\ldots ,x_{n}^{\ast }\in \mathcal{X}^{\ast }\ .
\label{grassmana anticommute}
\end{equation}%
\medskip

\noindent \underline{(ii):} Let $\wedge ^{\ast 0}\mathcal{X}\doteq {\mathbb{C%
}}$, while, for all $n\in {\mathbb{N}}$, we use the definition 
\begin{equation}
\wedge ^{\ast n}\mathcal{X}\doteq \mathrm{lin}\{x_{1}^{\ast }\wedge \cdots
\wedge x_{n}^{\ast }:x_{1}^{\ast },\ldots ,x_{n}^{\ast }\in \mathcal{X}%
^{\ast }\}\ .  \label{directsum0}
\end{equation}%
We then define the vector space%
\begin{equation}
\wedge ^{\ast }\mathcal{X}\doteq \bigoplus\limits_{n=0}^{\infty }\wedge
^{\ast n}\mathcal{X}\ .  \label{direct sums}
\end{equation}%
Recall that the infinite direct sum of a family $\{\mathcal{X}_{n}\}_{n\in {%
\mathbb{N}}}$ of vector spaces, like in the above definition, is the
subspace of the product space $\prod\limits_{n=0}^{\infty }\mathcal{X}_{n}$,
the elements of which are sequences that eventually vanish.\medskip

\noindent \underline{(iii):} For $n,m\in {\mathbb{N}}_{0}$, $\xi \in \wedge
^{\ast n}\mathcal{X}$ and $\zeta \in \wedge ^{\ast m}\mathcal{X}$, their 
\emph{exterior product} $\xi \wedge \zeta \in \wedge ^{\ast n+m}\mathcal{X}$
is defined by%
\begin{equation}
\xi \wedge \zeta \left( x_{1},\ldots ,x_{n+m}\right) \doteq \frac{1}{n!m!}%
\sum_{\pi \in \mathcal{S}_{n+m}}\left( -1\right) ^{\pi }\xi \left( x_{\pi
(1)},\ldots ,x_{\pi (n)}\right) \zeta \left( x_{\pi (n+1)},\ldots ,x_{\pi
(n+m)}\right) \ ,  \label{product}
\end{equation}%
where $\mathcal{S}_{N}$ is the set of all permutations of $N\in \mathbb{N}$
elements. This prescription uniquely defines an associative product on $%
\wedge ^{\ast }\mathcal{X}$, which is consistent with the definition of the $%
n$-linear form (\ref{eq mulit linear}):%
\begin{equation*}
x_{1}^{\ast }\wedge \left( x_{2}^{\ast }\wedge \cdots \wedge x_{n}^{\ast
}\right) =x_{1}^{\ast }\wedge \cdots \wedge x_{n}^{\ast }\ ,\qquad n\in {%
\mathbb{N}},\ x_{1}^{\ast },\ldots ,x_{n}^{\ast }\in \mathcal{X}^{\ast }\ .
\end{equation*}%
Compare (\ref{product}) with (\ref{eq mulit linear}), using the Leibniz
formula to compute the determinant.

\begin{definition}[Grassmann algebra]
\label{Definition Grassmann}\mbox{
}\newline
The Grassmann algebra on a topological vector space\emph{\ }$\mathcal{X}$ is
the (associative and distributive)\emph{\ }algebra $(\wedge ^{\ast }\mathcal{%
X},+,\wedge )$.
\end{definition}

\begin{notation}
\label{Notation product}\mbox{
}\newline
When there is no risk of ambiguity, we use $x_{1}^{\ast }\wedge \cdots
\wedge x_{n}^{\ast }\equiv x_{1}^{\ast }\cdots x_{n}^{\ast }$ to denote
exterior products.
\end{notation}

For $n\in {\mathbb{N}}_{0}$, $\wedge ^{\ast n}\mathcal{X}$ is precisely the
subspace of elements of degree $n$ of the \emph{graded} algebra $\wedge
^{\ast }\mathcal{X}$. The unit of the Grassmann algebra $\wedge ^{\ast }%
\mathcal{X}$ is denoted by 
\begin{equation*}
\mathfrak{1}\doteq 1\in \wedge ^{\ast 0}\mathcal{X}\subset \wedge ^{\ast }%
\mathcal{X}
\end{equation*}%
and $[\xi ]_{0}$ stands for the zero-degree component of any element $\xi $
of $\wedge ^{\ast }\mathcal{X}$.

The exponential function in Grassmann algebra is important in the sequel. It
is defined, as is usual, by 
\begin{equation}
\exp (\xi )\doteq \mathfrak{1}+\sum\limits_{k=1}^{\infty }\frac{\xi ^{k}}{k!}%
\ ,\qquad \xi \in \wedge ^{\ast }\mathcal{X}\ .  \label{exponentiel}
\end{equation}%
This function is well-defined because, by definition of $\wedge ^{\ast }%
\mathcal{X}$, $\xi \in \wedge ^{\ast }\mathcal{X}$ is contained in some
finite-dimensional subalgebra of $\wedge ^{\ast }\mathcal{X}$. Indeed, by
finite dimensionality, for any norm on the space $\wedge ^{\ast }\mathcal{X}$
and any $\xi \in \wedge ^{\ast }\mathcal{X}$, there is a constant $D_{\xi
}<\infty $ such that%
\begin{equation}
\left\Vert \xi ^{k}\right\Vert _{\wedge ^{\ast }\mathcal{X}}\leq (D_{\xi
})^{k}\ ,\qquad k\in \mathbb{N}\ .  \label{toto eq}
\end{equation}%
Hence, the series defining $\exp (\xi )$ is absolutely convergent in the
normed space $\wedge ^{\ast }\mathcal{X}$.

In the sequel, we use $\mathcal{X=\mathcal{H}}$. In this case, the linear
spaces $\mathrm{sCAR}(\mathcal{H},\mathfrak{A})$ and $\wedge ^{\ast }%
\mathcal{\mathcal{H}}$ are isomorphic to each other because they have
exactly the same dimension:%
\begin{equation}
\mathrm{dim}\left( \mathrm{sCAR}(\mathcal{H},\mathfrak{A})\right) =2^{%
\mathrm{dim}\mathcal{\mathcal{H}}}=\mathrm{dim}\left( \wedge ^{\ast }%
\mathcal{\mathcal{H}}\right) <\infty \ .  \label{dimension fock}
\end{equation}%
However, $(\wedge ^{\ast }\mathcal{\mathcal{H}},+,\wedge )$ is not
isomorphic to a self-dual CAR algebra over $\mathcal{\mathcal{H}}$, because
of the CAR (\ref{CAR Grassmann III}). We thus define below a new product and
an involution to make $\wedge ^{\ast }\mathcal{\mathcal{H}}$ a self-dual CAR
algebra $\mathrm{sCAR}(\mathcal{H},\mathfrak{A})$. This explicit
construction uses Berezin derivatives and integrals, which are meanwhile
pivotal to obtain the Brydges-Kennedy tree expansion of the G\"{a}%
rtner-Ellis generating functions.

\subsection{Berezin Integral}

As a preliminary step, we introduce the notion of \emph{Berezin derivatives}%
. For any $\varphi \in \mathcal{H}$, it is the linear operator $\delta
/\delta \varphi $ acting on the Grassmann algebra $\wedge ^{\ast }\mathcal{%
\mathcal{H}}$ that is uniquely defined by the conditions 
\begin{equation}
\frac{\delta }{\delta \varphi }\tilde{\varphi}=\left\langle \varphi ,\tilde{%
\varphi}\right\rangle _{\mathcal{H}}\mathfrak{1}\qquad \text{and}\qquad 
\frac{\delta }{\delta \varphi }\xi _{1}\xi _{2}=\left( \frac{\delta }{\delta
\varphi }\xi _{1}\right) \wedge \xi _{2}+\left( -1\right) ^{n}\xi _{1}\wedge
\left( \frac{\delta }{\delta \varphi }\xi _{2}\right) \ ,
\label{derivation plus1}
\end{equation}%
for any $\tilde{\varphi}\in \mathcal{H}$ and element $\xi _{1}\in \wedge
^{\ast n}\mathcal{H}$ of degree $n\in \mathbb{N}$, and all $\xi _{2}\in
\wedge ^{\ast }\mathcal{H}$. In particular, the map $\varphi \mapsto \delta
/\delta \varphi $ from $\mathcal{\mathcal{H}}$ to $\mathcal{B}(\wedge ^{\ast
}\mathcal{\mathcal{H)}}$ is antilinear and satisfies 
\begin{equation}
\frac{\delta }{\delta \varphi _{1}}\frac{\delta }{\delta \varphi _{2}}=-%
\frac{\delta }{\delta \varphi _{2}}\frac{\delta }{\delta \varphi _{1}}\
,\qquad \varphi _{1},\varphi _{2}\in \mathcal{H}\ .  \label{derivation plus2}
\end{equation}%
Note that $\delta /\delta \varphi $ is an annihilation operator on $\wedge
^{\ast }\mathcal{\mathcal{H}}$, viewed as the fermionic Fock space $\mathcal{%
F}_{\mathcal{H}^{\ast }}$. See Section \ref{Fock}. Viewed this way, $\delta
/\delta \varphi \doteq (\mathfrak{A}\varphi )\lrcorner $ for any $\varphi
\in \mathcal{H}$.

Recall that, for each $k\in \mathbb{N}_{0}$, $\mathcal{X}^{(k)}$ denotes a
copy of some vector space $\mathcal{X}$ and the corresponding copy of $\xi
\in \mathcal{X}$ is written as $\xi ^{(k)}$, see Notation \ref{remark
constant} (ii). For any $K\subset \{0,\ldots ,N\}$ with $N\in \mathbb{N}_{0}$%
, we identify $\wedge ^{\ast }(\oplus _{k\in K}\mathcal{H}^{(k)})$ with the
Grassmann subalgebra of $\wedge ^{\ast }(\oplus _{k=0}^{N}\mathcal{H}^{(k)})$
generated by the union%
\begin{equation*}
\bigcup\limits_{k\in K}\left\{ \varphi ^{(k)}:\varphi \in \mathcal{H}%
\right\} \ .
\end{equation*}%
We meanwhile identify $\wedge ^{\ast }\mathcal{H}^{(0)}$ with the Grassmann
algebra $\wedge ^{\ast }\mathcal{H}$, i.e.,%
\begin{equation}
\wedge ^{\ast }\mathcal{H}^{(0)}\equiv \wedge ^{\ast }\mathcal{H}\ .
\label{identification}
\end{equation}%
We are now in a position to define the so-called \emph{Berezin integral} 
\cite[Section I.3]{Berezin}:

\begin{definition}[Berezin integral]
\label{Grassmann Integral}\mbox{ }\newline
Fix $N\in \mathbb{N}_{0}$ and a basis projection $P$\ (Definition \ref{def
basis projection}) with $\{\psi _{i}\}_{i\in J}$ being any orthonormal basis
of its range $\mathfrak{h}_{P}$. For all $k\in \{0,\ldots ,N\}$, we define
the linear map 
\begin{equation*}
\int_{P}\mathrm{d}\left( \mathcal{H}^{(k)}\right) :\wedge ^{\ast }\left(
\oplus _{q=0}^{N}\mathcal{H}^{(q)}\right) \rightarrow \wedge ^{\ast }\left(
\oplus _{q\in \{0,\ldots ,N\}\backslash \{k\}}\mathcal{H}^{(q)}\right)
\end{equation*}%
by 
\begin{equation*}
\int_{P}\mathrm{d}\left( \mathcal{H}^{(k)}\right) \doteq \prod\limits_{i\in
J}\left( \frac{\delta }{\delta \psi _{i}^{(k)}}\frac{\delta }{\delta ((%
\mathfrak{A}\psi _{i})^{(k)})}\right) \ .
\end{equation*}
\end{definition}

\noindent For $N=0$, the Berezin integral defines a linear form from $\wedge
^{\ast }\mathcal{H}^{(0)}\equiv \wedge ^{\ast }\mathcal{H}$ to $\mathbb{C}%
\mathfrak{1}\equiv \mathbb{C}$. Recall that usual integrals can be seen as
linear forms on an algebra of continuous functions. This partially justifies
the use of the term \textquotedblleft integral\textquotedblright\ for the
map $\int_{P}\mathrm{d}\left( \mathcal{H}^{(k)}\right) $.

The product defining the Berezin integral, with index set $J$, \emph{does
not depend} on the order of its terms, because of (\ref{derivation plus2}).
For any fixed basis projection $P$, the Berezin integral \emph{also does not
depend} on the particular choice of the orthonormal basis $\{\psi
_{i}\}_{i\in J}$ of $\mathfrak{h}_{P}$: Considering the case $N=0$ without
loss of generality, for any unitary transformation $U$ on $\mathfrak{h}_{P}$%
, observe that, for all $i\in J$,%
\begin{equation}
\frac{\delta }{\delta (U\psi _{i})}=\sum\limits_{j\in J}\langle U\psi
_{i},\psi _{j}\rangle _{\mathcal{H}}\frac{\delta }{\delta \psi _{j}}\qquad 
\text{and}\qquad \frac{\delta }{\delta (\mathfrak{A}U\psi _{i})}%
=\sum\limits_{j\in J}\langle \psi _{j},U\psi _{i}\rangle _{\mathcal{H}}\frac{%
\delta }{\delta \left( \mathfrak{A}\psi _{j}\right) }\ .
\label{inequality facile}
\end{equation}%
Therefore, by using the fact that Berezin derivatives anticommute with each
other, one directly computes that%
\begin{equation}
\prod\limits_{i\in J}\left( \frac{\delta }{\delta (U\psi _{i})}\frac{\delta 
}{\delta (\mathfrak{A}U\psi _{i})}\right) =\prod\limits_{i\in J}\left( \frac{%
\delta }{\delta (\psi _{i})}\frac{\delta }{\delta (\mathfrak{A}\psi _{i})}%
\right) \ .  \label{equality}
\end{equation}

For two different basis projections, the corresponding Berezin integrals
can, at most, differ by a sign:

\begin{lemma}[Orientation of Berezin integrals]
\label{Lemma Berezin orientation}\mbox{
}\newline
For any basis projections $P_{1},P_{2}\in \mathcal{B}(\mathcal{H})$, there
exists $\varepsilon \equiv \varepsilon (P_{1},P_{2})\in \left\{ -1,1\right\} 
$\ such that, for all $N\in \mathbb{N}_{0}$ and $k\in \{0,\ldots ,N\}$, 
\begin{equation*}
\int_{P_{1}}\mathrm{d}\left( \mathcal{H}^{(k)}\right) =\varepsilon
\int_{P_{2}}\mathrm{d}\left( \mathcal{H}^{(k)}\right) \text{ }.
\end{equation*}
\end{lemma}

\begin{proof}
By \cite[Lemma 3.6]{A68}, for any pair of basis projections $P_{1},P_{2}\in 
\mathcal{B}(\mathcal{H})$ there is a (generally not unique) Bogoliubov
transformation $U$ such that $P_{2}=UP_{1}U^{\ast }$. In particular, by
using orthonormal bases $\{\psi _{i}^{(n)}\}_{i\in J}$ of $\mathfrak{h}%
_{P_{n}}$, $n=1,2$, and identities similar to (\ref{inequality facile}) one
gets 
\begin{equation*}
\int_{P_{1}}\mathrm{d}\left( \mathcal{H}^{(k)}\right) =\mathrm{det}%
(U)\int_{P_{2}}\mathrm{d}\left( \mathcal{H}^{(k)}\right)
\end{equation*}%
for some Bogoliubov transformation $U$. Therefore, by (\ref{orientation}),
the assertion follows.
\end{proof}

\noindent We therefore say that the corresponding Berezin integrals, and
hence the associated Basis projections, have the same \emph{orientation} if $%
\varepsilon =1$. Note that this statement shows that the orientation of any
Bogoliubov transformation $U$ such that $P_{2}=UP_{1}U^{\ast }$ only depends
on the basis projections $P_{1},P_{2}$.

\subsection{Bilinear Elements of Grassmann Algebra}

By using Notation \ref{remark constant} (ii), we define bilinear elements of
Grassmann algebra in a similar way to bilinear elements of self-dual CAR
algebra (Definition \ref{def trace state copy(1)}):

\begin{definition}[Bilinear elements of Grassmann algebras]
\label{def1+}\mbox{
}\newline
Let $\mathcal{H}$ be any finite-dimensional self-dual Hilbert space with
antiunitary involution $\mathfrak{A}$. \newline
\emph{(i)} Fix an orthonormal basis $\{\psi _{i}\}_{i\in I}$ of $\mathcal{H}$
and define, for all $H\in \mathcal{B}(\mathcal{H})$, a bilinear element of
the Grassmann algebra $\wedge ^{\ast }\mathcal{\mathcal{H}}$ by 
\begin{equation*}
\langle \mathcal{H},H\mathcal{H}\rangle \doteq \sum\limits_{i,j\in
I}\left\langle \psi _{i},H\psi _{j}\right\rangle _{\mathcal{H}}\left( 
\mathfrak{A}\psi _{j}\right) \wedge \psi _{i}\ .
\end{equation*}%
\emph{(ii)} Fix a basis projection $P$ associated with $(\mathcal{H},%
\mathfrak{A})$ and an orthonormal basis $\{\psi _{j}\}_{j\in J}$ of its
range $\mathfrak{h}_{P}$. Given $k,l\in \mathbb{N}_{0}$, 
\begin{equation*}
\langle \mathfrak{h}_{P}^{(k)},\mathfrak{h}_{P}^{(l)}\rangle \doteq
\sum\limits_{j\in J}\left( \mathfrak{A}\psi _{j}\right) ^{(k)}\wedge \psi
_{j}^{(l)}\ .
\end{equation*}
\end{definition}

\noindent This definition \emph{does not depend} on the particular choice of
the orthonormal basis. Compare Definition \ref{def1+} with Definition \ref%
{def trace state copy(1)}. Additionally, note that 
\begin{equation*}
\langle \mathcal{H},H\mathcal{H}\rangle =\sum\limits_{i,j\in I}\left\langle
\psi _{i},H\mathfrak{A}\psi _{j}\right\rangle _{\mathcal{H}}\psi _{j}\wedge
\psi _{i}=\sum\limits_{i,j\in I}\left\langle \mathfrak{A}\psi _{i},H\psi
_{j}\right\rangle _{\mathcal{H}}\left( \mathfrak{A}\psi _{j}\right) \wedge 
\mathfrak{A}\psi _{i}
\end{equation*}%
for all $H\in \mathcal{B}(\mathcal{H})$.

For any basis projection $P$ diagonalizing a self-dual operator $H$ on $(%
\mathcal{H},\mathfrak{A})$ (see Definition \ref{def one particle hamiltinian}
and (\ref{kappabisbiskappabisbis})) and any orthonormal basis $\{\psi
_{j}\}_{j\in J}$ of $\mathfrak{h}_{P}$,%
\begin{equation}
\langle \mathcal{H},H\mathcal{H}\rangle =2\sum\limits_{i,j\in J}\left\langle
\psi _{i},H\psi _{j}\right\rangle _{\mathcal{H}}\left( \mathfrak{A}\psi
_{j}\right) \wedge \psi _{i}\ .  \label{remark bilinear}
\end{equation}%
Compare this assertion with the first term of the right-hand side of
Equation (\ref{second quantizzation00}). Notice also that 
\begin{equation*}
\langle \mathcal{H},\mathbf{K}\left( P\right) \mathcal{H}\rangle =\frac{1}{2}%
\langle \mathcal{H},\left( P-P^{\bot }\right) \mathcal{H}\rangle =\langle 
\mathfrak{h}_{P},\mathfrak{h}_{P}\rangle \ .
\end{equation*}%
See Equation (\ref{equation idiote 3bis}).

Bilinear Hamiltonians of self-dual CAR algebra are used to fix a quasi-free
state, as explained in Lemma \ref{Lemma quasi free state}. Similarly, we
employ the exponential function (\ref{exponentiel}) and bilinear elements of
Grassmann algebra to define \emph{Gaussian Berezin integrals}:

For any basis projection $P$ diagonalizing a self-dual operator $H$ on $(%
\mathcal{H},\mathfrak{A})$, elementary computations yield%
\begin{equation}
\int_{P}\mathrm{d}\left( \mathcal{H}\right) \mathrm{e}^{\langle \mathcal{H},H%
\mathcal{H}\rangle }=\mathrm{det}\left( H_{P}\right) \ .
\label{equation a la con}
\end{equation}%
In particular, $\mathrm{det}\left( H_{P}\right) $ only depends on $H$ and
the orientation of $P$, by Lemma \ref{Lemma Berezin orientation}. Gaussian
Berezin integrals are then defined as follows:

\begin{definition}[Gaussian Berezin integrals]
\label{gaussian integral}\mbox{ }\newline
Let $C\in \mathcal{B}(\mathcal{H})$ be any invertible self-dual operator.
The Gaussian Berezin integral with covariance $C\in \mathcal{B}(\mathcal{H})$
is the linear map $\int \mathrm{d\mu }_{C}\left( \mathcal{H}\right) $ from $%
\wedge ^{\ast }\mathcal{H}$ to $\mathbb{C}\mathfrak{1}$ defined by 
\begin{equation*}
\int \mathrm{d\mu }_{C}\left( \mathcal{H}\right) \xi \doteq \mathrm{det}%
\left( \frac{C_{P}}{2}\right) \int_{P}\mathrm{d}\left( \mathcal{H}\right) 
\mathrm{e}^{\frac{1}{2}\langle \mathcal{H},C^{-1}\mathcal{H}\rangle }\wedge
\xi \text{ },\quad \xi \in \wedge ^{\ast }\mathcal{H}\,,
\end{equation*}%
where $P$ is any basis projection diagonalizing $C$.
\end{definition}

Note that $C^{-1}$ is a self-dual operator whenever $C$ is invertible and
self-dual. In this case, if $P$ diagonalizes $C$ then it also diagonalizes $%
C^{-1}$. Observe in the above definition that $\left( C_{P}/2\right)
^{-1}=(C^{-1})_{P}/2$.

Remark further that, by Lemma \ref{Lemma Berezin orientation} and Equation (%
\ref{equation a la con}), Gaussian Berezin integrals \emph{do not depend} on
the particular choice of the basis projection $P$ in Definition \ref%
{gaussian integral}. They have properties that are reminiscent of those of
quasi-free states (see, e.g., (\ref{ass O0-00})-(\ref{ass O0-00bis})):

\begin{proposition}[Gaussian Berezin integrals as Pfaffians]
\label{gaussian integral properties}\mbox{ }\newline
Let $C\in \mathcal{B}(\mathcal{H})$ be any invertible self-dual operator.
Then, $\int \mathrm{d\mu }_{C}\left( \mathcal{H}\right) \mathfrak{1}=%
\mathfrak{1}$ while, for all $N\in \mathbb{N}_{0}$ and $\varphi _{0},\ldots
,\varphi _{2N}\in \mathcal{H}$, 
\begin{equation}
\int \mathrm{d\mu }_{C}\left( \mathcal{H}\right) \varphi _{0}\cdots \varphi
_{2N}=0\ \ \text{and}\ \ \int \mathrm{d\mu }_{C}\left( \mathcal{H}\right)
\varphi _{1}\cdots \varphi _{2N}=\mathrm{Pf}\left[ \left\langle \mathfrak{A}%
\varphi _{k},C\varphi _{l}\right\rangle _{\mathcal{H}}\right] _{k,l=1}^{2N}%
\mathfrak{1}\ ,  \label{(holds)}
\end{equation}%
where $\mathrm{Pf}$ is the Pfaffian defined, for any skew-symmetric matrix,
by (\ref{Pfaffian}).
\end{proposition}

\begin{proof}
This kind of identity is well-known, see, e.g., \cite[Proposition 1.19]%
{FKT02}. We give its proof for completeness. Observe that $\left[
\left\langle \mathfrak{A}\varphi _{k},C\varphi _{l}\right\rangle _{\mathcal{H%
}}\right] _{k,l=1}^{2N}$ is a skew-symmetric matrix because $C\in \mathcal{B}%
(\mathcal{H})$ is a self-dual operator. $\int \mathrm{d\mu }_{C}\left( 
\mathcal{H}\right) \mathfrak{1}=\mathfrak{1}$ is an obvious consequence of (%
\ref{equation a la con}), while the first identity of (\ref{(holds)})
directly follows from the definition of Berezin integrals. The proof of the
second identity of (\ref{(holds)}) is less direct. It is performed in three
steps for any invertible self-dual operator $C\in \mathcal{B}(\mathcal{H})$
and basis projection $P$ diagonalizing $C$: \medskip

\noindent \underline{Step 1:} For any orthonormal basis $\{\psi _{i}\}_{i\in
I}$ of $\mathcal{H}$, let 
\begin{equation}
\langle \mathcal{H}^{(0)},\mathcal{H}^{(1)}\rangle \doteq \sum\limits_{i\in
I}(\mathfrak{A}\psi _{i}^{(0)})\wedge \psi _{i}^{(1)}=-\sum\limits_{i\in I}(%
\mathfrak{A}\psi _{i}^{(1)})\wedge \psi _{i}^{(0)}\ .  \label{lement o,1}
\end{equation}%
Observe that this Grassmann algebra element does not depend on the special
choice of the orthonormal basis. Moreover, 
\begin{equation*}
\varphi ^{(1)}=\frac{\delta }{\delta (\mathfrak{A}\varphi ^{(0)})}\langle 
\mathcal{H}^{(0)},\mathcal{H}^{(1)}\rangle \ ,\qquad \varphi \in \mathcal{H}%
\ .
\end{equation*}%
It follows that, for all $N\in \mathbb{N}$ and $\varphi _{1},\ldots ,\varphi
_{2N}\in \mathcal{H}$, 
\begin{equation}
\int \mathrm{d\mu }_{C}\left( \mathcal{H}\right) \varphi _{1}\cdots \varphi
_{2N}=\left[ \left( \prod\limits_{k=1}^{2N}\frac{\delta }{\delta (\mathfrak{A%
}\varphi _{k}^{(0)})}\right) \int \mathrm{d\mu }_{C}\left( \mathcal{H}%
^{(1)}\right) \mathrm{e}^{\langle \mathcal{H}^{(0)},\mathcal{H}^{(1)}\rangle
}\right] _{0}\ ,  \label{B1}
\end{equation}%
recalling that $[\xi ]_{0}$ denotes the zero-degree component of any element 
$\xi $ of the Grassmann (graded) algebra $\wedge ^{\ast }\mathcal{\mathcal{H}%
}\equiv \wedge ^{\ast }\mathcal{H}^{(0)}$. \medskip

\noindent \underline{Step 2:} For $s\in \left\{ 0,1\right\} $, define the
Grassmann algebra element%
\begin{equation*}
\mathbf{A}_{s}\doteq \frac{1}{2}\sum\limits_{i,j\in I}\left\langle \psi
_{i},C^{-1}\psi _{j}\right\rangle _{\mathcal{H}}\Phi _{s,j}^{(1,0)}\wedge
\Psi _{s,i}^{(1,0)}+\sum\limits_{i\in I}\left( \mathfrak{A}\psi
_{i}^{(0)}\right) \wedge \Psi _{s,i}^{(1,0)}\ ,
\end{equation*}%
where, for any $i\in I$,%
\begin{equation*}
\Psi _{s,i}^{(1,0)}\doteq \psi _{i}^{(1)}+s\sum\limits_{j\in I}\left\langle
\psi _{j},C\psi _{i}\right\rangle _{\mathcal{H}}\psi _{j}^{(0)}\ ,\quad \Phi
_{s,i}^{(1,0)}\doteq \mathfrak{A}\psi _{i}^{(1)}-s\sum\limits_{j\in
I}\left\langle \psi _{i},C\psi _{j}\right\rangle _{\mathcal{H}}\mathfrak{A}%
\psi _{j}^{(0)}\ .
\end{equation*}%
Using Definition \ref{def1+} (i) and (\ref{lement o,1}), for $s\in \left\{
0,1\right\} $,%
\begin{equation*}
\mathbf{A}_{s}=\frac{1}{2}\langle \mathcal{H}^{(1)},C^{-1}\mathcal{H}%
^{(1)}\rangle +\left( 1-s\right) \langle \mathcal{H}^{(0)},\mathcal{H}%
^{(1)}\rangle +\frac{s}{2}\langle \mathcal{H}^{(0)},C\mathcal{H}%
^{(0)}\rangle \ .
\end{equation*}%
Direct computations then show that, for any basis projection $P$,%
\begin{equation*}
\int_{P}\mathrm{d}\left( \mathcal{H}^{(1)}\right) \mathrm{e}^{\mathbf{A}%
_{0}}=\int_{P}\mathrm{d}\left( \mathcal{H}^{(1)}\right) \mathrm{e}^{\mathbf{A%
}_{1}}\ .
\end{equation*}%
Combined with (\ref{equation a la con}) and (\ref{B1}), these last two
equations yield 
\begin{equation}
\int \mathrm{d\mu }_{C}\left( \mathcal{H}\right) \varphi _{1}\cdots \varphi
_{2N}=\frac{1}{2^{N}N!}\left( \prod\limits_{k=1}^{2N}\frac{\delta }{\delta (%
\mathfrak{A}\varphi _{k})}\right) \langle \mathcal{H},C\mathcal{H}\rangle
^{N}  \label{B2}
\end{equation}%
for any $N\in \mathbb{N}$ and $\varphi _{1},\ldots ,\varphi _{2N}\in 
\mathcal{H}$.

\noindent \underline{Step 3:} To compute the right-hand side of (\ref{B2}),
it is convenient to take $\varphi _{1},\ldots ,\varphi _{2N}$ from an
orthonormal basis of $\mathcal{H}$. It suffices to consider this special
case because, for any $N\in \mathbb{N}_{0}$, the quantities%
\begin{equation}
\int \mathrm{d\mu }_{C}\left( \mathcal{H}\right) \varphi _{1}\cdots \varphi
_{2N}\quad \text{and}\quad \mathrm{Pf}\left[ \left\langle \mathfrak{A}%
\varphi _{k},C\varphi _{l}\right\rangle _{\mathcal{H}}\right] _{k,l=1}^{2N}%
\mathfrak{1}  \label{(expression)}
\end{equation}%
are linear with respect to $\varphi _{1},\ldots ,\varphi _{2N}\in \mathcal{H}
$. So, fix an orthonormal basis $\{\psi _{i}\}_{i\in I}$ of $\mathcal{H}$
and any index set $\{i_{1},\ldots ,i_{2N}\}\subset I$ for $N\in \mathbb{N}$.
Then, we infer from (\ref{lement o,1}) and (\ref{B2}) that 
\begin{eqnarray*}
&&\int \mathrm{d\mu }_{C}\left( \mathcal{H}\right) \psi _{i_{1}}\cdots \psi
_{i_{2N}} \\
&=&\frac{1}{2^{N}N!}\left( \prod\limits_{k=1}^{2N}\frac{\delta }{\delta (%
\mathfrak{A}\psi _{i_{k}})}\right) \sum_{\pi \in \mathcal{S}%
_{2N}}\prod\limits_{k=1}^{N}\left\langle \mathfrak{A}\psi _{i_{\pi \left(
2k-1\right) }},C\psi _{i_{\pi \left( 2k\right) }}\right\rangle _{\mathcal{H}%
}(\mathfrak{A}\psi _{i_{\pi \left( 2k\right) }})\wedge (\mathfrak{A}\psi
_{i_{\pi \left( 2k-1\right) }}) \\
&=&\left( \prod\limits_{k=1}^{2N}\frac{\delta }{\delta (\mathfrak{A}\psi
_{i_{k}})}\right) \sum_{\pi \in \mathcal{S}_{2N}}\frac{\left( -1\right)
^{\pi }}{2^{N}N!}\prod\limits_{k=1}^{N}\left\langle \mathfrak{A}\psi
_{i_{\pi \left( 2k-1\right) }},C\psi _{i_{\pi \left( 2k\right)
}}\right\rangle _{\mathcal{H}}(\mathfrak{A}\psi _{i_{2\left( N-k+1\right)
}})\wedge (\mathfrak{A}\psi _{i_{2\left( N-k+1\right) -1}})\ ,
\end{eqnarray*}%
which yields the Pfaffian we are looking for.
\end{proof}

\begin{corollary}[Gaussian Berezin integrals as determinants]
\label{gaussian integral properties copy(1)}\mbox{ }\newline
Let $C\in \mathcal{B}(\mathcal{H})$ be any invertible self-dual operator and
fix any basis projection $P$ diagonalizing $C$. Then, for all $N\in \mathbb{N%
}$ and $\varphi _{1},\ldots ,\varphi _{2N}\in \mathfrak{h}_{P}$,%
\begin{equation*}
\int \mathrm{d\mu }_{C}\left( \mathcal{H}\right) \left( \mathfrak{A}\varphi
_{1}\right) \cdots \left( \mathfrak{A}\varphi _{N}\right) \varphi
_{2N}\cdots \varphi _{N+1}=\mathrm{det}\left[ \left\langle \varphi
_{k},C\varphi _{N+l}\right\rangle _{\mathcal{H}}\right] _{k,l=1}^{N}%
\mathfrak{1}\ .
\end{equation*}
\end{corollary}

\begin{proof}
As, by assumption, $P$ is a basis projection diagonalizing the self-dual
operator $C$, this follows from Proposition \ref{gaussian integral
properties} together with the following identity for the Pfaffian: 
\begin{equation*}
\mathrm{Pf}\left( 
\begin{array}{cc}
0 & M \\ 
-M^{\mathrm{t}} & 0%
\end{array}%
\right) =\left( -1\right) ^{N\left( N-1\right) /2}\mathrm{det}\left(
M\right) \ ,
\end{equation*}%
for any matrix $M\in \mathrm{Mat}\left( N,\mathbb{C}\right) $, $N\in \mathbb{%
N}$, where $M^{\mathrm{t}}$ denotes its transpose.
\end{proof}

Compare Proposition \ref{gaussian integral properties} with Equations (\ref%
{ass O0-00})-(\ref{ass O0-00bis}) and Lemma \ref{Lemma quasi free state}. In
fact, the strong analogy between bilinear elements of self-dual CAR algebra
together with their associated quasi-free states and bilinear elements of
Grassmann algebra together with their Gaussian Berezin integrals is
essential to derive Pfaffian bounds in Section \ref{sect det bounds}.

\subsection{From Grassmann Algebra to Self-Dual CAR Algebra\label{Section
realisation explicit}}

We can now define a \textquotedblleft circle\textquotedblright\ product $%
\circ $ and an involution $^{\ast }$ on $\wedge ^{\ast }\mathcal{\mathcal{H}}
$ to make a self-dual CAR algebra (Definition \ref{def Self--dual CAR
Algebras}) out of this space . To this end, we take any basis projection $P$
with range $\mathfrak{h}_{P}$. See Definition \ref{def basis projection} and
Equation (\ref{definition H bar}). Then, for all $i,j,k,l\in \mathbb{N}_{0}$%
, we define by 
\begin{equation}
\varkappa _{(i,j)}^{(k,l)}:\wedge ^{\ast }(\mathfrak{h}_{P}^{(i)}\oplus 
\mathfrak{h}_{P}^{\ast (j)})\rightarrow \wedge ^{\ast }(\mathfrak{h}%
_{P}^{(k)}\oplus \mathfrak{h}_{P}^{\ast (l)})  \label{def kappa1}
\end{equation}%
the unique isomorphism of linear spaces such that $\varkappa
_{(i,j)}^{(k,l)}(z\mathfrak{1})=z\mathfrak{1}$ for $z\in {\mathbb{C}}$ and,
for any $m,n\in \mathbb{N}_{0}$ so that $m+n\geq 1$, and all $\varphi
_{1},\ldots ,\varphi _{m+n}\in \mathfrak{h}_{P}$, 
\begin{equation}
\varkappa _{(i,j)}^{(k,l)}\left( (\mathfrak{A}\varphi _{1})^{(i)}\cdots (%
\mathfrak{A}\varphi _{m})^{(i)}\varphi _{m+1}^{(j)}\cdots \varphi
_{m+n}^{(j)}\right) =(\mathfrak{A}\varphi _{1})^{(k)}\cdots (\mathfrak{A}%
\varphi _{m})^{(k)}\varphi _{m+1}^{(l)}\cdots \varphi _{m+n}^{(l)}
\label{def kappa2}
\end{equation}%
with $\varphi _{1}\wedge \varphi _{2}\equiv \varphi _{1}\varphi _{2}$ (see
Notation \ref{remark constant} (ii) and \ref{Notation product}). Here, if $%
m=0$, then there is no $\mathfrak{A}\varphi $ in the above equation. Mutatis
mutandis for $n=0$. Note that $\varkappa _{(i,j)}^{(k,l)}$ strongly depends
on the choice of the basis projection $P$\ even if we overlook the symbol $P$
in the notation of this isomorphism.

By using the objects defined by (\ref{exponentiel}), Definition \ref{def1+}
and (\ref{def kappa1})-(\ref{def kappa2}), as well as Berezin integrals
(Definition \ref{Grassmann Integral}), we introduce a new product on $\wedge
^{\ast }\mathcal{\mathcal{H}}$ depending on a basis projection:

\begin{definition}[Circle products with respect to basis projections]
\label{definition star}\mbox{
}\newline
Fix a basis projection $P$ with range $\mathfrak{h}_{P}$ and recall (\ref%
{definition H bar}), that is, $\mathcal{\mathcal{H}}\equiv \mathfrak{h}%
_{P}\oplus \mathfrak{h}_{P}^{\ast }$. For any $\xi _{0},\xi _{1}\in \wedge
^{\ast }\mathcal{\mathcal{H}}$, we define their circle product by%
\begin{equation*}
\xi _{0}\circ _{P}\xi _{1}\doteq \left( -1\right) ^{\frac{\mathrm{dim}%
\mathcal{H}}{2}}\int_{P}\mathrm{d}\left( \mathcal{H}^{(1)}\right) \varkappa
_{(0,0)}^{(0,1)}(\xi _{0})\varkappa _{(0,0)}^{(1,0)}(\xi _{1})\mathrm{e}%
^{-\langle \mathfrak{h}_{P}^{(0)},\mathfrak{h}_{P}^{(0)}\rangle }\mathrm{e}%
^{\langle \mathfrak{h}_{P}^{(0)},\mathfrak{h}_{P}^{(1)}\rangle }\mathrm{e}%
^{-\langle \mathfrak{h}_{P}^{(1)},\mathfrak{h}_{P}^{(1)}\rangle }\mathrm{e}%
^{\langle \mathfrak{h}_{P}^{(1)},\mathfrak{h}_{P}^{(0)}\rangle }\ .
\end{equation*}
\end{definition}

It directly follows from this definition that the circle product is
associative and distributive on the space $\wedge ^{\ast }\mathcal{\mathcal{H%
}}$ with the same unit $\mathfrak{1}$ as with the exterior product $\wedge $:

\begin{lemma}[Elementary properties of the circle product]
\label{Lemma --product}\mbox{
}\newline
Fix a basis projection $P$ with range $\mathfrak{h}_{P}$. Then, the circle
product $\circ _{P}$ has the following properties:\newline
\emph{(i)} Associativity: For all $N\in \mathbb{N}$, $N\geq 2$, and $\xi
_{0},\ldots ,\xi _{N-1}\in \wedge ^{\ast }\mathcal{\mathcal{H}}$ ,%
\begin{multline*}
\xi _{0}\circ _{P}\cdots \circ _{P}\xi _{N-1}=\left( -1\right) ^{\left(
N-1\right) \frac{\mathrm{dim}\mathcal{H}}{2}}\left(
\prod\limits_{k=1}^{N-1}\int_{P}\mathrm{d}\left( \mathcal{H}^{(k)}\right)
\right) \left( \prod\limits_{k=0}^{N-1}\varkappa _{(0,0)}^{(k,k+1\,\mathrm{%
mod}\,N)}(\xi _{k})\right) \\
\prod\limits_{k=0}^{N-1}\mathrm{e}^{-\langle \mathfrak{h}_{P}^{(k)},%
\mathfrak{h}_{P}^{(k)}\rangle }\mathrm{e}^{\langle \mathfrak{h}_{P}^{(k)},%
\mathfrak{h}_{P}^{(k+1\,\mathrm{mod}\,N)}\rangle }\ .
\end{multline*}%
\newline
\emph{(ii)} Distributivity over the addition: For $n_{1},n_{2}\in \mathbb{N}$%
, $z_{1},\ldots ,z_{n_{1}+n_{2}}\in \mathbb{C}$, and $\xi _{1},\ldots ,\xi
_{n_{1}+n_{2}}\in \wedge ^{\ast }\mathcal{\mathcal{H}}$,%
\begin{equation*}
\left( \sum_{j=1}^{n_{1}}z_{j}\xi _{j}\right) \circ _{P}\left(
\sum_{j=1}^{n_{2}}z_{j+n_{1}}\xi _{j+n_{1}}\right)
=\sum_{j_{1}=1}^{n_{1}}\sum_{j_{2}=1}^{n_{2}}\left(
z_{j_{1}}z_{j_{2}+n_{1}}\right) \xi _{j_{1}}\circ _{P}\xi _{j_{2}+n_{1}}\ .
\end{equation*}%
\emph{(iii)} The unit of the circle product coincides with the unit $%
\mathfrak{1}$ of the exterior product $\wedge $:%
\begin{equation*}
z\mathfrak{1}\circ _{P}\xi =\xi \circ _{P}z\mathfrak{1}=z\xi \ ,\qquad z\in {%
\mathbb{C}},\ \xi \in \wedge ^{\ast }\mathcal{\mathcal{H}}\ .
\end{equation*}%
\emph{(iv)} Clustering property: For any subspace $\mathcal{Y}\subset 
\mathcal{H}$ with orthogonal complement $\mathcal{Y}^{\perp }$,%
\begin{equation*}
\xi _{0}\circ _{P}\xi _{1}=\xi _{0}\wedge \xi _{1},\qquad \xi _{0}\in \wedge
^{\ast }\mathfrak{A}\left( \mathcal{Y}\right) ,\ \xi _{1}\in \wedge ^{\ast }%
\mathcal{Y}^{\perp }.
\end{equation*}%
\emph{(v)} Canonical Anti-commutation Relations (CAR): For any $N\in \mathbb{%
N}$ and $\varphi _{1},\ldots ,\varphi _{N+1}\in \mathcal{H}$, one has%
\footnote{%
Here we use the convention $\prod\limits_{m\in \emptyset }=1$.} 
\begin{equation*}
\left( \varphi _{1}\wedge \cdots \wedge \varphi _{N}\right) \circ
_{P}\varphi _{N+1}=\varphi _{1}\wedge \cdots \wedge \varphi
_{N+1}+\sum_{q=1}^{N}\left( -1\right) ^{N-q}\left\langle \mathfrak{A}\varphi
_{q},P^{\bot }\varphi _{N+1}\right\rangle _{\mathcal{H}}\prod\limits_{m\in
\left\{ 1,\ldots ,N\right\} \backslash \left\{ q\right\} }\varphi _{m}\ .
\end{equation*}%
\emph{(vi)} Wedge expansion of the antisymmetrized circle product: For $N\in 
\mathbb{N}\backslash \{1\}$ and $\varphi _{1},\ldots ,\varphi _{N}\in 
\mathcal{H}$,%
\begin{multline*}
\frac{1}{N!}\sum_{\pi \in \mathcal{S}_{N}}\left( -1\right) ^{\pi }\varphi
_{\pi \left( 1\right) }\circ _{P}\cdots \circ _{P}\varphi _{\pi \left(
N\right) }=\sum_{\mathcal{N\subseteq }\left\{ 1,\ldots ,N\right\} ,\
\left\vert \mathcal{N}\right\vert \in 2\mathbb{N}_{0}}2^{-\frac{\left\vert 
\mathcal{N}\right\vert }{2}}\mathrm{Pf}\left[ \left\langle \mathfrak{A}%
\varphi _{k},\frac{P^{\bot }-P}{2}\varphi _{l}\right\rangle _{\mathcal{H}}%
\right] _{k,l\in \mathcal{N}} \\
\mathrm{sign}\left( \mathcal{N},\left\{ 1,\ldots ,N\right\} \right)
\prod_{m\in \left\{ 1,\ldots ,N\right\} \backslash \mathcal{N}}\varphi _{m}\
,
\end{multline*}%
where the product on the right-hand side is $\wedge $ (see Notation \ref%
{Notation product}), 
\begin{equation*}
\mathrm{Pf}\left[ \left\langle \mathfrak{A}\varphi _{k},\frac{P^{\bot }-P}{2}%
\varphi _{l}\right\rangle _{\mathcal{H}}\right] _{k,l\in \emptyset }\doteq 1
\end{equation*}%
and, for any finite subsets $\mathcal{M}\subseteq \mathbb{N}$ and $\mathcal{N%
}\subseteq \mathcal{M}$\textrm{, }$\mathrm{sign}\left( \mathcal{N},\mathcal{M%
}\right) $ is the signature of the unique permutation of $\mathcal{M}$ which
takes all elements of $\mathcal{N}$ to the left in such a way that they are
ordered.
\end{lemma}

\begin{proof}
Assertions (i)-(iv) are deduced from straightforward, albeit sometimes
cumbersome, computations using Definitions \ref{def basis projection}, \ref%
{Grassmann Integral}, \ref{def1+} and \ref{definition star}, Equations (\ref%
{def kappa1})-(\ref{def kappa2}), together with the properties given by (\ref%
{grassmana anticommute}) and (\ref{derivation plus1})-(\ref{derivation plus2}%
). We now prove Assertion (v), starting with elementary observations: By
using (\ref{grassmana anticommute}), observe from (\ref{exponentiel}) and
Definition \ref{def1+} that, for any $k,l\in \mathbb{N}_{0}$ and basis
projection $P$,%
\begin{equation*}
\mathrm{e}^{\pm \langle \mathfrak{h}_{P}^{(k)},\mathfrak{h}_{P}^{(l)}\rangle
}=\prod\limits_{j\in J}\mathrm{e}^{\pm \left( \mathfrak{A}\psi _{j}\right)
^{(k)}\wedge \psi _{j}^{(l)}}=\prod\limits_{j\in J}\left( \mathfrak{1}\pm
\left( \mathfrak{A}\psi _{j}\right) ^{(k)}\wedge \psi _{j}^{(l)}\right) \ .
\end{equation*}%
Here, $\{\psi _{j}\}_{j\in J}$ is any orthonormal basis of the range $%
\mathfrak{h}_{P}$ of $P$. In particular,%
\begin{eqnarray}
&&\mathrm{e}^{-\langle \mathfrak{h}_{P}^{(0)},\mathfrak{h}_{P}^{(0)}\rangle }%
\mathrm{e}^{\langle \mathfrak{h}_{P}^{(0)},\mathfrak{h}_{P}^{(1)}\rangle }%
\mathrm{e}^{-\langle \mathfrak{h}_{P}^{(1)},\mathfrak{h}_{P}^{(1)}\rangle }%
\mathrm{e}^{\langle \mathfrak{h}_{P}^{(1)},\mathfrak{h}_{P}^{(0)}\rangle }
\label{blablao} \\
&=&\prod\limits_{j\in J}\left( \mathfrak{1}+\left( \mathfrak{A}\psi
_{j}\right) ^{(1)}\psi _{j}^{(0)}-\left( \mathfrak{A}\psi _{j}\right)
^{(1)}\psi _{j}^{(1)}+\left( \mathfrak{A}\psi _{j}\right) ^{(0)}\psi
_{j}^{(1)}-\left( \mathfrak{A}\psi _{j}\right) ^{(0)}\psi _{j}^{(0)}\right)
\ ,  \notag
\end{eqnarray}%
while, for all $j\in J$, 
\begin{equation}
\frac{\delta }{\delta \psi _{j}^{(1)}}\frac{\delta }{\delta \left( \mathfrak{%
A}\psi _{j}\right) ^{(1)}}\left( \mathfrak{1}+\left( \mathfrak{A}\psi
_{j}\right) ^{(1)}\psi _{j}^{(0)}-\left( \mathfrak{A}\psi _{j}\right)
^{(1)}\psi _{j}^{(1)}+\left( \mathfrak{A}\psi _{j}\right) ^{(0)}\psi
_{j}^{(1)}-\left( \mathfrak{A}\psi _{j}\right) ^{(0)}\psi _{j}^{(0)}\right)
=-\mathfrak{1}\ .  \label{blabla}
\end{equation}%
See \cite{LD2} for more details. (Note that we use Notation \ref{Notation
product}.). Fix next $N\in \mathbb{N}$ and $j_{1},\ldots ,j_{N+1}\in J$.
Then, using Equation (\ref{blablao}), we obtain that 
\begin{eqnarray*}
&&\left( \psi _{j_{1}}\wedge \cdots \wedge \psi _{j_{N}}\right) \circ
_{P}\left( \mathfrak{A}\psi _{j_{N+1}}\right) \\
&=&\left( -1\right) ^{\frac{\mathrm{dim}\mathcal{H}}{2}}\int_{P}\mathrm{d}%
\left( \mathcal{H}^{(1)}\right) \psi _{j_{1}}^{(1)}\wedge \cdots \wedge \psi
_{j_{N}}^{(1)}\wedge \left( \mathfrak{A}\psi _{j_{N+1}}\right) ^{(1)} \\
&&\prod\limits_{j\in J}\left( \mathfrak{1}+\left( \mathfrak{A}\psi
_{j}\right) ^{(1)}\psi _{j}^{(0)}-\left( \mathfrak{A}\psi _{j}\right)
^{(1)}\psi _{j}^{(1)}+\left( \mathfrak{A}\psi _{j}\right) ^{(0)}\psi
_{j}^{(1)}-\left( \mathfrak{A}\psi _{j}\right) ^{(0)}\psi _{j}^{(0)}\right) ,
\end{eqnarray*}%
which, together with (\ref{blabla}) and $|J|=\dim (\mathcal{H})/2$, implies
that%
\begin{eqnarray*}
&&\left( \psi _{j_{1}}\wedge \cdots \wedge \psi _{j_{N}}\right) \circ
_{P}\left( \mathfrak{A}\psi _{j_{N+1}}\right) \\
&=&\left( -1\right) ^{\left\vert \left\{ j_{1},\ldots ,j_{N+1}\right\}
\right\vert }\left( \prod\limits_{j\in \left\{ j_{1},\ldots ,j_{N+1}\right\}
}\frac{\delta }{\delta \psi _{j}^{(1)}}\frac{\delta }{\delta \left( 
\mathfrak{A}\psi _{j}\right) ^{(1)}}\right) \psi _{j_{1}}^{(1)}\wedge \cdots
\wedge \psi _{j_{N}}^{(1)}\wedge \mathfrak{A}\psi _{j_{N+1}}^{(1)} \\
&&\prod\limits_{j\in \left\{ j_{1},\ldots ,j_{N+1}\right\} }\left( \mathfrak{%
1}+\left( \mathfrak{A}\psi _{j}\right) ^{(1)}\psi _{j}^{(0)}-\left( 
\mathfrak{A}\psi _{j}\right) ^{(1)}\psi _{j}^{(1)}+\left( \mathfrak{A}\psi
_{j}\right) ^{(0)}\psi _{j}^{(1)}-\left( \mathfrak{A}\psi _{j}\right)
^{(0)}\psi _{j}^{(0)}\right) \ .
\end{eqnarray*}%
It follows that 
\begin{eqnarray*}
&&\left( \psi _{j_{1}}\wedge \cdots \wedge \psi _{j_{N}}\right) \circ
_{P}\left( \mathfrak{A}\psi _{j_{N+1}}\right) \\
&=&\psi _{j_{1}}\wedge \cdots \wedge \psi _{j_{N}}\wedge \mathfrak{A}\psi
_{j_{N+1}}+\sum_{q=1}^{N}\left( -1\right) ^{N-q}\left\langle \mathfrak{A}%
\psi _{j_{q}},P^{\bot }\mathfrak{A}\psi _{j_{N+1}}\right\rangle _{\mathcal{H}%
}\prod\limits_{j\in \left\{ j_{1},\ldots ,j_{N}\right\} \backslash \left\{
j_{q}\right\} }\psi _{j}\ .
\end{eqnarray*}%
By Definition of the product $\circ _{P}$\ and multilinearity of all the
products appearing in the above expression, ones deduces Assertion (v) from
this last equation.

It remains to prove Assertion (vi). We proceed by induction. The case $N=2$
follows from Assertion (v) and assume that Assertion (vi) holds true at some 
$N\in \mathbb{N}\backslash \{1\}$. Using the notation%
\begin{equation*}
\mathbb{\aleph }_{P}\left( \varphi _{1},\ldots ,\varphi _{m}\right) \doteq 
\frac{1}{m!}\sum_{\pi \in \mathcal{S}_{m}}\left( -1\right) ^{\pi }\varphi
_{\pi \left( 1\right) }\circ _{P}\cdots \circ _{P}\varphi _{\pi \left(
m\right) }
\end{equation*}%
for any $m\in \left\{ 1,\ldots ,N+1\right\} $, observe that%
\begin{equation*}
\mathbb{\aleph }_{P}\left( \varphi _{1},\ldots ,\varphi _{N+1}\right) =\frac{%
1}{\left( N+1\right) !}\sum_{\pi \in \mathcal{S}_{N+1}}\left( -1\right)
^{\pi }\mathbb{\aleph }_{P}\left( \varphi _{\pi \left( 1\right) },\ldots
,\varphi _{\pi \left( N\right) }\right) \circ _{P}\varphi _{\pi \left(
N+1\right) }.
\end{equation*}%
Thus, if Assertion (vi) holds true at some $N\in \mathbb{N}\backslash \{1\}$%
, then, for any $\varphi _{1},\ldots ,\varphi _{N+1}\in \mathcal{H}$ and
permutation $\pi \in \mathcal{S}_{N+1}$,%
\begin{eqnarray*}
\mathbb{\aleph }_{P}\left( \varphi _{\pi \left( 1\right) },\ldots ,\varphi
_{\pi \left( N\right) }\right) \circ _{P}\varphi _{\pi \left( N+1\right) }
&=&\sum_{\mathcal{N\subseteq }\left\{ 1,\ldots ,N\right\} ,\ \left\vert 
\mathcal{N}\right\vert \in 2\mathbb{N}_{0}}2^{-\frac{\left\vert \mathcal{N}%
\right\vert }{2}}\mathrm{Pf}\left[ \left\langle \mathfrak{A}\varphi _{\pi
\left( k\right) },\frac{P^{\bot }-P}{2}\varphi _{\pi \left( l\right)
}\right\rangle _{\mathcal{H}}\right] _{k,l\in \mathcal{N}} \\
&&\mathrm{sign}\left( \mathcal{N},\left\{ 1,\ldots ,N\right\} \right) \left(
\prod_{m\in \left\{ 1,\ldots ,N\right\} \backslash \mathcal{N}}\varphi _{\pi
\left( m\right) }\right) \circ _{P}\varphi _{\pi \left( N+1\right) }
\end{eqnarray*}%
Thus, by applying Assertion (v), 
\begin{eqnarray}
&&\frac{1}{\left( N+1\right) !}\sum_{\pi \in \mathcal{S}_{N+1}}\left(
-1\right) ^{\pi }\mathbb{\aleph }_{P}\left( \varphi _{\pi \left( 1\right)
},\ldots ,\varphi _{\pi \left( N\right) }\right) \circ _{P}\varphi _{\pi
\left( N+1\right) }  \label{sympachiant0} \\
&=&\mathbf{X}+\frac{1}{\left( N+1\right) !}\sum_{\pi \in \mathcal{S}%
_{N+1}}\left( -1\right) ^{\pi }\sum_{\mathcal{N\subseteq }\left\{ 1,\ldots
,N\right\} ,\ \left\vert \mathcal{N}\right\vert \in 2\mathbb{N}_{0}}2^{-%
\frac{\left\vert \mathcal{N}\right\vert }{2}}\mathrm{Pf}\left[ \left\langle 
\mathfrak{A}\varphi _{\pi \left( k\right) },\frac{P^{\bot }-P}{2}\varphi
_{\pi \left( l\right) }\right\rangle _{\mathcal{H}}\right] _{k,l\in \mathcal{%
N}}  \notag \\
&&\qquad \qquad \qquad \qquad \qquad \qquad \qquad \qquad \times \mathrm{sign%
}\left( \mathcal{N},\left\{ 1,\ldots ,N+1\right\} \right) \prod_{m\in
\left\{ 1,\ldots ,N+1\right\} \backslash \mathcal{N}}\varphi _{\pi \left(
m\right) },  \notag
\end{eqnarray}%
where 
\begin{multline*}
\mathbf{X}\doteq \frac{1}{\left( N+1\right) !}\sum_{\pi \in \mathcal{S}%
_{N+1}}\left( -1\right) ^{\pi }\sum_{\mathcal{N\subseteq }\left\{ 1,\ldots
,N\right\} ,\ \left\vert \mathcal{N}\right\vert \in 2\mathbb{N}%
_{0}}\sum_{q\in \left\{ 1,\ldots ,N\right\} \backslash \mathcal{N}}\left(
-1\right) ^{\left\vert \left\{ 1,\ldots ,N\right\} \backslash \mathcal{N}%
\right\vert -\left\vert \left\{ 1,\ldots ,q\right\} \cap \left\{ 1,\ldots
,N\right\} \backslash \mathcal{N}\right\vert } \\
\times \mathrm{sign}\left( \mathcal{N},\left\{ 1,\ldots ,N+1\right\} \right)
2^{-\frac{\left\vert \mathcal{N}\right\vert }{2}-1}\left\langle \mathfrak{A}%
\varphi _{\pi \left( q\right) },\frac{P^{\bot }-P}{2}\varphi _{\pi \left(
N+1\right) }\right\rangle _{\mathcal{H}} \\
\times \mathrm{Pf}\left[ \left\langle \mathfrak{A}\varphi _{\pi \left(
k\right) },\frac{P^{\bot }-P}{2}\varphi _{\pi \left( l\right) }\right\rangle
_{\mathcal{H}}\right] _{k,l\in \mathcal{N}}\prod\limits_{m\in \left\{
1,\ldots ,N\right\} \backslash \left( \mathcal{N}\cup \left\{ q\right\}
\right) }\varphi _{\pi \left( m\right) }.
\end{multline*}%
Now, by elementary manipulations, one gets that 
\begin{eqnarray}
\mathbf{X} &=&\frac{1}{\left( N+1\right) !}\sum_{\pi \in \mathcal{S}%
_{N+1}}\left( -1\right) ^{\pi }\sum_{\mathcal{M\subseteq }\left\{ 1,\ldots
,N+1\right\} ,\ \mathcal{M}\supseteq \{N+1\},\ \left\vert \mathcal{M}%
\right\vert \in 2\mathbb{N}_{0}}\sum_{q\in \mathcal{M}\backslash \left\{
N+1\right\} }\mathbf{Y}_{\mathcal{M},q,N}  \notag \\
&&\times 2^{-\frac{\left\vert \mathcal{M}\right\vert }{2}}\left\langle 
\mathfrak{A}\varphi _{\pi \left( N+1\right) },\frac{P^{\bot }-P}{2}\varphi
_{\pi \left( q\right) }\right\rangle _{\mathcal{H}}\mathrm{Pf}\left[
\left\langle \mathfrak{A}\varphi _{\pi \left( k\right) },\frac{P^{\bot }-P}{2%
}\varphi _{\pi \left( l\right) }\right\rangle _{\mathcal{H}}\right] _{k,l\in
\left( \mathcal{M}\backslash \{q,N+1\}\right) }  \notag \\
&&\times \prod\limits_{m\in \left\{ 1,\ldots ,N+1\right\} \backslash 
\mathcal{M}}\varphi _{\pi \left( m\right) }\ ,  \label{dfsfsdf}
\end{eqnarray}%
where, for any $N\in \mathbb{N}$, $\mathcal{M\subseteq }\left\{ 1,\ldots
,N+1\right\} $ with $\left\vert \mathcal{M}\right\vert \in 2\mathbb{N}_{0}$, 
$\mathcal{M}\supseteq \{N+1\}$, and $q\in \mathcal{M}\backslash \left\{
N+1\right\} $,%
\begin{multline*}
\mathbf{Y}_{\mathcal{M},q,N}\doteq \left( -1\right) ^{\left\vert \left\{
1,\ldots ,N+1\right\} \backslash \left( \mathcal{M}\backslash
\{q,N+1\}\right) \right\vert -\left\vert \left\{ 1,\ldots ,q\right\} \cap
\left\{ 1,\ldots ,N+1\right\} \backslash \left( \mathcal{M}\backslash
\{q,N+1\}\right) \right\vert } \\
\times \mathrm{sign}\left( \left( \mathcal{M}\backslash \{q,N+1\}\right)
,\left\{ 1,\ldots ,N+1\right\} \right) \ .
\end{multline*}%
It is straightforward to check that 
\begin{multline*}
\mathrm{sign}\left( \mathcal{M},\left\{ 1,\ldots ,N+1\right\} \right) =%
\mathrm{sign}\left( \mathcal{M}\backslash \{q,N+1\},\left\{ 1,\ldots
,N+1\right\} \right) \\
\times \left( -1\right) ^{\left\vert \left\{ 1,\ldots ,q\right\} \cap
\left\{ 1,\ldots ,N+1\right\} \backslash \left( \mathcal{M}\backslash
\{q,N+1\}\right) \right\vert +\left\vert \left\{ 1,\ldots ,q\right\} \cap 
\mathcal{M}\right\vert +N+1},
\end{multline*}%
that is,%
\begin{equation*}
\mathbf{Y}_{\mathcal{M},q,N}=\left( -1\right) ^{\left\vert \left\{ 1,\ldots
,q\right\} \cap \mathcal{M}\right\vert }\mathrm{sign}\left( \mathcal{M}%
,\left\{ 1,\ldots ,N+1\right\} \right) \ .
\end{equation*}%
It follows from (\ref{dfsfsdf}) that 
\begin{multline*}
\mathbf{X}=\frac{1}{\left( N+1\right) !}\sum_{\pi \in \mathcal{S}%
_{N+1}}\left( -1\right) ^{\pi }\sum_{\mathcal{M\subseteq }\left\{ 1,\ldots
,N+1\right\} ,\ \mathcal{M}\supseteq \{N+1\},\ \left\vert \mathcal{M}%
\right\vert \in 2\mathbb{N}_{0}}\mathrm{sign}\left( \mathcal{M},\left\{
1,\ldots ,N+1\right\} \right) \\
\times \sum_{q\in \mathcal{M}\backslash \left\{ N+1\right\} }\left(
-1\right) ^{\left\vert \left\{ 1,\ldots ,q\right\} \cap \mathcal{M}%
\right\vert }2^{-\frac{\left\vert \mathcal{M}\right\vert }{2}}\left\langle 
\mathfrak{A}\varphi _{\pi \left( N+1\right) },\frac{P^{\bot }-P}{2}\varphi
_{\pi \left( q\right) }\right\rangle _{\mathcal{H}} \\
\times \mathrm{Pf}\left[ \left\langle \mathfrak{A}\varphi _{\pi \left(
k\right) },\frac{P^{\bot }-P}{2}\varphi _{\pi \left( l\right) }\right\rangle
_{\mathcal{H}}\right] _{k,l\in \left( \mathcal{M}\backslash \{q,N+1\}\right)
}\prod\limits_{m\in \left\{ 1,\ldots ,N+1\right\} \backslash \mathcal{M}%
}\varphi _{\pi \left( m\right) }\ ,
\end{multline*}%
which, by the well-known cofactor expansion for Pfaffians \cite[Proposition
B.2]{FKT02}, lead to%
\begin{multline*}
\mathbf{X}=\frac{1}{\left( N+1\right) !}\sum_{\pi \in \mathcal{S}%
_{N+1}}\left( -1\right) ^{\pi }\sum_{\mathcal{M\subseteq }\left\{ 1,\ldots
,N+1\right\} ,\ \left\vert \mathcal{M}\cap \{N+1\}\right\vert =1,\
\left\vert \mathcal{M}\right\vert \in 2\mathbb{N}_{0}}2^{-\frac{\left\vert 
\mathcal{M}\right\vert }{2}} \\
\times \mathrm{Pf}\left[ \left\langle \mathfrak{A}\varphi _{\pi \left(
k\right) },\frac{P^{\bot }-P}{2}\varphi _{\pi \left( l\right) }\right\rangle
_{\mathcal{H}}\right] _{k,l\in \mathcal{M}}\mathrm{sign}\left( \mathcal{M}%
,\left\{ 1,\ldots ,N+1\right\} \right) \prod\limits_{m\in \left\{ 1,\ldots
,N+1\right\} \backslash \mathcal{M}}\varphi _{\pi \left( m\right) }\ .
\end{multline*}%
By combining this with (\ref{sympachiant0}), we conclude that Assertion (vi)
holds true for the case $N+1$, provided it holds true for $N\in \mathbb{N}$.
\end{proof}

\begin{remark}
\label{bound norm a CAR copy(3)}\mbox{
}\newline
By Lemma \ref{Lemma --product}, for any basis projection $P$, $m,n\in 
\mathbb{N}_{0}$ so that $m+n\geq 1$, and all $\varphi _{1},\ldots ,\varphi
_{n+m}\in \mathfrak{h}_{P}$, 
\begin{equation*}
\left( \mathfrak{A}\varphi _{1}\right) \circ _{P}\cdots \circ _{P}\left( 
\mathfrak{A}\varphi _{m}\right) \circ _{P}\varphi _{m+1}\circ _{P}\cdots
\circ _{P}\varphi _{m+n}=\left( \mathfrak{A}\varphi _{1}\right) \wedge
\cdots \wedge \left( \mathfrak{A}\varphi _{m}\right) \wedge \varphi
_{m+1}\wedge \cdots \wedge \varphi _{m+n}\ .
\end{equation*}%
If $m=0$, then there is no \textquotedblleft $\mathfrak{A}\varphi $%
\textquotedblright\ in the above equation. Mutatis mutandis for $n=0$.
\end{remark}

\begin{remark}
\label{different product}\mbox{
}\newline
For any pair of basis projections $P_{1},P_{2}\in \mathcal{B}(\mathcal{H})$,
recall the existence of a Bogoliubov transformation $U$ such that $%
P_{2}=UP_{1}U^{\ast }$, by \cite[Lemma 3.6]{A68}. Then, one deduces from
Lemma \ref{Lemma --product} (v) that 
\begin{equation*}
\left( U\varphi _{1}\right) \circ _{P_{2}}\left( U\varphi _{2}\right)
-\varphi _{1}\circ _{P_{1}}\varphi _{2}=\left( U\varphi _{1}\right) \wedge
\left( U\varphi _{2}\right) -\varphi _{1}\wedge \varphi _{2}\ ,\qquad
\varphi _{1},\varphi _{2}\in \mathcal{\mathcal{H}}\ .
\end{equation*}
\end{remark}

The space $(\wedge ^{\ast }\mathcal{\mathcal{H}},+)$ endowed with the circle
product $\circ _{P}$ is thus an (associative and distributive)\emph{\ }%
algebra, like $(\wedge ^{\ast }\mathcal{\mathcal{H}},+,\wedge )$, for \emph{%
any} basis projection $P$. By using the antiunitary involution $\mathfrak{A}$
on $\mathcal{\mathcal{H}}$, we can define an involution on $\wedge ^{\ast }%
\mathcal{\mathcal{H}}$, which turns $(\wedge ^{\ast }\mathcal{\mathcal{H}}%
,+,\circ _{P})$ into a $\ast $-algebra:

\begin{definition}[Involution on $\wedge ^{\ast }\mathcal{\mathcal{H}}$]
\label{definition involution}\mbox{
}\newline
The antilinear map $\xi \mapsto \xi ^{\ast }$ from $\wedge ^{\ast }\mathcal{%
\mathcal{H}}$ to itself is uniquely defined for any basis projection $P$ by
the conditions $\mathfrak{1}^{\ast }=\mathfrak{1}$ and%
\begin{equation*}
(\varphi _{1}\circ _{P}\cdots \circ _{P}\varphi _{n})^{\ast }=\left( 
\mathfrak{A}\varphi _{n}\right) \circ _{P}\cdots \circ _{P}\left( \mathfrak{A%
}\varphi _{1}\right) \ ,\qquad n\in \mathbb{N},\ \varphi _{1},\ldots
,\varphi _{n}\in \mathcal{H}\ .
\end{equation*}
\end{definition}

\noindent The involution on $\wedge ^{\ast }\mathcal{\mathcal{H}}$ \emph{%
does not depend} on the choice of the basis projection $P$ and makes the
Grassmann algebra $(\wedge ^{\ast }\mathcal{\mathcal{H}},+,\wedge )$ a $\ast 
$-algebra:

\begin{lemma}[Grassmann algebra as $^{\ast }$--algebra]
\label{Lemma line Laplace copy(1)}\mbox{ }\newline
$(\wedge ^{\ast }\mathcal{\mathcal{H}},+,\wedge )$ equipped with the
involution $^{\ast }$ defined from Definition \ref{definition involution} is
a $\ast $-algebra, i.e.,%
\begin{equation}
\left( \xi _{0}\wedge \xi _{1}\right) ^{\ast }=\xi _{1}^{\ast }\wedge \xi
_{0}^{\ast }\ ,\qquad \xi _{0},\xi _{1}\in \wedge ^{\ast }\mathcal{\mathcal{H%
}}\ .  \label{star algebra}
\end{equation}
\end{lemma}

\begin{proof}
Fix a basis projection $P$ with range $\mathfrak{h}_{P}$. It suffices to
show (\ref{star algebra}) for monomials $\xi _{0},\xi _{1}$ in $\varphi \in 
\mathfrak{h}_{P}$ and $\phi ^{\ast }\in \mathfrak{h}_{P}^{\ast }$, by
linearity and antilinearity. This special case can be shown with
straightforward computations. See for instance Remark \ref{bound norm a CAR
copy(3)}.
\end{proof}

For any basis projection $P$, the algebra $(\wedge ^{\ast }\mathcal{\mathcal{%
H}},+,\circ _{P})$ endowed with this involution shares, by construction, all
properties of self-dual CAR algebra except for the norm:

\begin{theorem}[From Grassmann algebra to self-dual CAR algebra]
\label{satz.star.car}\mbox{ }\newline
Given a self-dual Hilbert space $\mathcal{\mathcal{H}}$ with even dimension $%
\mathrm{dim}\mathcal{\mathcal{H}}\in 2\mathbb{N}$ and antiunitary involution 
$\mathfrak{A}$, and a basis projection $P$ associated with $(\mathcal{%
\mathcal{H}},\mathfrak{A})$, $(\wedge ^{\ast }\mathcal{\mathcal{H}},+,\circ
_{P},^{\ast })$ is a $\ast $-algebra generated by $\mathfrak{1}$ and the
family $\{\varphi ^{\ast }\}_{\varphi \in \mathcal{\mathcal{H}}}$ of
elements satisfying Conditions \emph{(a)-(c)} of Definition \ref{def
Self--dual CAR Algebras}.
\end{theorem}

\begin{proof}
Conditions (a)-(b) of Definition \ref{def Self--dual CAR Algebras} applied
to the family $\{\varphi ^{\ast }\}_{\varphi \in \mathcal{\mathcal{H}}}$ are
easily verified. Only Condition (c), that is, 
\begin{equation*}
\varphi _{1}^{\ast }\circ _{P}\varphi _{2}+\varphi _{2}\circ _{P}\varphi
_{1}^{\ast }=\left( \mathfrak{A}\varphi _{1}\right) \circ _{P}\varphi
_{2}+\varphi _{2}\circ _{P}\left( \mathfrak{A}\varphi _{1}\right)
=\left\langle \varphi _{1},\varphi _{2}\right\rangle _{\mathcal{H}}\,%
\mathfrak{1}\ ,\qquad \varphi _{1},\varphi _{2}\in \mathcal{\mathcal{H}}\ ,
\end{equation*}%
has to be checked: Condition (c) results from Lemma \ref{Lemma --product},
in particular Assertion (v), together with properties of the antiunitary
involution $\mathfrak{A}$, Definition \ref{definition involution} and
elementary computations.
\end{proof}

For two different basis projections $P_{1},P_{2}\in \mathcal{B}(\mathcal{H})$%
, $(\wedge ^{\ast }\mathcal{\mathcal{H}},+,\circ _{P_{1}},^{\ast })$ and $%
(\wedge ^{\ast }\mathcal{\mathcal{H}},+,\circ _{P_{2}},^{\ast })$ differ as
algebras because their respective circle products do not coincide, in
general. See Remark \ref{different product}. They are, nevertheless, $\ast $%
-isomorphic to each other. In the same way, there is a canonical $\ast $%
-isomorphism between a self-dual CAR algebra constructed from $\left( 
\mathcal{H},\mathfrak{A}\right) $ and $\wedge ^{\ast }\mathcal{\mathcal{H}}$:

\begin{definition}[Canonical isomorphism of $\ast $--algebra]
\label{definition isomorphism}\mbox{
}\newline
For any basis projection $P$, we define the canonical isomorphism 
\begin{equation*}
\varkappa _{P}:(\mathrm{sCAR}\left( \mathcal{H},\mathfrak{A}\right) ,+,\cdot
,^{\ast })\rightarrow (\wedge ^{\ast }\mathcal{\mathcal{H}},+,\circ
_{P},^{\ast })
\end{equation*}%
via the conditions $\varkappa _{P}(z\mathfrak{1})=z\mathfrak{1}$ and $%
\varkappa _{P}\left( \mathrm{B}(\varphi )\right) =\varphi ^{\ast }$ for all $%
\varphi \in \mathcal{\mathcal{H}}$.
\end{definition}

Bilinear elements of self-dual CAR algebra (Definition \ref{def trace state
copy(1)}) are mapped via $\varkappa _{P}$ (up to some constant) to bilinear
elements of Grassmann algebra, as\ stated in Definition \ref{def1+}: By (\ref%
{equation facilissimo}), Lemma \ref{Lemma --product} (v) and Definition \ref%
{definition isomorphism}, for any basis projection $P$,%
\begin{equation}
\varkappa _{P}\left( \langle \mathrm{B},H\mathrm{B}\rangle \right) =\langle 
\mathcal{H},H\mathcal{H}\rangle +\mathrm{Tr}_{\mathcal{H}}\left( P^{\bot
}HP^{\bot }\right) \mathfrak{1}\ ,\qquad H\in \mathcal{B}(\mathcal{H})\ .
\label{petit calcul2}
\end{equation}%
Compare with Equations (\ref{second quantizzation00}) and (\ref{second quant
II}).

To make $(\wedge ^{\ast }\mathcal{\mathcal{H}},+,\circ _{P},^{\ast })$ a
self-dual CAR ($C^{\ast }$-) algebra for a basis projection $P$, it suffices
to equip this $\ast $-algebra with the following norm: 
\begin{equation}
\left\Vert \xi \right\Vert _{\wedge ^{\ast }\mathcal{\mathcal{H}}}\doteq
\left\Vert \varkappa _{P}^{-1}\left( \xi \right) \right\Vert _{\mathrm{sCAR}%
\left( \mathcal{H},\mathfrak{A}\right) }\ ,\qquad \xi \in \wedge ^{\ast }%
\mathcal{\mathcal{H}}\ .  \label{def norm}
\end{equation}%
See Definition \ref{def Self--dual CAR Algebras} and Theorem \ref%
{satz.star.car}. In this case, $\varkappa _{P}$ is an isometry.

\begin{remark}
\mbox{
}\newline
For any basis projection $P$ with range $\mathfrak{h}_{P}$, $\wedge ^{\ast }%
\mathcal{H}$ can be equipped with a scalar product and then seen as the
fermionic Fock space $\mathcal{F}_{\mathcal{H}}$, as explained in Section %
\ref{Fock}. By finite dimensionality of $\mathcal{H}$, the norm (\ref{def
norm}) is equivalent, albeit different, to the one coming from the scalar
product. In fact, $\wedge ^{\ast }\mathcal{H}$ can be viewed as the $C^{\ast
}$-algebra $\mathcal{B}(\mathcal{F}_{\mathfrak{h}_{P}})$, the norm (\ref{def
norm}) being the operator norm.
\end{remark}

\section{Generating Functions as Berezin Integrals\label{Fermionic Path
Integral}}

First, we use a Chernoff product formula for the exponentials $\mathrm{e}%
^{-\beta H}$ and $\mathrm{e}^{sK}$, see Equation (\ref{trace}). Using the
tracial state formula (Theorem \ref{satz.spurformel}) we then deduce a
representation of generating functions as Gaussian Berezin integrals
(Definition \ref{gaussian integral}). More precisely, we prove Equation (\ref%
{trace2}). This latter equation refers to a kind of fermionic path integral
for generating functions, which is one of the objectives of the paper.

Note that in all the sections, $P$ is always a basis projection (Definition %
\ref{def basis projection}), which determines the product $\circ _{P}$ of
the self-dual CAR algebra $(\wedge ^{\ast }\mathcal{\mathcal{H}},+,\circ
_{P},^{\ast })$.

\subsection{The Berezin Integral Representation of the Tracial State}

Through the isomorphism $\varkappa _{P}$, defined for any basis projection $%
P $ by Definition \ref{definition isomorphism}, the tracial state $\mathrm{tr%
}\in \mathrm{sCAR}(\mathcal{H},\mathfrak{A})^{\ast }$ (Definition \ref{def
trace state}) can be represented in terms of Berezin integrals. To explain
this, we introduce the following definitions:\medskip

\noindent \underline{(i):} For any basis projection $P$ and integers $n\in 
\mathbb{N}_{0}$ and $k\in \{0,\ldots ,n\}$, we define the map 
\begin{equation}
\varkappa _{P}^{(k)}\doteq \varkappa _{(0,0)}^{(k,k)}\circ \varkappa _{P}
\label{definition xi-k}
\end{equation}%
from $\mathrm{sCAR}(\mathcal{H},\mathfrak{A})$ to the $k$-copy $\wedge
^{\ast }\mathcal{H}^{(k)}$ of $\wedge ^{\ast }\mathcal{H}$. Note that $%
\varkappa _{P}^{(0)}=\varkappa _{P}$, see (\ref{identification}) and
Definition \ref{definition isomorphism}. \medskip

\noindent \underline{(ii):} Using Notation \ref{remark constant} (ii), we
define the finite-dimensional Hilbert spaces 
\begin{equation}
\mathfrak{H}^{(n)}\doteq \bigoplus\limits_{k=0}^{n-1}\mathcal{H}^{(k)}\
,\qquad n\in \mathbb{N}\ .  \label{hilbert etendu}
\end{equation}%
The corresponding scalar product is the usual one for direct sums of Hilbert
spaces. In particular, for any orthonormal basis $\{\psi _{i}\}_{i\in I}$ of 
$\mathcal{H}$, the union $\cup _{k=0}^{n-1}\{\psi _{i}^{(k)}\}_{i\in I}$ is
an orthonormal basis of $\mathfrak{H}^{(n)}$. \medskip

\noindent \underline{(iii):} For any $n\in \mathbb{N}$, a map $A$ from $%
\mathcal{H}$ to itself, like the antiunitary involution $\mathfrak{A}$, is
canonically extended to a map $\hat{A}$ from $\mathfrak{H}^{(n)}$ to itself
by the conditions%
\begin{equation}
\hat{A}\varphi ^{(k)}\doteq \left( A\varphi \right) ^{(k)}\ ,\qquad k\in
\{0,\ldots ,n-1\}\ .  \label{definition extensino0}
\end{equation}%
Notice that $\mathfrak{\hat{A}}$ is an antiunitary involution, and any basis
projection $P$ associated with $(\mathcal{H},\mathfrak{A})$ yields a basis
projection $\hat{P}$ associated with $(\mathfrak{H}^{(n)},\mathfrak{\hat{A}}%
) $ for any $n\in \mathbb{N}$. \medskip

\noindent \underline{(iv):} For any $n\in \mathbb{N}$, we introduce the
linear map $\partial \in \mathcal{B}(\mathfrak{H}^{(n)})$ defined by the
identities%
\begin{equation}
\partial \varphi ^{(0)}=\varphi ^{(0)}+\varphi ^{(n-1)}\qquad \text{and}%
\qquad \partial \varphi ^{(k)}=\varphi ^{(k)}-\varphi ^{(k-1)}
\label{discrete derivative}
\end{equation}%
for all $\varphi \in \mathcal{H}$ and $k\in \{1,\ldots ,n-1\}$. Note that
its adjoint $\partial ^{\ast }$ satisfies%
\begin{equation*}
\partial ^{\ast }\varphi ^{(n-1)}=\varphi ^{(0)}+\varphi ^{(n-1)}\qquad 
\text{and}\qquad \partial ^{\ast }\varphi ^{(k)}=\varphi ^{(k)}-\varphi
^{(k+1)}
\end{equation*}%
for all $\varphi \in \mathcal{H}$ and $k\in \{0,\ldots ,n-2\}$. The operator 
$\partial $\ is normal but not self-adjoint. \medskip

\noindent \underline{(v):} For every $n\in \mathbb{N}$ and any basis
projection $P$ with range $\mathfrak{h}_{P}$, we then define from (\ref%
{definition extensino0}) and (\ref{discrete derivative}) the operator 
\begin{equation}
\boldsymbol{\partial }_{P}\doteq \hat{P}\partial \hat{P}-\mathfrak{\hat{A}}%
\hat{P}\partial ^{\ast }\hat{P}\mathfrak{\hat{A}}\in \mathcal{B}(\mathfrak{H}%
^{(n)})\ .  \label{def Ah1new0}
\end{equation}%
It is a self-dual operator on $(\mathfrak{H}^{(n)},\mathfrak{\hat{A}})$
because $\boldsymbol{\partial }_{P}^{\ast }=-\mathfrak{\hat{A}}\boldsymbol{%
\partial }_{P}\mathfrak{\hat{A}}$. See Definitions \ref{def basis projection}
and \ref{def one particle hamiltinian}. \medskip

\noindent Then, we get the well-known \emph{tracial state formula} in the
context of Grassmann algebra:

\begin{theorem}[Tracial state formula]
\label{satz.spurformel}\mbox{ }\newline
Fix a basis projection $P$. Then, for all $n\in \mathbb{N}$ and $%
A_{0},\ldots ,A_{n-1}\in \mathrm{sCAR}(\mathcal{H},\mathfrak{A})$, 
\begin{equation*}
\mathrm{tr}(A_{0}\cdots A_{n-1})\mathfrak{1}=2^{-\frac{\mathrm{dim}\mathcal{H%
}}{2}}\int_{P}\mathrm{d}\left( \mathfrak{H}^{(n)}\right) \mathrm{e}^{\frac{1%
}{2}\langle \mathfrak{H}^{(n)},\boldsymbol{\partial }_{P}\mathfrak{H}%
^{(n)}\rangle }\left( \prod\limits_{k=0}^{n-1}\varkappa
_{P}^{(k)}(A_{k})\right) \ .
\end{equation*}
\end{theorem}

\begin{proof}
Observe that Remark \ref{bound norm a CAR copy(2)} holds true for $\mathfrak{%
h=h}_{P}$ and any basis projection $P$. Then, by using linearity properties
and explicit computations on arbitrary, normally-ordered monomials, we prove
that 
\begin{equation*}
\mathrm{tr}(A)\mathfrak{1}=2^{-\frac{\mathrm{dim}\mathcal{H}}{2}}\int_{P}%
\mathrm{d}\left( \mathcal{H}\right) \varkappa _{P}(A)\mathrm{e}^{2\langle 
\mathfrak{h}_{P},\mathfrak{h}_{P}\rangle }\ ,\qquad A\in \mathrm{sCAR}\left( 
\mathcal{H},\mathfrak{A}\right) \ .
\end{equation*}%
Next, we use the equality 
\begin{equation*}
\varkappa _{P}\left( A_{0}\cdots A_{n-1}\right) =\varkappa _{P}\left(
A_{0}\right) \circ _{P}\cdots \circ _{P}\varkappa _{P}\left( A_{n-1}\right)
\ ,
\end{equation*}%
together with Lemma \ref{Lemma --product} (i) and a renumbering, to get the
assertion. Note that 
\begin{equation*}
\int_{P}\mathrm{d}\left( \mathfrak{H}^{(n)}\right)
=\prod\limits_{k=0}^{n-1}\int_{P}\mathrm{d}\left( \mathcal{H}^{(k)}\right) \
,\qquad n\in \mathbb{N}\ ,
\end{equation*}%
by Definition \ref{Grassmann Integral}, while, for any basis projection $P$, 
\begin{equation}
\frac{1}{2}\langle \mathfrak{H}^{(n)},\boldsymbol{\partial }_{P}\mathfrak{H}%
^{(n)}\rangle =\langle \mathfrak{h}_{P}^{(0)},\mathfrak{h}_{P}^{(0)}\rangle
+\langle \mathfrak{h}_{P}^{(0)},\mathfrak{h}_{P}^{(n-1)}\rangle
+\sum\limits_{k=1}^{n-1}\left( \langle \mathfrak{h}_{P}^{(k)},\mathfrak{h}%
_{P}^{(k)}\rangle -\langle \mathfrak{h}_{P}^{(k)},\mathfrak{h}%
_{P}^{(k-1)}\rangle \right) \ ,  \label{def Ah1new1}
\end{equation}%
by Definition \ref{def1+}, Equations (\ref{definition extensino0}) and (\ref%
{def Ah1new0}). For more details we recommend \cite{LD2}.
\end{proof}

\subsection{Chernoff Product Formula}

We now define a Chernoff product approximation in self-dual algebra as
follows:

\begin{definition}[A Chernoff product approximation]
\label{def XA}\mbox{
}\newline
Fix a basis projection $P$. For every element $A\in \mathrm{sCAR}(\mathcal{H}%
,\mathfrak{A})$ and all $n\in {\mathbb{N}}$, 
\begin{equation*}
X_{A}^{(n)}\doteq \left[ \varkappa _{P}^{-1}\left( \exp \left( \frac{1}{n}%
\varkappa _{P}(A)\right) \right) \right] ^{n}\in \mathrm{sCAR}\left( 
\mathcal{H},\mathfrak{A}\right) \ ,
\end{equation*}%
where we recall that $\varkappa _{P}$ is the canonical isomorphism of $\ast $%
-algebras from Definition \ref{definition isomorphism}, while $\exp \left(
\xi \right) $ is defined by (\ref{exponentiel})\footnote{%
In other words, we use the exterior product $\wedge $ (and \emph{not} the
circle product from Definition \ref{definition star}) of the Grassmann
algebra $\wedge ^{\ast }\mathcal{\mathcal{H}}$ (see (\ref{product}))\ to
define the exponential.} for all $\xi \in \wedge ^{\ast }\mathcal{H}$.
\end{definition}

\noindent In the limit $n\rightarrow \infty $, $X_{A}^{(n)}$ gives the usual
exponential of $A$:

\begin{theorem}[Chernoff product formula]
\label{lemma exp}\mbox{ }\newline
For any basis projection $P$ and $A\in \mathrm{sCAR}(\mathcal{H},\mathfrak{A}%
)$,%
\begin{equation*}
(X_{A}^{(n)})^{\ast }=X_{A^{\ast }}^{(n)}\qquad \text{and}\qquad
\lim\limits_{n\rightarrow \infty }X_{A}^{(n)}=\exp \left( A\right) \ .
\end{equation*}%
This convergence is uniform on bounded subsets of the $C^{\ast }$-algebra $%
\mathrm{sCAR}(\mathcal{H},\mathfrak{A})$.
\end{theorem}

\begin{proof}
Since 
\begin{equation*}
\varkappa _{P}^{-1}\left( \exp \left( \varkappa _{P}(0)\right) \right) =%
\mathfrak{1}\in \mathrm{sCAR}(\mathcal{H},\mathfrak{A})\quad \text{and}\quad
\partial _{s}\left\{ \varkappa _{P}^{-1}\left( \exp \left( s\varkappa
_{P}(A)\right) \right) \right\} _{s=0}=A\ ,
\end{equation*}%
the assertion is a usual Chernoff product formula. See, e.g., \cite[5.2
Theorem]{EngelNagel}. Although general results proving the Chernoff product
formula are available, we provide here an explicit proof to extract
properties that are particular to our choice of the Chernoff product
approximation.

By finite dimensionality of $\mathcal{H}$, the involution of $\wedge ^{\ast }%
\mathcal{\mathcal{H}}$ is continuous. By Lemma \ref{Lemma line Laplace
copy(1)} and Equation (\ref{exponentiel}), 
\begin{equation*}
\left[ \exp (\xi )\right] ^{\ast }=\exp (\xi ^{\ast })\ ,\qquad \xi \in
\wedge ^{\ast }\mathcal{\mathcal{H}}\ .
\end{equation*}%
Since the map $\varkappa _{P}$ is a $\ast $-isomorphism (Definition \ref%
{definition isomorphism}), for every $n\in {\mathbb{N}}$ it follows that $%
X_{A}^{(n)\ast }=X_{A^{\ast }}^{(n)}$ for any $A\in \mathrm{sCAR}(\mathcal{H}%
,\mathfrak{A})$.

For every $n\in {\mathbb{N}}$ and $A\in \mathrm{sCAR}(\mathcal{H},\mathfrak{A%
})$ observe next that%
\begin{equation}
X_{A}^{(n)}=\left( 1+\frac{A}{n}\right) ^{n}+\sum\limits_{k=1}^{n}\binom{n}{k%
}\tilde{X}_{A}^{k}\left( 1+\frac{1}{n}A\right) ^{n-k},  \label{totototo0}
\end{equation}%
where 
\begin{equation*}
\tilde{X}_{A}\doteq \sum\limits_{m=2}^{\infty }\frac{1}{m!}\left( \frac{1}{n}%
\right) ^{m}\varkappa _{P}^{-1}(\varkappa _{P}(A)^{m})\ .
\end{equation*}%
So, it suffices to show that the last term of (\ref{totototo0}) converges to
zero when $n\rightarrow \infty $ to prove the Chernoff product formula. By
finite dimensionality of $\mathcal{H}$, note that the linear map $\varkappa
_{P}$ from $\mathrm{sCAR}(\mathcal{H},\mathfrak{A})$ to $\wedge ^{\ast }%
\mathcal{\mathcal{H}}$ is continuous. Moreover, its inverse $\varkappa
_{P}^{-1}$ is also continuous, by Definition (\ref{def norm}). From (\ref%
{toto eq}), we obtain that 
\begin{align}
\left\Vert \tilde{X}_{A}\right\Vert _{\mathrm{sCAR}\left( \mathcal{H},%
\mathfrak{A}\right) }& \leq \sum\limits_{m=2}^{\infty }\frac{1}{m!}\left( 
\frac{1}{n}\right) ^{m}\left\Vert \varkappa _{P}(A)^{m}\right\Vert _{\wedge
^{\ast }\mathcal{\mathcal{H}}}  \notag \\
& \leq \frac{1}{n^{2}}D_{\varkappa _{P}(A)}^{2}\sum\limits_{m=2}^{\infty }%
\frac{1}{m!}\left( \frac{1}{n}\right) ^{m-2}D_{\varkappa _{P}(A)}^{m-2} 
\notag \\
& \leq \frac{1}{n^{2}}D_{\varkappa _{P}(A)}^{2}\exp \left( \frac{1}{n}%
D_{\varkappa _{P}(A)}\right) \ .  \label{totoplus}
\end{align}%
Now, by using the definition 
\begin{equation*}
D_{A}\doteq \max \left\{ D_{\varkappa _{P}(A)}\mathrm{e}^{D_{\varkappa
_{P}(A)}},\left\Vert A\right\Vert _{\mathrm{sCAR}\left( \mathcal{H},%
\mathfrak{A}\right) }\right\} \ ,
\end{equation*}%
we infer from (\ref{totoplus}) that%
\begin{eqnarray}
\left\Vert \sum\limits_{k=1}^{n}\binom{n}{k}\tilde{X}_{A}^{k}\left( 1+\frac{1%
}{n}A\right) ^{n-k}\right\Vert _{\mathrm{sCAR}\left( \mathcal{H},\mathfrak{A}%
\right) } &\leq &\sum\limits_{k=1}^{n}\binom{n}{k}\left( \frac{D_{A}^{2}}{%
n^{2}}\right) ^{k}\left( 1+\frac{D_{A}}{n}\right) ^{n-k}  \notag \\
&=&\left( 1+\frac{D_{A}}{n}+\frac{D_{A}^{2}}{n^{2}}\right) ^{n}-\left( 1+%
\frac{D_{A}}{n}\right) ^{n}  \notag \\
&\leq &\left( 1+\frac{D_{A}}{n}\right) ^{n}\left( \left( 1+\frac{D_{A}^{2}}{%
n^{2}}\right) ^{n}-1\right)  \notag \\
&\leq &\mathrm{e}^{D_{A}}\Big(\mathrm{e}^{\frac{D_{A}^{2}}{n}}-1\Big)\ .
\label{totototo}
\end{eqnarray}%
By finite dimensionality of $\mathcal{H}$, for any bounded subset $\mathfrak{%
B}$ of $\mathrm{sCAR}\left( \mathcal{H},\mathfrak{A}\right) $, 
\begin{equation*}
\sup_{A\in \mathfrak{B}}D_{A}<\infty \qquad \text{and}\qquad
\lim_{n\rightarrow \infty }\sup_{A\in \mathfrak{B}}\left\Vert \left( 1+\frac{%
A}{n}\right) ^{n}-\exp (A)\right\Vert _{\mathrm{sCAR}\left( \mathcal{H},%
\mathfrak{A}\right) }=0\ .
\end{equation*}%
As a consequence, by (\ref{totototo0}) and (\ref{totototo}), the
convergence, as $n\rightarrow \infty $, of $X_{A}^{(n)}$ to $\exp (A)$ is
uniform on bounded subsets of $\mathrm{sCAR}(\mathcal{H},\mathfrak{A})$.
\end{proof}

\begin{remark}[Self--adjoint case]
\label{lemma exp copy(2)}\mbox{ }\newline
If $\mathfrak{B}\subset \mathrm{sCAR}(\mathcal{H},\mathfrak{A})$ is a
bounded subset of self-adjoint elements, then there is $N_{\mathfrak{B}}\in {%
\mathbb{N}}$ such that, for all $n\geq N_{\mathfrak{B}}$, $s\in \mathbb{R}$
and $A\in \mathfrak{B}$, the element $%
%TCIMACRO{\TeXButton{\big(}{\big(}}%
%BeginExpansion
\big(%
%EndExpansion
X_{A}^{(n)}%
%TCIMACRO{\TeXButton{\big)}{\big)}}%
%BeginExpansion
\big)%
%EndExpansion
^{s}$ is well-defined and self-adjoint, by Theorem \ref{lemma exp} and the
spectral theorem. Like in Theorem \ref{lemma exp}, it tends to $\exp (sA)$
in the limit $n\rightarrow \infty $. Moreover, in this case, again by the
spectral theorem, the family of maps $s\mapsto 
%TCIMACRO{\TeXButton{\big(}{\big(}}%
%BeginExpansion
\big(%
%EndExpansion
X_{A}^{(n)}%
%TCIMACRO{\TeXButton{\big)}{\big)}}%
%BeginExpansion
\big)%
%EndExpansion
^{s}$, $n\geq N_{\mathfrak{B}}$, $A\in \mathfrak{B}$, is equicontinuous.
\end{remark}

If one applies the above Chernoff product formula to bilinear elements of
self-dual CAR algebra (Definition \ref{def trace state copy(1)}) then one
gets an expression similar to a \emph{Bernoulli or Euler limit}: Since, by
Definition \ref{def1+}, for any $H\in \mathcal{B}(\mathcal{H})$, 
\begin{equation}
\mathrm{e}^{\langle \mathcal{H},H\mathcal{H}\rangle }=\prod\limits_{i,j\in I}%
\mathrm{e}^{\left\langle \psi _{i},H\psi _{j}\right\rangle _{\mathcal{H}%
}\left( \mathfrak{A}\psi _{j}\right) \wedge \psi _{i}}=\prod\limits_{i,j\in
I}\left( \mathfrak{1}+\left\langle \psi _{i},H\psi _{j}\right\rangle _{%
\mathcal{H}}\left( \mathfrak{A}\psi _{j}\right) \wedge \psi _{i}\right) \ ,
\label{petit calcul}
\end{equation}%
for any basis projection $P$ and operator $H\in \mathcal{B}(\mathcal{H})$,
by (\ref{petit calcul2}),%
\begin{equation}
X_{\langle \mathrm{B},H\mathrm{B}\rangle }^{(n)}=\mathrm{e}^{\mathrm{Tr}_{%
\mathcal{H}}\left( P^{\bot }HP^{\bot }\right) }\left[ \varkappa
_{P}^{-1}\left( \prod\limits_{i,j\in I}\left( \mathfrak{1}+\frac{1}{n}%
\left\langle \psi _{i},H\psi _{j}\right\rangle _{\mathcal{H}}\psi _{j}^{\ast
}\wedge \psi _{i}\right) \right) \right] ^{n}\ ,\qquad n\in {\mathbb{N}}\ ,
\label{trotter approx}
\end{equation}%
where we recall that $\varphi ^{\ast }\doteq \mathfrak{A}\varphi $, by
Definition \ref{definition involution}. Because of Theorem \ref{lemma exp}, $%
X_{\langle \mathrm{B},H\mathrm{B}\rangle }^{(n)}$ converges to $\exp
(\langle \mathrm{B},H\mathrm{B}\rangle )$ in the limit $n\rightarrow \infty $%
.

When $H$ is a self-dual Hamiltonian on $(\mathcal{H},\mathfrak{A})$
(Definition \ref{def one particle hamiltinian}), there is a basis projection 
$P$ diagonalizing $H$ (see (\ref{kappabisbiskappabisbis})), which implies
that the Chernoff product approximation (\ref{trotter approx}) has an
explicit formulation in terms of the generators $\{\mathrm{B}(\varphi
)\}_{\varphi \in \mathcal{\mathcal{H}}}$ of $\mathrm{sCAR}(\mathcal{H},%
\mathfrak{A})$:

\begin{lemma}[Chernoff product approximation and bilinear elements]
\label{chi second II}\mbox{ }\newline
Let $H$ be a self-dual Hamiltonian on $(\mathcal{H},\mathfrak{A})$ and take
a basis projection $P$ diagonalizing $H$. Let $\{\psi _{j}\}_{j\in J}\subset 
\mathfrak{h}_{P}$ be any orthonormal basis of eigenvectors of $h\doteq
2PHP\in \mathcal{B}(\mathfrak{h}_{P})$. Then, for all $n\in {\mathbb{N}}$
with $n>2\left\Vert H\right\Vert _{\mathcal{B}(\mathcal{H})}$ and any $s\in 
\mathbb{R}$, $X_{\langle \mathrm{B},H\mathrm{B}\rangle }^{(n)}>0$ and 
\begin{equation*}
\left( X_{\langle \mathrm{B},H\mathrm{B}\rangle }^{(n)}\right) ^{s}=\mathrm{e%
}^{s\mathrm{Tr}_{\mathcal{H}}\left( P^{\bot }HP^{\bot }\right)
}\prod\limits_{j\in J}\left( \mathfrak{1}+\frac{2}{n}\left\langle \psi
_{j},H\psi _{j}\right\rangle _{\mathcal{H}}\mathrm{B}\left( \psi _{j}\right) 
\mathrm{B}\left( \psi _{j}\right) ^{\ast }\right) ^{sn}.
\end{equation*}
\end{lemma}

\begin{proof}
Under the assumptions of the lemma, the Chernoff product approximations (\ref%
{trotter approx}) are now equal to%
\begin{equation*}
X_{\langle \mathrm{B},H\mathrm{B}\rangle }^{(n)}=\mathrm{e}^{\mathrm{Tr}_{%
\mathcal{H}}\left( P^{\bot }HP^{\bot }\right) }\left[ \varkappa
_{P}^{-1}\left( \prod\limits_{j\in J}\left( \mathfrak{1}+\frac{2}{n}%
\left\langle \psi _{j},H\psi _{j}\right\rangle _{\mathcal{H}}\psi _{j}^{\ast
}\wedge \psi _{j}\right) \right) \right] ^{n}
\end{equation*}%
for all $n\in {\mathbb{N}}$. In particular, by Lemma \ref{Lemma --product}
(iv)-(v) and Definition \ref{definition isomorphism}, 
\begin{equation}
X_{\langle \mathrm{B},H\mathrm{B}\rangle }^{(n)}=\mathrm{e}^{\mathrm{Tr}_{%
\mathcal{H}}\left( P^{\bot }HP^{\bot }\right) }\prod\limits_{j\in J}\left( 
\mathfrak{1}+\frac{2}{n}\left\langle \psi _{j},H\psi _{j}\right\rangle _{%
\mathcal{H}}\mathrm{B}\left( \psi _{j}\right) \mathrm{B}\left( \psi
_{j}\right) ^{\ast }\right) ^{n}\ .  \label{trotter approx2}
\end{equation}%
Obviously, 
\begin{equation*}
\sup_{j\in J}\left\vert \left\langle \psi _{j},H\psi _{j}\right\rangle _{%
\mathcal{H}}\right\vert \leq \left\Vert H\right\Vert _{\mathcal{B}(\mathcal{H%
})}\ .
\end{equation*}%
Therefore, by Definition \ref{def Self--dual CAR Algebras} (c), Remark \ref%
{Remark bounded generator}, and Equation (\ref{trotter approx2}), for any
integer $n>2\left\Vert H\right\Vert _{\mathcal{B}(\mathcal{H})}$, the
Chernoff product approximation $X_{\langle \mathrm{B},H\mathrm{B}\rangle
}^{(n)}$ is the product of commuting and strictly positive elements 
\begin{equation*}
\mathfrak{1}+\frac{2}{n}\left\langle \psi _{j},H\psi _{j}\right\rangle _{%
\mathcal{H}}\mathrm{B}\left( \psi _{j}\right) \mathrm{B}\left( \psi
_{j}\right) ^{\ast }\ ,\qquad j\in J\ .
\end{equation*}%
The assertion then follows.
\end{proof}

\subsection{Feynman-Kac-like Formula for Generating Functions\label%
{Feynman--Kac--like Formula}}

We are now in a position to derive a Feynman-Kac-like formula for terms of
the form (\ref{trace}), that is, 
\begin{equation*}
\mathrm{tr}(\mathrm{e}^{-\beta \mathbf{H}}\mathrm{e}^{sK})\ ,\qquad \beta
\in \mathbb{R}^{+},\ s\in \mathbb{R}\quad \text{and}\quad \mathbf{H}=\mathbf{%
H}^{\ast },K=K^{\ast }\in \mathrm{sCAR}(\mathcal{H},\mathfrak{A})\ ,
\end{equation*}%
with the tracial state $\mathrm{tr}\in \mathrm{sCAR}(\mathcal{H},\mathfrak{A}%
)^{\ast }$ of Definition \ref{def trace state}. They are related to
logarithmic moment generating functions associated with thermodynamic
sequences at equilibrium. See Equations (\ref{trace2})-(\ref{equation
generating funct}).

We restrict our analysis to even elements, that is, elements $A\in \mathrm{%
sCAR}(\mathcal{H},\mathfrak{A})$ satisfying the property $\mathbf{\chi }_{-%
\mathbf{1}_{\mathcal{H}}}(A)=A$, see (\ref{Bogoliubov automorphism}). Then,
the Feynman-Kac-like formula we are seeking is a straightforward consequence
of the Chernoff product formula (Theorem \ref{lemma exp}) and the tracial
state formula (Theorem \ref{satz.spurformel}):

\begin{theorem}[Feynman-Kac-like formula for the tracial state]
\label{coro equation cool2}\mbox{ }\newline
Let $\mathbf{H},K\in \mathrm{sCAR}(\mathcal{H},\mathfrak{A})$ be two even
elements, $K$ being also self-adjoint. Then, for all $\beta \in \mathbb{R}%
^{+}$, $s\in \mathbb{R}$ and any basis projection $P$, 
\begin{multline*}
\mathrm{tr}\left( \mathrm{e}^{-\beta \mathbf{H}}\mathrm{e}^{sK}\right) 
\mathfrak{1}=2^{-\frac{\mathrm{dim}\mathcal{H}}{2}}\lim_{n\rightarrow \infty
}\int_{P}\mathrm{d}\left( \mathfrak{H}^{(n_{\beta })}\right) \mathrm{e}^{%
\frac{1}{2}\langle \mathfrak{H}^{(n_{\beta })},\boldsymbol{\partial }_{P}%
\mathfrak{H}^{(n_{\beta })}\rangle } \\
\wedge \exp \left( \frac{\beta s}{n}\sum_{k=n}^{n_{\beta }-1}\varkappa
_{P}^{(k)}\left( K\right) -\frac{\beta }{n}\sum_{k=0}^{n-1}\varkappa
_{P}^{(k)}(\mathbf{H})\right)
\end{multline*}%
with $\varkappa _{P}^{(k)}$, $\mathfrak{H}^{(n_{\beta })}$ and $\boldsymbol{%
\partial }_{P}$ respectively defined by (\ref{definition xi-k}), (\ref%
{hilbert etendu}) and (\ref{def Ah1new0}). Here, $n_{\beta }\doteq
n+\left\lfloor n/\beta \right\rfloor $, with $\left\lfloor x\right\rfloor $
being the largest natural number smaller than $x\in \mathbb{R}^{+}$.
\end{theorem}

\begin{proof}
Fix $\beta \in \mathbb{R}^{+}$, $s\in \mathbb{R}$, two elements $\mathbf{H}%
,K=K^{\ast }\in \mathrm{sCAR}(\mathcal{H},\mathfrak{A})$ and any basis
projection $P$. We then infer from Theorem \ref{lemma exp} and Remark \ref%
{lemma exp copy(2)} that 
\begin{equation*}
\mathrm{tr}\left( \mathrm{e}^{-\beta \mathbf{H}}\mathrm{e}^{sK}\right)
=\lim_{n\rightarrow \infty }\mathrm{tr}\left( X_{-\beta \mathbf{H}%
}^{(n)}\left( X_{s\beta K}^{(n)}\right) ^{n^{-1}\left\lfloor n/\beta
\right\rfloor }\right) \ ,
\end{equation*}%
which, combined with (\ref{definition xi-k}), Theorem \ref{satz.spurformel}
and Definition \ref{def XA}, in turn implies that%
\begin{align}
\mathrm{tr}\left( \mathrm{e}^{-\beta \mathbf{H}}\mathrm{e}^{sK}\right) &
=2^{-\frac{\mathrm{dim}\mathcal{H}}{2}}\lim_{n\rightarrow \infty }\int_{P}%
\mathrm{d}\left( \mathfrak{H}^{(n_{\beta })}\right) \mathrm{e}^{\frac{1}{2}%
\langle \mathfrak{H}^{(n_{\beta })},\boldsymbol{\partial }_{P}\mathfrak{H}%
^{(n_{\beta })}\rangle }\left[ \prod\limits_{k=0}^{n-1}\exp \left( -\frac{%
\beta }{n}\varkappa _{P}^{(k)}(\mathbf{H})\right) \right]
\label{equation general} \\
& \qquad \qquad \qquad \qquad \qquad \qquad \qquad \wedge
\prod\limits_{k=n}^{n_{\beta }-1}\exp \left( \frac{\beta s}{n}\varkappa
_{P}^{(k)}\left( K\right) \right) \ .  \notag
\end{align}%
Now, if $\mathbf{H}$ and $K$ are even, then this last equation implies the
assertion. Observe that the products involved in the right-hand side of
Equation (\ref{equation general}) all refer to the exterior product $\wedge $
of $\wedge ^{\ast }\mathcal{\mathcal{H}}$. Thus, this space is viewed as
Grassmann algebra (Definition \ref{Definition Grassmann}).
\end{proof}

\begin{remark}[Non-even and non-self-adjoint case]
\label{coro equation cool2 copy(1)}\mbox{ }\newline
Equation (\ref{equation general}) gives a Feynman-Kac-like formula for
traces of the form (\ref{trace}) for all $\mathbf{H},K=K^{\ast }\in \mathrm{%
sCAR}(\mathcal{H},\mathfrak{A})$. We focus our study on the cases for which $%
\mathbf{H}$ and $K$ are even, which is the typical case for fermionic models
in theoretical physics. Note also that the self-adjointness of $K$ could be
relaxed. We omit the details.
\end{remark}

We focus on bilinear elements (Definition \ref{def trace state copy(1)})
perturbed by an even,\ self-adjoint element $W=W^{\ast }\in \mathrm{sCAR}(%
\mathcal{H},\mathfrak{A})$, that is, fermionic Hamiltonians of the form%
\begin{equation}
\mathbf{H}=-\frac{1}{2}\langle \mathrm{B},H\mathrm{B}\rangle +W\ ,\qquad
H=H^{\ast }\in \mathcal{B}(\mathcal{H})\ .  \label{form hamilonian}
\end{equation}%
In this case, the Feynman-Kac-like formula for traces can be written in
terms of a limit of a Gaussian Berezin integral (Definition \ref{gaussian
integral}):

\begin{corollary}[Feynman-Kac-like formula for generating functions]
\label{satz.spurformel copy(1)}\mbox{ }\newline
Let $H\in \mathcal{B}(\mathcal{H})$ be any self-dual Hamiltonian, and let $%
W,K\in \mathrm{sCAR}(\mathcal{H},\mathfrak{A})$ be even elements, with $K$
being self-adjoint. Then, for all $\beta \in \mathbb{R}^{+}$, $s\in \mathbb{R%
}$ and any basis projection $P$ diagonalizing $H$, 
\begin{equation*}
\frac{\mathrm{tr}\left( \mathrm{e}^{\beta \left( \frac{1}{2}\langle \mathrm{B%
},H\mathrm{B}\rangle -W\right) }\mathrm{e}^{sK}\right) }{\mathrm{tr}\left( 
\mathrm{e}^{\frac{\beta }{2}\langle \mathrm{B},H\mathrm{B}\rangle }\right) }%
\mathfrak{1}=\lim_{n\rightarrow \infty }\int \mathrm{d\mu }_{C_{H,P}^{(n)}}(%
\mathfrak{H}^{(n_{\beta })})\exp \left( \frac{\beta s}{n}\sum_{k=n}^{n_{%
\beta }-1}\varkappa _{P}^{(k)}\left( K\right) -\frac{\beta }{n}%
\sum_{k=0}^{n-1}\varkappa _{P}^{(k)}(W)\right) \ .
\end{equation*}%
Here, for any integer $n>\beta \left\Vert H\right\Vert _{\mathcal{B}(%
\mathcal{H})}$,%
\begin{equation}
C_{H,P}^{(n)}\doteq \left( \boldsymbol{\partial }_{P}+\frac{\beta }{n}%
\mathbf{1}\left[ k<n\right] \hat{H}\right) ^{-1}\in \mathcal{B}(\mathfrak{H}%
^{(n_{\beta })})  \label{covariance def}
\end{equation}%
with $\hat{H}$ and $\boldsymbol{\partial }_{P}$ respectively defined by (\ref%
{definition extensino0}) and (\ref{def Ah1new0}), while $\mathbf{1}\left[ k<n%
\right] $ is the orthogonal projection on the subspace $\mathfrak{H}%
^{(n)}\subset \mathfrak{H}^{(n_{\beta })}$ (\ref{hilbert etendu}). Recall
that $n_{\beta }\doteq n+\left\lfloor n/\beta \right\rfloor $.
\end{corollary}

\begin{proof}
Fix all the parameters of the corollary. On the one hand, we deduce from
Lemmata \ref{Lemma --product} (iv)-(v) and \ref{chi second II}, together
with Definition \ref{definition isomorphism} and elementary computations
like (\ref{remark bilinear}) and (\ref{petit calcul}), that, for all $n\in {%
\mathbb{N}}$ such that $n>\beta \left\Vert H\right\Vert _{\mathcal{B}(%
\mathcal{H})}$,%
\begin{equation}
\mathrm{e}^{\frac{\beta }{2n}\langle \mathfrak{H}^{(n)},\hat{H}\mathfrak{H}%
^{(n)}\rangle }=\mathrm{e}^{-\frac{\beta }{2}\mathrm{Tr}_{\mathcal{H}}\left(
P^{\perp }HP^{\perp }\right) }\prod\limits_{k=0}^{n-1}\varkappa
_{P}^{(k)}\left( \left[ X_{\beta \langle \mathrm{B},H\mathrm{B}\rangle
/2}^{(n)}\right] ^{\frac{1}{n}}\right) \ .  \label{step1}
\end{equation}%
The product here refers to the exterior one, $\wedge $ (Notation \ref%
{Notation product}). On the other hand, we deduce from (\ref{equation a la
con}) that%
\begin{equation}
\mathrm{det}\left[ \hat{P}B_{H,P}^{(n)}\hat{P}\right] \mathfrak{1}=\int_{%
\hat{P}}\mathrm{d}\left( \mathfrak{H}^{(n_{\beta })}\right) \mathrm{e}^{%
\frac{1}{2}\langle \mathfrak{H}^{(n_{\beta })},\boldsymbol{\partial }_{P}%
\mathfrak{H}^{(n_{\beta })}\rangle }\mathrm{e}^{\frac{\beta }{2n}\langle 
\mathfrak{H}^{(n)},\hat{H}\mathfrak{H}^{(n)}\rangle }\ ,\qquad n\in \mathbb{N%
}\ ,  \label{equatino a la conbis}
\end{equation}%
with 
\begin{equation*}
B_{H,P}^{(n)}\doteq \frac{1}{2}\left( \boldsymbol{\partial }_{P}+\frac{\beta 
}{n}\mathbf{1}\left[ k<n\right] \hat{H}\right) \in \mathcal{B}(\mathfrak{H}%
^{(n_{\beta })})\ .
\end{equation*}

Therefore, we infer from (\ref{step1}) and (\ref{equatino a la conbis}),
together with Theorem \ref{satz.spurformel}, that 
\begin{equation}
\mathrm{det}\left[ \hat{P}B_{H,P}^{(n)}\hat{P}\right] =2^{\frac{\mathrm{dim}%
\mathcal{H}}{2}}\mathrm{e}^{-\frac{\beta }{2}\mathrm{Tr}_{\mathcal{H}}\left(
P^{\perp }HP^{\perp }\right) }\mathrm{tr}\left( X_{\beta \langle \mathrm{B},H%
\mathrm{B}\rangle /2}^{(n)}\right) >0  \label{step1bis}
\end{equation}%
for any integer $n>\beta \left\Vert H\right\Vert _{\mathcal{B}(\mathcal{H})}$%
, since the tracial state $\mathrm{tr}$ is faithful and $X_{\beta \langle 
\mathrm{B},H\mathrm{B}\rangle /2}^{(n)}>0$ (Lemma \ref{chi second II}). In
particular, for $n>\beta \left\Vert H\right\Vert _{\mathcal{B}(\mathcal{H})}$%
, the operator $\hat{P}B_{H,P}^{(n)}\hat{P}$ is invertible in the range $%
\mathfrak{h}_{\hat{P}}$. Note that $B_{H,P}^{(n)}$ is self-dual on $(%
\mathfrak{H}^{(n_{\beta })},\mathfrak{\hat{A}})$ because $H\in \mathcal{B}(%
\mathcal{H})$ is a self-dual operator by assumption. Moreover, if $P$ is a
basis projection diagonalizing $H$, then $\hat{P}$ is a basis projection
diagonalizing $B_{H,P}^{(n)}$. Hence, $B_{H,P}^{(n)}$ is invertible for $%
n>\beta \left\Vert H\right\Vert _{\mathcal{B}(\mathcal{H})}$ and the
covariance $C_{H,P}^{(n)}$ is, in this case, well-defined.

Now, we infer from Theorem \ref{coro equation cool2} and Equation (\ref%
{petit calcul2}) that%
\begin{eqnarray}
&&2^{\frac{\mathrm{dim}\mathcal{H}}{2}}\mathrm{e}^{-\frac{\beta }{2}\mathrm{%
Tr}_{\mathcal{H}}\left( P^{\perp }HP^{\perp }\right) }\mathrm{tr}\left( 
\mathrm{e}^{\beta \left( \frac{1}{2}\langle \mathrm{B},H\mathrm{B}\rangle
-W\right) }\mathrm{e}^{sK}\right) \mathfrak{1}  \label{det2} \\
&=&\lim_{n\rightarrow \infty }\int_{\hat{P}}\mathrm{d}\left( \mathfrak{H}%
^{(n_{\beta })}\right) \mathrm{e}^{\frac{1}{2}\langle \mathfrak{H}%
^{(n_{\beta })},\boldsymbol{\partial }_{P}\mathfrak{H}^{(n_{\beta })}\rangle
+\frac{\beta }{2n}\langle \mathfrak{H}^{(n)},\hat{H}\mathfrak{H}%
^{(n)}\rangle }\mathrm{e}^{\frac{\beta s}{n}\sum_{k=n}^{n_{\beta
}-1}\varkappa _{P}^{(k)}\left( K\right) -\frac{\beta }{n}\sum_{k=0}^{n-1}%
\varkappa _{P}^{(k)}(W)}\ .  \notag
\end{eqnarray}%
Therefore, by Definition \ref{gaussian integral}, it suffices to combine (%
\ref{equatino a la conbis}) with (\ref{det2}) to get the assertion.
\end{proof}

Corollary \ref{satz.spurformel copy(1)} refers to a Feynman-Kac-like formula
by viewing the variable $k\in \{0,\ldots ,n-1\}$ in the definition (\ref%
{hilbert etendu}) of the finite-dimensional Hilbert space $\mathfrak{H}%
^{(n)} $, $n\in \mathbb{N}$, as a \textquotedblleft \emph{time coordinate}%
\textquotedblright . Each element of a (copied) orthonormal basis $\{\psi
_{i}^{(k)}\}_{i\in I}$ of $\mathcal{H}^{(k)}$ is associated with a
\textquotedblleft \emph{space coordinate}\textquotedblright\ $i\in I$. In
this context, the normal operator $\partial $ defined by (\ref{discrete
derivative}) is a \emph{discrete-time derivative operator}, which, by
Equation (\ref{def Ah1new0}), has the self-dual version $\boldsymbol{%
\partial }_{P}$.

\section{Properties of the Covariance\label{sect det bounds}}

\subsection{Pfaffian Bounds\label{sect det bounds copy(1)}}

Having in mind further applications of the Feynman-Kac-like formula of
Corollary \ref{satz.spurformel copy(1)}, we aim to bound Pfaffians of the
form%
\begin{equation}
\mathrm{Pf}\left[ \mathfrak{M}_{j_{q},j_{l}}\left\langle \mathfrak{A}\varphi
_{q}^{(k_{q})},C_{H,P}^{(n)}\varphi _{l}^{(k_{l})}\right\rangle _{\mathfrak{H%
}^{(n_{\beta })}}\right] _{q,l=1}^{2N}  \label{det to be bounded}
\end{equation}%
for $\beta \in \mathbb{R}^{+}$, $n,m,N\in \mathbb{N}$ with $n$ being
sufficiently large, $\mathfrak{M}\in \mathrm{Mat}\left( m,\mathbb{R}\right) $
with $\mathfrak{M}\geq 0$, and all%
\begin{equation*}
\{(k_{q},\varphi _{q},j_{q})\}_{q=1}^{2N}\subset \{0,\ldots ,n_{\beta
}-1\}\times \mathcal{H}\times \{1,\ldots ,m\}\ .
\end{equation*}%
In (\ref{det to be bounded}), $P$ is a basis projection diagonalizing a
self-dual Hamiltonian $H$ on $(\mathcal{H},\mathfrak{A})$. Recall that $%
n_{\beta }\doteq n+\left\lfloor n/\beta \right\rfloor $ and $C_{H,P}^{(n)}$
is the covariance defined by (\ref{covariance def}) for integers $n>\beta
\left\Vert H\right\Vert _{\mathcal{B}(\mathcal{H})}$. Observe also that the
matrix $\mathfrak{M}$ is symmetric, for it is positive and, hence, the $%
2N\times 2N$ matrix in (\ref{det to be bounded}) is skew-symmetric.

The positive, real matrix $\mathfrak{M}$ appears in the so-called
Brydges-Kennedy tree expansions. See, e.g., \cite[Section 1.3]{universal}
and \cite{LD2} for more details. For previous applications of this
expansion, see also \cite[Section 3]{BGPS}, \cite[Section 3.2]{GM}, and more
recently \cite[Section 5.A.]{GMP}. In \emph{all} previous results, the
covariance refers to the special case $k<n$ in Corollary \ref%
{satz.spurformel copy(1)}. Moreover, as explained in Section \ref{Section
intro LD}, in our approach, multiscale analysis is not necessary to treat
the Matsubara UV problem.

We start by bounding determinants of the form 
\begin{equation}
\mathrm{det}\left[ \left\langle \varphi _{q}^{(k_{q})},C_{H,P}^{(n)}\varphi
_{N+l}^{(k_{N+l})}\right\rangle _{\mathfrak{H}^{(n_{\beta })}}\right]
_{q,l=1}^{N}  \label{det to be bounded1}
\end{equation}%
for $\beta \in \mathbb{R}^{+}$, $n,N\in \mathbb{N}$ with $n$ being
sufficiently large, and all%
\begin{equation*}
\{(k_{q},\varphi _{q})\}_{q=1}^{2N}\subset \{0,\ldots ,n_{\beta }-1\}\times 
\mathfrak{h}_{P}\ .
\end{equation*}%
Then, we show below that the general case $m\in \mathbb{N}$ can always be
reduced to the situation $m=1$ and $\mathfrak{M=}1$, by redefining the
self-dual Hilbert space $\mathcal{H}$, while, for $m=1$ and $\mathfrak{M=}1$%
, the Pfaffian (\ref{det to be bounded}) can be written as the sum of $%
2^{2N} $ determinants of the form (\ref{det to be bounded1}). See
Proposition \ref{gaussian integral properties}.

The ingredients to bound (\ref{det to be bounded1}) are: (I) Corollary \ref%
{gaussian integral properties copy(1)}, which states that%
\begin{equation}
\mathrm{det}\left[ \left\langle \varphi _{q}^{(k_{q})},C_{H,P}^{(n)}\varphi
_{N+l}^{(k_{N+l})}\right\rangle _{\mathfrak{H}^{(n_{\beta })}}\right]
_{q,l=1}^{N}\mathfrak{1}=\int \mathrm{d\mu }_{C_{H,P}^{(n)}}(\mathfrak{H}%
^{(n_{\beta })})\left( \varphi _{1}^{\ast }\right) ^{(k_{1})}\cdots \left(
\varphi _{N}^{\ast }\right) ^{(k_{N})}\varphi _{2N}^{(k_{2N})}\cdots \varphi
_{N+1}^{(k_{N+1})}\ ,  \label{equation cool}
\end{equation}%
where $\varphi ^{\ast }\doteq \mathfrak{A}\varphi $ for all $\varphi \in 
\mathcal{H}$ (Definition \ref{definition involution}); (II) the relationship
between the exterior product $\wedge $ and the circle product $\circ _{P}$
as stated in Lemma \ref{Lemma --product} (iv)-(v); (III) the Chernoff
product approximation of bilinear elements of the self-dual CAR algebra
explained in Lemma \ref{chi second II}; (IV) the tracial state formula
(Theorem \ref{satz.spurformel}), which allows us to represent determinants (%
\ref{equation cool}) as traces; and (V) H\"{o}lder inequalities for Schatten
norms. This approach based on Grassmann-algebra methods turns out to be more
efficient, in the finite-dimensional case, than the construction done in 
\cite{universal}, which involves quasi-free states in suitably-chosen CAR
algebras.

We define Schatten norms in the context of self-dual CAR\ algebra by 
\begin{equation}
\left\Vert A\right\Vert _{s}\doteq \left( \mathrm{tr}\left( \left\vert
A\right\vert ^{s}\right) \right) ^{\frac{1}{s}}\ ,\qquad A\in \mathrm{sCAR}(%
\mathcal{H},\mathfrak{A})\ ,\ s\geq 1\ ,  \label{Holder jean sans bras0}
\end{equation}%
and 
\begin{equation}
\left\Vert A\right\Vert _{\infty }\doteq \lim_{s\rightarrow \infty }\left( 
\mathrm{tr}\left( \left\vert A\right\vert ^{s}\right) \right) ^{\frac{1}{s}%
}=\left\Vert A\right\Vert _{\mathrm{sCAR}(\mathcal{H},\mathfrak{A})}\
,\qquad A\in \mathrm{sCAR}(\mathcal{H},\mathfrak{A})\ ,
\label{Holder jean sans bras00}
\end{equation}%
where $|A|\doteq (A^{\ast }A)^{1/2}$ and $\mathrm{tr}\in \mathrm{sCAR}(%
\mathcal{H},\mathfrak{A})^{\ast }$ is the tracial state of Definition \ref%
{def trace state}. See also Remark \ref{bound norm a CAR copy(2)}. H\"{o}%
lder inequalities for Schatten norms then refer to the following bounds: For
any $m\in \mathbb{N}$, $r,s_{1},\ldots ,s_{m}\in \lbrack 1,\infty ]$ such
that $\sum_{j=1}^{m}1/s_{j}=1/r$, and all elements $A_{1},\ldots ,A_{m}\in 
\mathrm{sCAR}(\mathcal{H},\mathfrak{A})$,%
\begin{equation}
\left\Vert A_{1}\cdots A_{m}\right\Vert _{r}\leq \prod_{j=1}^{m}\left\Vert
A_{j}\right\Vert _{s_{j}}\ .  \label{Holder jean sans bras}
\end{equation}%
This type of inequality, combined with (\ref{equation cool}) and Lemmata \ref%
{Lemma --product} and \ref{chi second II}, yields a \emph{sharp} bound on
the determinant (\ref{det to be bounded1}):

\begin{theorem}[Determinant bounds]
\label{det bound I}\mbox{ }\newline
Let $H$ be a self-dual Hamiltonian on $(\mathcal{H},\mathfrak{A})$ and take
any basis projection $P$ diagonalizing $H$. Then, for $\beta \in \mathbb{R}%
^{+}$, $n,N\in \mathbb{N}$ with $n>\beta \left\Vert H\right\Vert _{\mathcal{B%
}(\mathcal{H})}$, all $k_{1},\ldots ,k_{2N}\in \{0,\ldots ,n_{\beta }-1\}$ ($%
n_{\beta }\doteq n+\left\lfloor n/\beta \right\rfloor $) and $\varphi
_{1},\ldots ,\varphi _{2N}\in \mathfrak{h}_{P}$, the following bound holds
true: 
\begin{equation*}
\left\vert \mathrm{det}\left[ \left\langle \varphi
_{q}^{(k_{q})},C_{H,P}^{(n)}\varphi _{N+l}^{(k_{N+l})}\right\rangle _{%
\mathfrak{H}^{(n_{\beta })}}\right] _{q,l=1}^{N}\right\vert \leq
\prod_{q=1}^{2N}\left\Vert \varphi _{q}\right\Vert _{\mathcal{H}}\text{ }.
\end{equation*}
\end{theorem}

\begin{proof}
Fix all parameters of the theorem. Pick any orthonormal basis $\{\psi
_{j}\}_{j\in J}\subset \mathfrak{h}_{P}$ of eigenvectors of $H_{P}\doteq
2PHP\in \mathcal{B}(\mathfrak{h}_{P})$. Similar to the derivation of (\ref%
{step1}), by Lemmata \ref{Lemma --product} (iv) and \ref{chi second II},
together with Condition (c) (Definition \ref{def Self--dual CAR Algebras}),
Definition \ref{definition isomorphism}, and Equation (\ref{grassmana
anticommute}), for any $n_{1},n_{2}\in \mathbb{N}$, and all indices $%
j_{1},\ldots ,j_{n_{1}+n_{2}}\in J$, 
\begin{eqnarray*}
&&\psi _{j_{1}}^{\ast }\cdots \psi _{j_{n_{1}}}^{\ast }\varkappa _{P}\left( 
\left[ X_{\beta \langle \mathrm{B},H\mathrm{B}\rangle /2}^{(n)}\right] ^{%
\frac{1}{n}}\right) \psi _{j_{n_{1}+1}}\cdots \psi _{j_{n_{1}+n_{2}}} \\
&=&\varkappa _{P}\left( \mathrm{B}(\psi _{j_{1}})\cdots \mathrm{B}(\psi
_{j_{n_{1}}})\left[ X_{\beta \langle \mathrm{B},H\mathrm{B}\rangle /2}^{(n)}%
\right] ^{\frac{1}{n}}\mathrm{B}(\psi _{j_{n_{1}+1}})^{\ast }\cdots \mathrm{B%
}(\psi _{j_{n_{1}+n_{2}}})^{\ast }\right) \ .
\end{eqnarray*}%
By taking linear combinations of the above equalities, we thus deduce that,
for any $n_{1},n_{2}\in \mathbb{N}$ and all $\varphi _{1},\ldots ,\varphi
_{n_{1}+n_{2}}\in \mathfrak{h}_{P}$,%
\begin{eqnarray}
&&\varphi _{1}^{\ast }\cdots \varphi _{n_{1}}^{\ast }\varkappa _{P}\left( 
\left[ X_{\beta \langle \mathrm{B},H\mathrm{B}\rangle /2}^{(n)}\right] ^{%
\frac{1}{n}}\right) \varphi _{n_{1}+1}\cdots \varphi _{n_{1}+n_{2}}
\label{step2} \\
&=&\varkappa _{P}\left( \mathrm{B}(\varphi _{1})\cdots \mathrm{B}(\varphi
_{n_{1}})\left[ X_{\beta \langle \mathrm{B},H\mathrm{B}\rangle /2}^{(n)}%
\right] ^{\frac{1}{n}}\mathrm{B}(\varphi _{n_{1}+1})^{\ast }\cdots \mathrm{B}%
(\varphi _{n_{1}+n_{2}})^{\ast }\right) \ .  \notag
\end{eqnarray}%
We are now in a position to write the determinant as a trace in order to
next use H\"{o}lder inequalities: By Definition \ref{gaussian integral} and
Equations (\ref{covariance def})-(\ref{step1bis}) and (\ref{equation cool}), 
\begin{equation}
\mathrm{det}\left[ \left\langle \varphi _{q}^{(k_{q})},C_{H,P}^{(n)}\varphi
_{N+l}^{(k_{N+l})}\right\rangle _{\mathfrak{H}^{(n_{\beta })}}\right]
_{q,l=1}^{N}\mathfrak{1}=\frac{2^{-\frac{\mathrm{dim}\mathcal{H}}{2}}}{%
\mathrm{tr}\left( X_{\beta \langle \mathrm{B},H\mathrm{B}\rangle
/2}^{(n)}\right) }\int_{P}\mathrm{d}\left( \mathfrak{H}^{(n_{\beta
})}\right) \mathrm{e}^{\frac{1}{2}\langle \mathfrak{H}^{(n_{\beta })},%
\boldsymbol{\partial }_{P}\mathfrak{H}^{(n_{\beta })}\rangle }\mathfrak{Q}
\label{step31}
\end{equation}%
with 
\begin{equation*}
\mathfrak{Q}\doteq \left( \prod\limits_{k=0}^{n-1}\varkappa _{P}^{(k)}\left( %
\left[ X_{\beta \langle \mathrm{B},H\mathrm{B}\rangle /2}^{(n)}\right] ^{%
\frac{1}{n}}\right) \right) \left( \varphi _{1}^{\ast }\right)
^{(k_{1})}\cdots \left( \varphi _{N}^{\ast }\right) ^{(k_{N})}\varphi
_{2N}^{(k_{2N})}\cdots \varphi _{N+1}^{(k_{N+1})}\ .
\end{equation*}%
To get the assertion, it suffices now to reorganize the exterior products $%
\wedge $ in $\mathfrak{Q}$ by regrouping all terms associated with the same
index $k\in \{0,\ldots ,n_{\beta }-1\}$. Note that an additional minus sign
can appear, because of the antisymmetry of $\wedge $. Doing this, we rewrite
the element $\mathfrak{Q}$ in the form $\pm \varkappa _{P}^{(0)}\left(
A_{0}\right) \wedge \cdots \wedge \varkappa _{P}^{(n_{\beta }-1)}\left(
A_{n_{\beta }-1}\right) $ and, by Lemma \ref{Lemma --product} (iv) and
Theorem \ref{satz.spurformel}, together with (\ref{step31}), 
\begin{equation}
\mathrm{det}\left[ \left\langle \varphi _{q}^{(k_{q})},C_{H,P}^{(n)}\varphi
_{N+l}^{(k_{N+l})}\right\rangle _{\mathfrak{H}^{(n_{\beta })}}\right]
_{q,l=1}^{N}=\pm \frac{1}{\mathrm{tr}\left( X_{\beta \langle \mathrm{B},H%
\mathrm{B}\rangle /2}^{(n)}\right) }\mathrm{tr}(A_{0}\cdots A_{n_{\beta
}-1})\ .  \label{step32}
\end{equation}%
More precisely, using Definition \ref{definition isomorphism} and (\ref%
{definition xi-k}), for any $k\notin \{k_{1},\ldots ,k_{2N}\}$, 
\begin{equation*}
A_{k}=\left[ X_{\beta \langle \mathrm{B},H\mathrm{B}\rangle /2}^{(n)}\right]
^{\frac{1}{n}}
\end{equation*}%
unless $k\in \{n,\ldots ,n_{\beta }-1\}$ in which case $A_{k}=\mathfrak{1}$.
If $k\in \{k_{1},\ldots ,k_{2N}\}$ then $\varkappa _{P}(A_{k})$ is of the
form (\ref{step2}), with $H$ being zero whenever $k\in \{n,\ldots ,n_{\beta
}-1\}$. Therefore, one can apply the inequality 
\begin{equation*}
\left\vert \mathrm{tr}\left( B\right) \right\vert \leq \mathrm{tr}\left(
\left\vert B\right\vert \right) \ ,\qquad B\in \mathrm{sCAR}(\mathcal{H},%
\mathfrak{A})\ ,
\end{equation*}%
and H\"{o}lder inequalities (\ref{Holder jean sans bras}) to the right-hand
side of (\ref{step32}), with $m=n+2N$, $r=1$, $s_{j}=n$ for each term $%
[X_{\beta \langle \mathrm{B},H\mathrm{B}\rangle /2}^{(n)}]^{\frac{1}{n}}$
and $s_{j}=\infty $ for every generator $\mathrm{B}(\varphi )$ (where $%
\varphi \in \mathfrak{h}_{P}$). Since, by Remark \ref{Remark bounded
generator}, Lemma \ref{chi second II} and Equation (\ref{Holder jean sans
bras00}), 
\begin{equation*}
\left\Vert \left[ X_{\beta \langle \mathrm{B},H\mathrm{B}\rangle /2}^{(n)}%
\right] ^{\frac{1}{n}}\right\Vert _{n}=\left( \mathrm{tr}\left( X_{\beta
\langle \mathrm{B},H\mathrm{B}\rangle /2}^{(n)}\right) \right) ^{\frac{1}{n}%
}\qquad \text{and}\qquad \left\Vert \mathrm{B}\left( \varphi \right)
\right\Vert _{\infty }\leq \left\Vert \varphi \right\Vert _{\mathcal{H}}\ ,
\end{equation*}%
the assertion then follows.
\end{proof}

\begin{corollary}[Pfaffian bounds -- I]
\label{Pfaffian bounds I}\mbox{ }\newline
Let $H$ be a self-dual Hamiltonian on $(\mathcal{H},\mathfrak{A})$ and take
any basis projection $P$ diagonalizing $H$. Then, for $\beta \in \mathbb{R}%
^{+}$, $n,N\in \mathbb{N}$ with $n>\beta \left\Vert H\right\Vert _{\mathcal{B%
}(\mathcal{H})}$, all $k_{1},\ldots ,k_{2N}\in \{0,\ldots ,n_{\beta }-1\}$
and $\varphi _{1},\ldots ,\varphi _{2N}\in \mathcal{H}$, the following bound
holds true: 
\begin{equation*}
\left\vert \mathrm{Pf}\left[ \left\langle \mathfrak{A}\varphi
_{q}^{(k_{q})},C_{H,P}^{(n)}\varphi _{l}^{(k_{l})}\right\rangle _{\mathfrak{H%
}^{(n_{\beta })}}\right] _{q,l=1}^{2N}\right\vert \leq
\prod_{q=1}^{2N}\left( \left\Vert P\varphi _{q}\right\Vert _{\mathcal{H}%
}+\left\Vert P^{\perp }\varphi _{q}\right\Vert _{\mathcal{H}}\right) \text{ }%
.
\end{equation*}
\end{corollary}

\begin{proof}
The proof is a direct consequence of Proposition \ref{gaussian integral
properties}, Theorem \ref{det bound I}, Equations (\ref{frakA and perp}), (%
\ref{definition H bar}) and (\ref{equation cool}), together with the fact
that $\hat{P}$ diagonalizes the self-dual operator $C_{H,P}^{(n)}$.
\end{proof}

Bounds for general Pfaffians of the form (\ref{det to be bounded}) can be
deduced from Corollary \ref{Pfaffian bounds I}, which corresponds to the
special case $m=1$ and $\mathfrak{M=}1$. This is done by extending the
self-dual Hilbert space $(\mathcal{H},\mathfrak{A})$, similar to what is
done in \cite[Section 1.3]{universal}: For any fixed $m\in \mathbb{N}$, a
(generic) non-vanishing real positive matrix $\mathfrak{M}\in \mathrm{Mat}%
\left( m,\mathbb{R}\right) $ gives rise to a positive sesquilinear form on $%
\mathbb{C}^{m}$ defined by 
\begin{equation}
\left\langle \left( x_{1},\ldots ,x_{m}\right) ,\left( y_{1},\ldots
,y_{m}\right) \right\rangle _{\mathbb{C}^{m}}^{\mathfrak{M}}\doteq
\sum_{j,l=1}^{m}\overline{x_{j}}\ y_{l}\mathfrak{M}_{j,l}\ .
\label{scalar product}
\end{equation}%
In general, this sesquilinear form is degenerated. Then, one can define a
Hilbert space 
\begin{equation*}
\mathbb{M}\doteq \mathbb{C}^{m}/\{x\in \mathbb{C}^{m}:\left\langle
x,x\right\rangle _{\mathbb{C}^{m}}^{\mathfrak{M}}=0\}
\end{equation*}%
whose scalar product is defined by 
\begin{equation*}
\left\langle \left[ x\right] ,\left[ y\right] \right\rangle _{\mathbb{M}%
}\doteq \left\langle x,y\right\rangle _{\mathbb{C}^{m}}^{\mathfrak{M}}\
,\qquad x,y\in \mathbb{C}^{m}\ .
\end{equation*}%
The map from $\mathbb{C}^{m}$ to itself defined by%
\begin{equation*}
x=\left( x_{1},\ldots ,x_{m}\right) \mapsto \bar{x}\doteq \left( \overline{%
x_{1}},\ldots ,\overline{x_{m}}\right)
\end{equation*}%
yields an antiunitary involution $[x]\mapsto \lbrack \bar{x}]$ on $\mathbb{M}
$ denoted by $\mathbb{A}$.

Then, it suffices to replace $(\mathcal{H},\mathfrak{A})$ with $(\mathcal{H}%
\otimes \mathbb{M},\mathfrak{A}\otimes \mathbb{A})$ to get, from Corollary %
\ref{Pfaffian bounds I}, bounds for Pfaffians of the form (\ref{det to be
bounded}):

\begin{corollary}[Pfaffian bounds -- II]
\label{Pfaffian bounds II}\mbox{ }\newline
Let $H$ be a self-dual Hamiltonian on $(\mathcal{H},\mathfrak{A})$ and take
any basis projection $P$ diagonalizing $H$. Then, for $\beta \in \mathbb{R}%
^{+}$, $n,m,N\in \mathbb{N}$ with $n>\beta \left\Vert H\right\Vert _{%
\mathcal{B}(\mathcal{H})}$, $\mathfrak{M}\in \mathrm{Mat}\left( m,\mathbb{R}%
\right) $ with $\mathfrak{M}\geq 0$, and all 
\begin{equation*}
\{(k_{q},\varphi _{q},j_{q})\}_{q=1}^{2N}\subset \{0,\ldots ,n_{\beta
}-1\}\times \mathcal{H}\times \{1,\ldots ,m\}\ ,
\end{equation*}%
the following bound holds true: 
\begin{eqnarray*}
\left\vert \mathrm{Pf}\left[ \mathfrak{M}_{j_{q},j_{l}}\left\langle 
\mathfrak{A}\varphi _{q}^{(k_{q})},C_{H,P}^{(n)}\varphi
_{l}^{(k_{l})}\right\rangle _{\mathfrak{H}^{(n_{\beta })}}\right]
_{q,l=1}^{2N}\right\vert &\leq &\prod_{q=1}^{2N}\left( \left\Vert P\varphi
_{q}\right\Vert _{\mathcal{H}}+\left\Vert P^{\perp }\varphi _{q}\right\Vert
_{\mathcal{H}}\right) \mathfrak{M}_{j_{q},j_{q}}^{1/2} \\
&\leq &2^{N}\prod_{q=1}^{2N}\left\Vert \varphi _{q}\right\Vert _{\mathcal{H}}%
\mathfrak{M}_{j_{q},j_{q}}^{1/2}\text{ }.
\end{eqnarray*}
\end{corollary}

\begin{proof}
Fix $m\in \mathbb{N}$ and $\mathfrak{M}\in \mathrm{Mat}\left( m,\mathbb{R}%
\right) $ with $\mathfrak{M}\geq 0$. Let $\mathcal{\tilde{H}}\doteq \mathcal{%
H}\otimes \mathbb{M}$ and $\mathfrak{\tilde{A}\doteq A}\otimes \mathbb{A}$.
Any operator $A\in \mathcal{B}(\mathcal{H})$ can be extended to an operator
acting on $\mathcal{\tilde{H}}$ by the definition $\tilde{A}\doteq A\otimes 
\mathbf{1}_{\mathbb{M}}\in \mathcal{B}(\mathcal{\tilde{H}})$. A basis
projection $P$ (range $\mathfrak{h}_{P}$) associated with $(\mathcal{H},%
\mathfrak{A})$ yields a basis projection $\tilde{P}$ associated with $(%
\mathcal{\tilde{H}},\mathfrak{\tilde{A}})$, whose range is $\mathfrak{h}_{%
\tilde{P}}=\mathfrak{h}_{P}\otimes \mathbb{M}$. See Definition \ref{def
basis projection}. If $H$ is a self-dual Hamiltonian on $(\mathcal{H},%
\mathfrak{A})$ (Definition \ref{def one particle hamiltinian}), then $\tilde{%
H}$ is also a self-dual Hamiltonian on $(\mathcal{\tilde{H}},\mathfrak{%
\tilde{A}})$, with the same norm. If the basis projection $P$ diagonalizes $%
H $ then $\tilde{P}$ also diagonalizes $\tilde{H}$.

Meanwhile, we use the canonical identification%
\begin{equation*}
\mathfrak{\tilde{H}}^{(n)}\doteq \bigoplus\limits_{k=0}^{n-1}\mathcal{\tilde{%
H}}^{(k)}\equiv \mathfrak{H}^{(n)}\otimes \mathbb{M}\text{ },\qquad n\in 
\mathbb{N}\ ,
\end{equation*}%
via the unitary map uniquely defined by the conditions 
\begin{equation*}
(\varphi \otimes \lbrack x])^{(k)}\mapsto \varphi ^{(k)}\otimes \lbrack x]%
\text{ },\qquad \varphi \in \mathcal{H},\ x\in \mathbb{C}^{m}\ ,
\end{equation*}%
for any $k\in \{0,\ldots ,n-1\}$. It is then straightforward to verify that%
\begin{equation*}
C_{\tilde{H},\tilde{P}}^{(n)}=C_{H,P}^{(n)}\otimes \mathbf{1}_{\mathbb{M}}\ .
\end{equation*}%
See Equations (\ref{hilbert etendu}), (\ref{def Ah1new0}) and (\ref%
{covariance def}).

Now, using the notation $\mathfrak{e}_{j}\doteq \left[ e_{j}\right] \in 
\mathbb{M}$, where $\left\{ e_{j}\right\} _{j=1}^{m}$ is the canonical basis
of $\mathbb{C}^{m}$, note that 
\begin{equation}
\mathfrak{M}_{j,l}=\left\langle \mathfrak{e}_{j},\mathfrak{e}%
_{l}\right\rangle _{\mathbb{M}}\ ,\qquad j,l\in \left\{ 1,\ldots ,m\right\}
\ .  \label{ej fract}
\end{equation}%
Therefore, if $H$ is a self-dual Hamiltonian on $(\mathcal{H},\mathfrak{A})$
and $P$ any basis projection $P$ diagonalizing $H$, then the assertion is a
direct consequence of Corollary \ref{Pfaffian bounds I} for $\beta \in 
\mathbb{R}^{+}$, $n,N\in \mathbb{N}$ with $n>\beta \left\Vert H\right\Vert _{%
\mathcal{B}(\mathcal{H})}$, $j_{1},\ldots ,j_{2N}\in \{1,\ldots ,m\}$, and
all elements 
\begin{equation*}
\{(k_{q},\varphi _{q}\otimes \mathfrak{e}_{j_{q}})\}_{q=1}^{2N}\subset
\{0,\ldots ,n_{\beta }-1\}\times \mathfrak{h}_{\tilde{P}}\ .
\end{equation*}%
Remark from (\ref{ej fract}) that 
\begin{equation*}
\Vert \tilde{P}(\varphi \otimes \mathfrak{e}_{j})\Vert _{\mathcal{\tilde{H}}%
}+\Vert \tilde{P}^{\perp }(\varphi \otimes \mathfrak{e}_{j})\Vert _{\mathcal{%
\tilde{H}}}=\left( \Vert P\varphi \Vert _{\mathcal{H}}+\Vert P^{\perp
}\varphi \Vert _{\mathcal{H}}\right) \mathfrak{M}_{j,j}^{1/2}\ ,\qquad
\varphi \in \mathcal{H},\ j\in \left\{ 1,\ldots ,m\right\} \ .
\end{equation*}
\end{proof}

\begin{remark}
\label{Remark bound}\mbox{ }\newline
Similar to \cite[Corollary 1.6]{universal} one can check that the bound on
Pfaffians obtained in Corollary \ref{Pfaffian bounds II} is sharp if $%
\varphi _{1},\ldots ,\varphi _{2N}\in \mathfrak{h}_{P}\cup \mathfrak{h}%
_{P^{\perp }}$, in the sense given in \cite[Equation (14)]{universal}. In
particular, the general Pfaffian bound of Corollary \ref{Pfaffian bounds II}
is sharp up to a factor of $2^{N}$.
\end{remark}

\subsection{Covariances as Correlations of Quasi-Free States\label{sec
Covariances as Correlations of Quasi--Free States}}

Fix, in all this subsection, an inverse temperature $\beta \in \mathbb{R}%
_{0}^{+}$, some self-dual Hamiltonian $H$ on $(\mathcal{H},\mathfrak{A})$, a
basis projection $P$ diagonalizing $H$, and $n\in \mathbb{N}$ so that $%
n>\beta \left\Vert H\right\Vert _{\mathcal{B}(\mathcal{H})}$. Using the
trace representation (\ref{step32}) of determinants associated with
covariances, one can rewrite the matrix elements%
\begin{equation}
\left\langle \hat{\varphi}_{1},C_{H,P}^{(n)}\hat{\varphi}_{2}\right\rangle _{%
\mathfrak{H}^{(n_{\beta })}}\ ,\qquad \hat{\varphi}_{1},\hat{\varphi}_{2}\in 
\mathfrak{H}^{(n_{\beta })},  \label{matrix element}
\end{equation}%
in terms of space-time correlations of a quasi-free state.

Using the map $\mathbf{K}:\mathcal{B}(\mathcal{\mathcal{H})\rightarrow B}(%
\mathcal{\mathcal{H})}$ (\ref{equation idiote 3bis}), we define a sequence
of self-dual Hamiltonians on $(\mathcal{H},\mathfrak{A})$ by%
\begin{equation}
H^{(n)}\doteq \beta ^{-1}n\mathbf{K}\left( \ln \left( 1+n^{-1}\beta H\right)
\right) =\frac{\beta ^{-1}n}{2}\ln \left( \frac{1+n^{-1}\beta H}{%
1-n^{-1}\beta H}\right)  \label{Hn}
\end{equation}%
for any $n>\beta \left\Vert H\right\Vert _{\mathcal{B}(\mathcal{H})}$. If
the basis projection $P$ diagonalizes $H$ then it also diagonalizes $H^{(n)}$%
. In view of the trace representation (\ref{step32}) and Lemma \ref{Lemma
quasi free state}, this Hamitonian is constructed such that the equality 
\begin{equation}
X_{\beta \langle \mathrm{B},H\mathrm{B}\rangle /2}^{(n)}=D\mathrm{e}^{\beta
\langle \mathrm{B},H^{(n)}\mathrm{B}\rangle /2}\ ,\qquad \beta \in \mathbb{R}%
_{0}^{+}\ ,  \label{eq sdjkmf}
\end{equation}%
holds true for some finite constant $D>0$. See (\ref{remark bilinear}) and
Lemma \ref{chi second II}. Observe meanwhile that $H^{(n)}$ is an
approximation of $H$: There is a finite constant $D>0$ such that, for $\beta
\in \mathbb{R}_{0}^{+}$ and $n>2\beta \left\Vert H\right\Vert _{\mathcal{B}(%
\mathcal{H})}$, 
\begin{equation}
\left\Vert H^{(n)}-H\right\Vert _{\mathcal{B}(\mathcal{H})}\leq Dn^{-2}\beta
^{2}\left\Vert H\right\Vert _{\mathcal{B}(\mathcal{H})}^{3}\ .  \label{Hnbis}
\end{equation}%
Compare also (\ref{Hn}) with \cite[Eqs. (16)-(17)]{universal}.

Then, using the definition $\left[ x\right] _{+}\doteq (x+|x|)/2$ for $x\in 
\mathbb{R}$ and%
\begin{equation}
\alpha _{n}\left( k,q\right) \doteq \beta n^{-1}\left[ \min \left\{
n-k,q-k\right\} \right] _{+}\in \left[ 0,\beta \right] \ ,\qquad k,q\in 
\mathbb{Z}\ ,  \label{alphanbis}
\end{equation}%
we rewrite the matrix elements (\ref{matrix element}), at fixed time
coordinates, in terms of space-time correlations of the quasi-free states $%
\rho _{H^{(n)}}$ with symbol $(1+\mathrm{e}^{-\beta H^{(n)}})^{-1}$ (see (%
\ref{ass O0-00})-(\ref{ass O0-00bis}) and (\ref{symbol})-(\ref{symbolbis})):

\begin{theorem}[Covariances as correlations of quasi-free states]
\label{covaraince sum}\mbox{ }\newline
Let $H$ be a self-dual Hamiltonian on $(\mathcal{H},\mathfrak{A})$ and take
any basis projection $P$ diagonalizing $H$. Then, for any $\beta \in \mathbb{%
R}^{+}$, $n\in \mathbb{N}$ with $n>\beta \left\Vert H\right\Vert _{\mathcal{B%
}(\mathcal{H})}$, $k_{1},k_{2}\in \{0,\ldots ,n_{\beta }-1\}$ and $\varphi
_{1},\varphi _{2}\in \mathcal{H}$, 
\begin{equation*}
\left\langle \varphi _{1}^{(k_{1})},C_{H,P}^{(n)}\varphi
_{2}^{(k_{2})}\right\rangle _{\mathfrak{H}^{(n_{\beta })}}=\left\langle
\varphi _{1},\mathfrak{C}_{H,P}^{(n)}\left( k_{1},k_{2}\right) \varphi
_{2}\right\rangle _{\mathcal{H}},
\end{equation*}%
where%
\begin{equation}
\mathfrak{C}_{H,P}^{(n)}\left( k_{1},k_{2}\right) \doteq \left\{ 
\begin{array}{lll}
P\frac{\mathrm{e}^{-\alpha _{n}\left( k_{1},k_{2}+1\right) H^{(n)}}}{1+%
\mathrm{e}^{-\beta H^{(n)}}}P+P^{\bot }\frac{\mathrm{e}^{-\alpha _{n}\left(
k_{1}+1,k_{2}\right) H^{(n)}}}{1+\mathrm{e}^{-\beta H^{(n)}}}P^{\bot } & 
\text{for} & k_{1}<k_{2}\ , \\ 
-P\frac{\mathrm{e}^{\alpha _{n}\left( k_{2}+1,k_{1}\right) H^{(n)}}}{1+%
\mathrm{e}^{\beta H^{(n)}}}P-P^{\bot }\frac{\mathrm{e}^{\alpha _{n}\left(
k_{2},k_{1}+1\right) H^{(n)}}}{1+\mathrm{e}^{\beta H^{(n)}}}P^{\bot } & 
\text{for} & k_{1}>k_{2}\ , \\ 
P\frac{\mathrm{e}^{-\alpha _{n}\left( k_{1},k_{1}+1\right) H^{(n)}}}{1+%
\mathrm{e}^{-\beta H^{(n)}}}P-P^{\bot }\frac{\mathrm{e}^{\alpha _{n}\left(
k_{1},k_{1}+1\right) H^{(n)}}}{1+\mathrm{e}^{\beta H^{(n)}}}P^{\bot } & 
\text{for} & k_{1}=k_{2}\ .%
\end{array}%
\right.  \label{alphanbis2}
\end{equation}
\end{theorem}

\begin{proof}
Fix all parameters of the theorem. By construction, the covariance $%
C_{H,P}^{(n)}$ is a self-dual operator on $(\mathfrak{H}^{(n_{\beta })},%
\mathfrak{\hat{A}})$, which is diagonalized by the extension $\hat{P}$ of $P$
(see (\ref{definition extensino0}) with $n_{\beta }$ replacing $n$). In
particular, for any $\varphi _{1},\varphi _{2}\in \mathcal{H}$\ and $%
k_{1},k_{2}\in \{0,\ldots ,n_{\beta }-1\}$,%
\begin{eqnarray}
\left\langle \varphi _{1}^{(k_{1})},C_{H,P}^{(n)}\varphi
_{2}^{(k_{2})}\right\rangle _{\mathfrak{H}^{(n_{\beta })}} &=&\left\langle
\left( P\varphi _{1}\right) ^{(k_{1})},C_{H,P}^{(n)}\left( P\varphi
_{2}\right) ^{(k_{2})}\right\rangle _{\mathfrak{H}^{(n_{\beta })}}
\label{eq a  la con} \\
&&-\left\langle \left( P\mathfrak{A}\varphi _{2}\right)
^{(k_{2})},C_{H,P}^{(n)}\left( P\mathfrak{A}\varphi _{1}\right)
^{(k_{1})}\right\rangle _{\mathfrak{H}^{(n_{\beta })}}.  \notag
\end{eqnarray}%
Therefore, it suffices to do computations for $\varphi _{1},\varphi _{2}\in 
\mathfrak{h}_{P}$ only. In this special case, it is straightforward, albeit
cumbersome, to derive the following equalities from Equations (\ref{step32})
and (\ref{eq sdjkmf}), together with Lemmata \ref{Lemma quasi-free} and \ref%
{Lemma quasi free state}: If $k_{1}\leq k_{2}$ then 
\begin{equation*}
\left\langle \varphi _{1}^{(k_{1})},C_{H,P}^{(n)}\varphi
_{2}^{(k_{2})}\right\rangle _{\mathfrak{H}^{(n_{\beta })}}=\rho
_{H^{(n)}}\left( \mathrm{B}(\mathrm{e}^{-\alpha _{n}\left(
k_{1},k_{2}+1\right) H^{(n)}}\varphi _{1})\mathrm{B}(\mathfrak{A}\varphi
_{2})\right) \ ,
\end{equation*}%
whereas if $k_{2}<k_{1}$, 
\begin{equation*}
\left\langle \varphi _{1}^{(k_{1})},C_{H,P}^{(n)}\varphi
_{2}^{(k_{2})}\right\rangle _{\mathfrak{H}^{(n_{\beta })}}=-\rho
_{H^{(n)}}\left( \mathrm{B}(\mathfrak{A}\varphi _{2})\mathrm{B}(\mathrm{e}%
^{\alpha _{n}\left( k_{2}+1,k_{1}\right) H}\varphi _{1})\right) \ .
\end{equation*}%
By combining these two equalities for $\varphi _{1},\varphi _{2}\in 
\mathfrak{h}_{P}$ with Equations (\ref{symbolbis}) and (\ref{eq a la con}),
we directly get the theorem. Note additionally that, $C_{H,P}^{(n)}$ being
self-dual, for all $k_{1},k_{2}\in \{0,\ldots ,n_{\beta }-1\}$,%
\begin{equation*}
\mathfrak{C}_{H,P}^{(n)}\left( k_{2},k_{1}\right) =-\mathfrak{AC}%
_{H,P}^{(n)}\left( k_{1},k_{2}\right) ^{\ast }\mathfrak{A}\text{ }.
\end{equation*}%
This relation can be used to obtain the first line in the right-hand side of
(\ref{alphanbis2}) from the second one and vice versa, or to check mutual
consistency of these two lines.
\end{proof}

\subsection{Summability\label{Section summability}}

In this section we derive bounds on the decay of the covariances (\ref%
{covariance def}) for general fermion systems on the lattice, from the
celebrated Combes-Thomas estimates. See also \cite{Roeck-Salmhofer} which
has simultaneously been done and uses Combes-Thomas estimates.

\subsubsection{Lattice Fermion Systems in the Self-dual Formalism}

We consider fermion systems on the lattice $\mathbb{Z}^{d}$ ($d\in \mathbb{N}
$) with arbitrary (finite) spin set $\mathrm{S}$. We define the Hilbert
spaces $\mathcal{H}_{\mathrm{S}}\doteq \ell ^{2}\left( \mathrm{S}\right)
\oplus \ell ^{2}\left( \mathrm{S}\right) ^{\ast }$ and $\mathcal{H}%
_{L}\doteq \ell ^{2}\left( \Lambda _{L};\mathcal{H}_{\mathrm{S}}\right) $
for all $L\in \mathbb{R}_{0}^{+}\cup \{\infty \}$, where 
\begin{equation}
\Lambda _{L}\doteq \{(x_{1},\ldots ,x_{d})\in \mathbb{Z}^{d}:|x_{1}|,\ldots
,|x_{d}|\leq L\}  \label{eq:boxesl1}
\end{equation}%
is a cubic box of side length $\mathcal{O(}L)$. See also Sections \ref%
{notation hilbert} and \ref{Spin or CAR Unital Algebras}, in particular
Equation (\ref{Hilbert self dual}).

Fix any antiunitary involution $\mathfrak{A}_{\mathrm{S}}$ on $\mathcal{H}_{%
\mathrm{S}}$. A canonical example is given by (\ref{antiunitary simple}) for 
$\mathcal{H}=\mathcal{H}_{\mathrm{S}}$. For any $L\in \mathbb{R}_{0}^{+}\cup
\{\infty \}$, we define an antiunitary involution $\mathfrak{A}_{L}$ on $%
\mathcal{H}_{L}$ by%
\begin{equation}
\left( \mathfrak{A}_{L}\varphi \right) \left( x\right) \doteq \mathfrak{A}_{%
\mathrm{S}}\left( \varphi \left( x\right) \right) \ ,\qquad x\in \Lambda
_{L},\ \varphi \in \mathcal{H}_{L}\ .  \label{involution}
\end{equation}%
Then, $(\mathcal{H}_{L},\mathfrak{A}_{L})$ is a self-dual Hilbert space for
any $L\in \mathbb{R}_{0}^{+}\cup \{\infty \}$. Note that $\mathcal{H}_{%
\mathrm{S}}$ and $\mathcal{H}_{L}$ are finite-dimensional, with even
dimension, whenever $L<\infty $: Let 
\begin{equation*}
\mathbb{X}_{L}\doteq \Lambda _{L}\times \mathrm{S}\times \{+,-\}\ ,\qquad
L\in \mathbb{R}_{0}^{+}\cup \{\infty \}\ .
\end{equation*}%
The canonical orthonormal basis $\left\{ \mathfrak{e}_{\mathbf{x}}\right\} _{%
\mathbf{x}\in \mathbb{X}_{L}}$ of $\mathcal{H}_{L}$, $L\in \mathbb{R}%
_{0}^{+}\cup \{\infty \}$, is defined by%
\begin{equation}
\mathfrak{e}_{\mathbf{x}}(y)\doteq \delta _{x,y}\mathfrak{f}_{\mathrm{s},v}\
,\qquad \mathbf{x=}(x,\mathrm{s},v)\in \mathbb{X}_{L},\quad y\in \Lambda
_{L}\ ,  \label{canonical onb1}
\end{equation}%
where $\mathfrak{f}_{\mathrm{s},+}\doteq \mathfrak{A}_{\mathrm{S}}\mathfrak{f%
}_{\mathrm{s},-}\in \mathcal{H}_{\mathrm{S}}$ and $\mathfrak{f}_{\mathrm{s}%
,-}(\mathrm{t})\doteq \delta _{\mathrm{s},\mathrm{t}}$ for any $\mathrm{s},%
\mathrm{t}\in \mathrm{S}$.

In the self-dual formalism, a lattice fermion system in infinite volume is
defined by a self-dual Hamiltonian $H_{\infty }\in \mathcal{B}(\mathcal{H}%
_{\infty })$ on $(\mathcal{H}_{\infty },\mathfrak{A}_{\infty })$, that is, $%
H_{\infty }=H_{\infty }^{\ast }=-\mathfrak{A}_{\infty }H_{\infty }\mathfrak{A%
}_{\infty }$. See Definition \ref{def one particle hamiltinian} which is
here extended to the infinite-dimensional case. For a fixed basis projection 
$P_{\infty }$ diagonalizing $H_{\infty }$, the operator $P_{\infty
}H_{\infty }P_{\infty }$ is the so-called one-particle Hamiltonian\
associated with the system, as already explained at the end of Section \ref%
{Sect Bilinear Elements}. See also Section \ref{notation hilbert copy(2)}.
To obtain the corresponding self-dual Hamiltonians in finite volume we use
the orthogonal projector $P_{\mathcal{H}_{L}}\in \mathcal{B}(\mathcal{H}%
_{\infty })$ on $\mathcal{H}_{L}$ and define%
\begin{equation}
H_{L}\doteq P_{\mathcal{H}_{L}}H_{\infty }P_{\mathcal{H}_{L}}\ ,\qquad L\in 
\mathbb{R}_{0}^{+}\ .  \label{definition finite volume hamiltonian}
\end{equation}%
By construction, if $H_{\infty }$ is a self-dual Hamiltonian on $(\mathcal{H}%
_{\infty },\mathfrak{A}_{\infty })$, then, for any $L\in \mathbb{R}_{0}^{+}$%
, $H_{L}$ is a self-dual Hamiltonian on $(\mathcal{H}_{L},\mathfrak{A}_{L})$.

\subsubsection{Combes-Thomas Estimate and Summability of the Fermi
Distribution}

We present a version of the Combes-Thomas estimate, first proven in \cite%
{CT73}, that is adapted to the present setting: Fix $\epsilon \in (0,1]$.
Given a self-adjoint $H=H^{\ast }\in \mathcal{B}(\mathcal{H}_{\infty })$
whose spectrum is denoted by $\mathrm{spec}(H)$, we define the constants%
\begin{equation*}
\mathbf{S}(H,\mu )\doteq \sup_{\mathbf{x}_{1}\mathbf{=}(x_{1},\mathrm{s}%
_{1},v_{1})\in \mathbb{X}_{\infty }}\sum_{\mathbf{x}_{2}\mathbf{=}(x_{2},%
\mathrm{s}_{2},v_{2})\in \mathbb{X}_{\infty }}\left( \mathrm{e}^{\mu
|x_{1}-x_{2}|^{\epsilon }}-1\right) \left\vert \left\langle \mathfrak{e}_{%
\mathbf{x}_{1}},H\mathfrak{e}_{\mathbf{x}_{2}}\right\rangle \right\vert \in 
\mathbb{R}_{0}^{+}\cup \left\{ \infty \right\} \ ,
\end{equation*}%
for $\mu \in \mathbb{R}_{0}^{+}$, and 
\begin{equation*}
\Delta (H,z)\doteq \inf \left\{ \left\vert z-\lambda \right\vert :\lambda
\in \mathrm{spec}(H)\right\} \ ,\qquad z\in \mathbb{C}\ ,
\end{equation*}%
as the distance from the point $z$ to the spectrum of $H$. Since the
function $x\mapsto (e^{xr}-1)/x$ is increasing on $\mathbb{R}^{+}$ for any
fixed $r\geq 0$, it follows that 
\begin{equation}
\mathbf{S}(H,\mu _{1})\leq \frac{\mu _{1}}{\mu _{2}}\mathbf{S}(H,\mu _{2})\
,\qquad \mu _{2}\geq \mu _{1}\geq 0.  \label{inequality combes easy}
\end{equation}%
The Combes-Thomas estimate we use is the following:

\begin{theorem}[Combes-Thomas]
\label{Combes-Thomas}\mbox{ }\newline
Let $\epsilon \in (0,1]$, $H=H^{\ast }\in \mathcal{B}(\mathcal{H}_{\infty })$%
, $\mu \in \mathbb{R}_{0}^{+}$ and $z\in \mathbb{C}$. If $\Delta (H,z)>%
\mathbf{S}(H,\mu )$ then 
\begin{equation*}
\left\vert \left\langle \mathfrak{e}_{\mathbf{x}},(z-H)^{-1}\mathfrak{e}_{%
\mathbf{y}}\right\rangle \right\vert \leq \frac{\mathrm{e}^{-\mu
|x-y|^{\epsilon }}}{\Delta (H,z)-\mathbf{S}(H,\mu )}\ ,\qquad \mathbf{x=}(x,%
\mathrm{s},v),\,\mathbf{y=}(y,\mathrm{t},w)\in \mathbb{X}_{\infty }.
\end{equation*}
\end{theorem}

\begin{proof}
This proposition is a version of the first part of \cite[Theorem 10.5]%
{AizenmanWarzel} and is proven in the same way, because, for $\epsilon \in
(0,1]$, $\mathbf{x=}(x,\mathrm{s},v)$ and $\mathbf{y}=(y,\mathrm{t},w)$, $%
d_{\epsilon }\left( \mathbf{x},\mathbf{y}\right) \doteq |x-y|^{\epsilon }$
defines a pseudometric on $\mathbb{X}_{\infty }$.
\end{proof}

\begin{corollary}[Bound on differences of resolvents]
\label{Lemma AG98}\mbox{ }\newline
Let $\epsilon \in (0,1]$, $H=H^{\ast }\in \mathcal{B}(\mathcal{H}_{\infty })$%
, $\mu \in \mathbb{R}_{0}^{+}$ and $\eta \in \mathbb{R}^{+}$ such that $%
\mathbf{S}(H,\mu )\leq \eta /2$. Then, for all $\mathbf{x=}(x,\mathrm{s},v),%
\mathbf{y=}(y,\mathrm{t},w)\in \mathbb{X}_{\infty }$ and $u\in \mathbb{R}$, 
\begin{eqnarray*}
&&\left\vert \left\langle \mathfrak{e}_{\mathbf{x}},((H-u)^{2}+\eta
^{2})^{-1}\mathfrak{e}_{\mathbf{y}}\right\rangle \right\vert \\
&\leq &12\mathrm{e}^{-\mu |x-y|^{\epsilon }}\left\langle \mathfrak{e}_{%
\mathbf{x}},((H-u)^{2}+\eta ^{2})^{-1}\mathfrak{e}_{\mathbf{x}}\right\rangle
^{1/2}\left\langle \mathfrak{e}_{\mathbf{y}},((H-u)^{2}+\eta ^{2})^{-1}%
\mathfrak{e}_{\mathbf{y}}\right\rangle ^{1/2}.
\end{eqnarray*}
\end{corollary}

\begin{proof}
This is proven like \cite[Lemma 3]{AG98}, by replacing the euclidean metric
of $\mathbb{Z}^{d}$ with the pseudometric of $\mathbb{X}_{\infty }$, as
above. See also \cite[Chapter 13]{AizenmanWarzel}.
\end{proof}

We are now in a position to prove the summability of the Fermi distribution:

\begin{lemma}[Summability of the Fermi distribution]
\label{Lemma fermi1}\mbox{ }\newline
Fix a (finite) spin set $\mathrm{S}$, $d\in \mathbb{N}$, $\beta \in \mathbb{R%
}^{+}$, $\epsilon \in (0,1]$, $\upsilon \in \mathbb{R}_{0}^{+}$ and $%
H=H^{\ast }\in \mathcal{B}(\mathcal{H}_{\infty })$. Then, for all $\mathbf{x=%
}(x,\mathrm{s},v)$ and $\mathbf{y=}(y,\mathrm{t},w)\in \mathbb{X}_{\infty }$%
, 
\begin{multline*}
\mathbf{D}_{H,\beta ,\upsilon ,\epsilon }\doteq \sup_{\alpha \in \left[
0,\beta \right] }\sup_{\mathbf{x}\in \mathbb{X}_{\infty }}\left\{
\sum\limits_{\mathbf{y}\in \mathbb{X}_{\infty }}\mathrm{e}^{\upsilon
|x-y|^{\epsilon }}\left\vert \left\langle \mathfrak{e}_{\mathbf{x}},\frac{%
\mathrm{e}^{\alpha H}}{1+\mathrm{e}^{\beta H}}\mathfrak{e}_{\mathbf{y}%
}\right\rangle _{\mathcal{H}_{\infty }}\right\vert \right\} \\
\leq 96\left\vert \mathrm{S}\right\vert \inf_{\mu \in \mathbb{R}%
_{0}^{+}}\inf_{\epsilon \in (0,1]}\sum_{x\in \mathbb{Z}^{d}}\mathrm{e}%
^{\left( \upsilon -\mu \min \left\{ 1,\frac{\pi }{4\beta \mathbf{S}(H,\mu )}%
\right\} \right) |x|^{\epsilon }}\ .
\end{multline*}%
In particular, for $\mu =2\upsilon $ and $\upsilon =\beta ^{-1}\mu _{0}\in 
\mathbb{R}^{+}$, where $\mu _{0}\in \mathbb{R}^{+}$ is such that $8\mathbf{S}%
(H,\mu _{0})\leq \pi $, we have $\mathbf{D}_{H,\beta ,\upsilon ,\epsilon }=%
\mathcal{O}((\beta +1)^{d\epsilon ^{-1}})$.
\end{lemma}

\begin{proof}
The proof is a simple adaptation of the one from \cite[Theorem 3]{AG98}: Fix
all parameters of the lemma and observe that Corollary \ref{Lemma AG98}
combined with Inequality (\ref{inequality combes easy}) yields%
\begin{eqnarray}
&&\left\vert \left\langle \mathfrak{e}_{\mathbf{x}},((H-u)^{2}+\eta
^{2})^{-1}\mathfrak{e}_{\mathbf{y}}\right\rangle \right\vert
\label{Lemma AG982} \\
&\leq &12\mathrm{e}^{-\frac{\mu \eta }{2\mathbf{S}(H,\mu )}|x-y|^{\epsilon
}}\left\langle \mathfrak{e}_{\mathbf{x}},((H-u)^{2}+\eta ^{2})^{-1}\mathfrak{%
e}_{\mathbf{x}}\right\rangle ^{1/2}\left\langle \mathfrak{e}_{\mathbf{y}%
},((H-u)^{2}+\eta ^{2})^{-1}\mathfrak{e}_{\mathbf{y}}\right\rangle ^{1/2} 
\notag
\end{eqnarray}%
for $\mathbf{x=}(x,\mathrm{s},v),\mathbf{y=}(y,\mathrm{t},w)\in \mathbb{X}%
_{\infty }$, $u\in \mathbb{R}$ and $\eta \in (0,2\mathbf{S}(H,\mu )]$. On
the other hand, at fixed $\alpha \in \lbrack 0,\beta ]$ and $\beta \in 
\mathbb{R}^{+}$ the function on the stripe 
\begin{equation*}
\mathbb{R}+\frac{\pi i}{2\beta }\left[ -1,1\right] \subset \mathbb{C}
\end{equation*}%
defined by 
\begin{equation*}
G\left( z\right) \doteq \frac{\mathrm{e}^{\alpha z}}{1+\mathrm{e}^{\beta z}}
\end{equation*}%
is analytic and uniformly bounded by $1$, because 
\begin{multline*}
\sup_{x\in \mathbb{R},y\in \lbrack -\pi /2\beta ,\pi /2\beta ]}\left\vert
G\left( x+iy\right) \right\vert =\sup_{x\in \mathbb{R},y\in \lbrack -\pi
/2\beta ,\pi /2\beta ]}\frac{\mathrm{e}^{\alpha x}}{\left( 1+2\mathrm{e}%
^{\beta x}\cos \beta y+\mathrm{e}^{2\beta x}\right) ^{1/2}} \\
\leq \sup_{x\in \mathbb{R}}\frac{\mathrm{e}^{\alpha x}}{\left( 1+\mathrm{e}%
^{2\beta x}\right) ^{1/2}}\leq 1
\end{multline*}%
for $\alpha \in \lbrack 0,\beta ]$ and $\beta \in \mathbb{R}^{+}$. Using
Cauchy's integral formula and some translation by $\pm i\eta $, we write
this function as%
\begin{eqnarray}
G\left( E\right) &=&\frac{1}{2\pi i}\int_{\mathbb{R}}\left( \frac{G\left(
u-i\eta \right) }{u-i\eta -E}-\frac{G\left( u+i\eta \right) }{u+i\eta -E}%
\right) \mathrm{d}u  \notag \\
&=&\frac{\eta }{\pi }\int_{\mathbb{R}}\frac{G\left( u-i\eta \right) +G\left(
u+i\eta \right) }{\left( E-u\right) ^{2}+\eta ^{2}}\mathrm{d}u-\frac{2\eta }{%
\pi }\int_{\mathbb{R}}\frac{G\left( u\right) }{\left( E-u\right) ^{2}+4\eta
^{2}}\mathrm{d}u  \label{FE}
\end{eqnarray}%
for all $E\in \mathbb{R}$ and $\eta \in (0,\pi /(2\beta )]$. By spectral
calculus, together with (\ref{Lemma AG982})-(\ref{FE}) and the
Cauchy-Schwarz inequality, it follows that%
\begin{equation*}
\sup_{\alpha \in \left[ 0,\beta \right] }\left\vert \left\langle \mathfrak{e}%
_{\mathbf{x}},\frac{\mathrm{e}^{\alpha H}}{1+\mathrm{e}^{\beta H}}\mathfrak{e%
}_{\mathbf{y}}\right\rangle _{\mathcal{H}_{\infty }}\right\vert \leq 48\exp
\left( -\mu \min \left\{ 1,\frac{\pi }{4\beta \mathbf{S}(H,\mu )}\right\}
|x-y|^{\epsilon }\right)
\end{equation*}%
for all $\mathbf{x=}(x,\mathrm{s},v),\mathbf{y=}(y,\mathrm{t},w)\in \mathbb{X%
}_{\infty }$, $\mu \in \mathbb{R}_{0}^{+}$, and $\epsilon \in (0,1]$. This
in turn implies the assertion. In particular, for $\upsilon =\beta ^{-1}\mu
_{0}\in \mathbb{R}^{+}$ and fixed $\epsilon \in (0,1]$, where $\mu _{0}\in 
\mathbb{R}^{+}$ is some constant such that $8\mathbf{S}(H,\mu _{0})\leq \pi $
(which always exists, because of (\textsf{\ref{inequality combes easy}})),
by taking $\mu =2\upsilon $ and using (\textsf{\ref{inequality combes easy}}%
), one gets that%
\begin{equation*}
\mathbf{D}_{H,\beta ,\upsilon ,\epsilon }\leq 96\left\vert \mathrm{S}%
\right\vert \sum_{x\in \mathbb{Z}^{d}}\mathrm{e}^{-\mu _{0}\beta
^{-1}|x|^{\epsilon }}=\mathcal{O}\left( \left( \beta +1\right) ^{d\epsilon
^{-1}}\right) \ .
\end{equation*}
\end{proof}

\subsubsection{Summability of Covariances -- General Bound}

We consider the following condition:

\begin{condition}
\label{Condition a la con}\mbox{ }\newline
There is a sequence $\mathcal{P}\doteq \left\{ P_{L}\right\} _{L\in \mathbb{R%
}_{0}^{+}}$ of basis projections associated with $(\mathcal{H}_{L},\mathfrak{%
A}_{L})$, diagonalizing $H_{L}$, such that: \newline
\emph{(a)} $P_{L}$ strongly converges, as $L\rightarrow \infty $, to an
orthogonal projection $P_{\infty }\in \mathcal{B}(\mathcal{H}_{\infty })$. 
\newline
\emph{(b)} The limit projection $P_{\infty }$ satisfies 
\begin{equation*}
\mathbf{D}_{\mathcal{P},\upsilon ,\epsilon }\doteq \sup_{\mathbf{x}_{1}%
\mathbf{=}(x_{1},\mathrm{s}_{1},v_{1})\in \mathbb{X}_{\infty }}\sum\limits_{%
\mathbf{x}_{2}\mathbf{=}(x_{2},\mathrm{s}_{2},v_{2})\in \mathbb{X}_{\infty }}%
\mathrm{e}^{\upsilon |x_{1}-x_{2}|^{\epsilon }}\left\vert \left\langle 
\mathfrak{e}_{\mathbf{x}_{1}},P_{\infty }\mathfrak{e}_{\mathbf{x}%
_{2}}\right\rangle _{\mathcal{H}_{\infty }}\right\vert <\infty
\end{equation*}%
for some fixed $\epsilon \in (0,1]$ and $\upsilon \in \mathbb{R}_{0}^{+}$.
\end{condition}

This condition is trivially fulfilled for \emph{all} gauge-invariant
self-dual Hamiltonians, that is, self-dual Hamiltonians $H_{\infty }$
satisfying 
\begin{equation*}
\left\langle \mathfrak{e}_{(x,\mathrm{s},v)},H_{\infty }\mathfrak{e}_{(y,%
\mathrm{t},v)}\right\rangle =0\ ,\qquad x,y\in \mathbb{Z}^{d},\ \mathrm{s},%
\mathrm{t}\in \mathrm{S},\mathrm{\ }v\in \{+,-\}\ .
\end{equation*}%
In this case, for any $L\in \mathbb{R}_{0}^{+}$, one can choose the sequence 
$\left\{ P_{L}\right\} _{L\in \mathbb{R}_{0}^{+}}$ of basis projections
defined by 
\begin{equation*}
P_{L}\mathfrak{e}_{(x,\mathrm{s},v)}=\delta _{v,-}\mathfrak{e}_{(x,\mathrm{s}%
,v)}\ ,\qquad (x,\mathrm{s},v)\in \Lambda _{L}\times \mathrm{S}\times
\{+,-\}\ ,
\end{equation*}%
where $\left\{ \mathfrak{e}_{\mathbf{x}}\right\} _{\mathbf{x}\in \mathbb{X}%
_{\infty }}$ is the orthonormal basis defined by (\ref{canonical onb1}).
This obviously yields $\mathbf{D}_{\mathcal{P}}=1$.

Condition \ref{Condition a la con} (a) is convenient because it allows us to
rewrite the three limits of the decay parameter%
\begin{eqnarray}
\mathbf{\omega }_{H_{\infty },\mathcal{P},\upsilon ,\epsilon ,\gimel }
&\doteq &\limsup_{L_{\mathrm{i}}\rightarrow \infty }\limsup_{L_{\mathrm{f}%
}\rightarrow \infty }\lim_{n\rightarrow \infty }\sup_{k_{1}\in \{0,\ldots
,n_{\beta }-1\}}\sup_{\mathbf{x}_{1}\mathbf{=}(x_{1},\mathrm{s}%
_{1},v_{1})\in \mathbb{X}_{L_{\mathrm{i}}}}  \label{decay parameter} \\
&&\left\{ \beta n^{-1}\sum\limits_{k_{2}=0}^{n_{\beta }-1}\sum\limits_{%
\mathbf{x}_{2}\mathbf{=}(x_{2},\mathrm{s}_{2},v_{2})\in \mathbb{X}_{L_{%
\mathrm{i}}}}\mathrm{e}^{\gimel \tilde{\alpha}\left( \beta ^{-1}k_{1},\beta
^{-1}k_{2}\right) +\upsilon |x_{1}-x_{2}|^{\epsilon }}\left\vert
\left\langle \mathfrak{e}_{\mathbf{x}_{1}}^{(k_{1})},C_{H_{L_{\mathrm{f}%
}},P_{L_{\mathrm{f}}}}^{(n)}\mathfrak{e}_{\mathbf{x}_{2}}^{(k_{2})}\right%
\rangle _{\mathfrak{H}^{(n_{\beta })}}\right\vert \right\}   \notag
\end{eqnarray}%
in a more tractable form, where $\epsilon \in (0,1]$, $\upsilon ,\gimel \in 
\mathbb{R}_{0}^{+}$ , $\tilde{\alpha}$ is the pseudometric on $[0,\beta +1]$
defined by%
\begin{equation}
\tilde{\alpha}\left( u_{1},u_{2}\right) \doteq \min \{\alpha \left(
u_{1},u_{2}\right) ,\beta -\alpha \left( u_{1},u_{2}\right) \}\text{ },\text{%
\qquad }u_{1},u_{2}\in \lbrack 0,\beta +1]\text{ },  \label{sdsdsdsdsdsd}
\end{equation}%
with%
\begin{equation*}
\alpha \left( u_{1},u_{2}\right) \doteq \left[ \min \left\{ \beta -\min
\{u_{1},u_{2}\},\max \{u_{1},u_{2}\}-\min \{u_{1},u_{2}\}\right\} \right]
_{+}\in \left[ 0,\beta \right] \ ,\quad u_{1},u_{2}\in \lbrack 0,\beta +1)\ .
\end{equation*}%
Recall that $C_{H_{L_{\mathrm{f}}},P_{L_{\mathrm{f}}}}^{(n)}$ is the
covariance (\ref{covariance def}) for $\mathcal{H=H}_{L_{\mathrm{f}}}$ and $%
n_{\beta }\doteq n+\left\lfloor n/\beta \right\rfloor $. Similarly to
Equations (\ref{alphanbis}) and (\ref{alphanbis2}), define the function 
\begin{equation}
F_{u_{1},u_{2}}\left( H,P\right) \doteq \left\{ 
\begin{array}{lll}
P\frac{\mathrm{e}^{-\alpha \left( u_{1},u_{2}\right) H}}{1+\mathrm{e}%
^{-\beta H}}P+P^{\bot }\frac{\mathrm{e}^{-\alpha \left( u_{1},u_{2}\right) H}%
}{1+\mathrm{e}^{-\beta H}}P^{\bot } & \text{for} & u_{1}<u_{2}\ , \\ 
-P\frac{\mathrm{e}^{\alpha \left( u_{1},u_{2}\right) H}}{1+\mathrm{e}^{\beta
H}}P-P^{\bot }\frac{\mathrm{e}^{\alpha \left( u_{1},u_{2}\right) H}}{1+%
\mathrm{e}^{\beta H}}P^{\bot } & \text{for} & u_{1}\geq u_{2}\ ,%
\end{array}%
\right. \text{ }  \label{function a la con encore}
\end{equation}%
for any $\beta \in \mathbb{R}^{+}$, $u_{1},u_{2}\in \lbrack 0,\beta +1)$, $%
H\in \mathcal{B}(\mathcal{H})$, and orthogonal projection $P$. Then, one
gets the following expression for the decay parameter $\mathbf{\omega }%
_{H_{\infty },\mathcal{P},\upsilon ,\epsilon ,0}$:

\begin{lemma}[Explicit expression of the decay parameter]
\label{Lemma fermi1 copy(1)}\mbox{ }\newline
Fix $d\in \mathbb{N}$, $\beta \in \mathbb{R}^{+}$, $\epsilon \in (0,1]$ and $%
\upsilon ,\gimel \in \mathbb{R}_{0}^{+}$. Let $H_{\infty }\in \mathcal{B}(%
\mathcal{H}_{\infty })$ be any self-dual Hamiltonian on $(\mathcal{H}%
_{\infty },\mathfrak{A}_{\infty })$. If Condition \ref{Condition a la con}
(a) holds true, then 
\begin{multline*}
\mathbf{\omega }_{H_{\infty },\mathcal{P},\upsilon ,\epsilon ,\gimel
}=\sup_{u_{1}\in \lbrack 0,\beta +1)}\sup_{\mathbf{x}_{1}\mathbf{=}(x_{1},%
\mathrm{s}_{1},v_{1})\in \mathbb{X}_{\infty }}\sum\limits_{\mathbf{x}_{2}%
\mathbf{=}(x_{2},\mathrm{s}_{2},v_{2})\in \mathbb{X}_{\infty }}\mathrm{e}%
^{\upsilon |x_{1}-x_{2}|^{\epsilon }} \\
\times \int_{\lbrack 0,\beta +1)}\mathrm{e}^{\gimel \tilde{\alpha}\left(
u_{1},u_{2}\right) }\left\vert \left\langle \mathfrak{e}_{\mathbf{x}%
_{1}},F_{u_{1},u_{2}}\left( H_{\infty },P_{\infty }\right) \mathfrak{e}_{%
\mathbf{x}_{2}}\right\rangle _{\mathcal{H}_{\infty }}\right\vert \mathrm{d}%
u_{2}.
\end{multline*}
\end{lemma}

\begin{proof}
Fix all parameters of the lemma. Explicit computations using the spectral
theorem and Inequality (\ref{Hnbis}) show that, for any $\beta \in \mathbb{R}%
^{+}$, $n>\beta \left\Vert H_{\infty }\right\Vert _{\mathcal{B}(\mathcal{H}%
_{\infty })}$, $L_{\mathrm{f}}\in \mathbb{R}_{0}^{+}\cup \{\infty \}$ and $%
\alpha \in \left[ 0,\beta \right] $,%
\begin{eqnarray*}
\left\Vert \frac{\mathrm{e}^{\pm \alpha H_{L_{\mathrm{f}}}}}{1+\mathrm{e}%
^{\pm \beta H_{L_{\mathrm{f}}}}}-\frac{\mathrm{e}^{\pm \alpha H_{L_{\mathrm{f%
}}}^{(n)}}}{1+\mathrm{e}^{\pm \beta H_{L_{\mathrm{f}}}^{(n)}}}\right\Vert _{%
\mathcal{B}(\mathcal{H}_{\infty })} &\leq &\sup_{\left\vert \lambda
\right\vert \leq \Vert H_{L_{\mathrm{f}}}\Vert _{\mathcal{B}(\mathcal{H}%
_{\infty })}}\left\vert \frac{\mathrm{e}^{\pm \alpha \lambda }}{1+\mathrm{e}%
^{\pm \beta \lambda }}-\frac{\mathrm{e}^{\pm \frac{\alpha \beta ^{-1}n}{2}%
\left( \ln \left( \frac{1+n^{-1}\beta \lambda }{1-n^{-1}\beta \lambda }%
\right) \right) }}{1+\mathrm{e}^{\pm \frac{n}{2}\left( \ln \left( \frac{%
1+n^{-1}\beta \lambda }{1-n^{-1}\beta \lambda }\right) \right) }}\right\vert 
\\
&\leq &2\beta \left\Vert H_{L_{\mathrm{f}}}-H_{L_{\mathrm{f}%
}}^{(n)}\right\Vert _{\mathcal{B}(\mathcal{H}_{\infty })}\leq Dn^{-2}\beta
^{3}\left\Vert H_{\infty }\right\Vert _{\mathcal{B}(\mathcal{H}_{\infty
})}^{3}\ ,
\end{eqnarray*}%
for some finite constant $D$. It follows that 
\begin{multline*}
\mathbf{\omega }_{H_{\infty },\mathcal{P},\upsilon ,\epsilon ,\gimel
}=\limsup_{L_{\mathrm{i}}\rightarrow \infty }\limsup_{L_{\mathrm{f}%
}\rightarrow \infty }\sup_{u_{1}\in \lbrack 0,\beta +1)}\sup_{\mathbf{x}%
_{1}\in \mathbb{X}_{L_{\mathrm{i}}}} \\
\left\{ \sum\limits_{\mathbf{x}_{2}\in \mathbb{X}_{L_{\mathrm{i}}}}\mathrm{e}%
^{\upsilon |x_{1}-x_{2}|^{\epsilon }}\int_{[0,\beta +1)}\mathrm{e}^{\gimel 
\tilde{\alpha}\left( u_{1},u_{2}\right) }\left\vert \left\langle \mathfrak{e}%
_{\mathbf{x}_{1}},F_{u_{1},u_{2}}\left( H_{L_{\mathrm{f}}},P_{L_{\mathrm{f}%
}}\right) \mathfrak{e}_{\mathbf{x}_{2}}\right\rangle _{\mathcal{H}_{\infty
}}\right\vert \mathrm{d}u_{2}\right\} 
\end{multline*}%
for $\epsilon \in (0,1]$ and $\upsilon ,\gimel \in \mathbb{R}_{0}^{+}$. By
Condition \ref{Condition a la con} (a), $P_{L_{\mathrm{f}}}$ strongly
converges, as $L_{\mathrm{f}}\rightarrow \infty $, to an orthogonal
projection $P_{\infty }$. Therefore, because the set $\mathbb{X}_{L_{\mathrm{%
i}}}$ is finite and the family 
\begin{equation*}
E\mapsto \int_{\lbrack 0,\beta +1)}\frac{\mathrm{e}^{\gimel \tilde{\alpha}%
\left( u_{1},u_{2}\right) \pm E\alpha \left( u_{1},u_{2}\right) }}{1+\mathrm{%
e}^{\pm \beta E}}\mathrm{d}u_{2}\ ,\qquad u_{1}\in \lbrack 0,\beta +1)\ ,
\end{equation*}%
of functions is uniformly equicontinuous on compact subsets of $\mathbb{R}$,
we deduce from the last equality that 
\begin{multline*}
\mathbf{\omega }_{H_{\infty },\mathcal{P},\upsilon ,\epsilon ,\gimel
}=\limsup_{L_{\mathrm{i}}\rightarrow \infty }\sup_{u_{1}\in \lbrack 0,\beta
+1)}\sup_{\mathbf{x}_{1}\in \mathbb{X}_{L_{\mathrm{i}}}} \\
\left\{ \sum\limits_{\mathbf{x}_{2}\in \mathbb{X}_{L_{\mathrm{i}}}}\mathrm{e}%
^{\upsilon |x_{1}-x_{2}|^{\epsilon }}\int_{[0,\beta +1)}\mathrm{e}^{\gimel 
\tilde{\alpha}\left( u_{1},u_{2}\right) }\left\vert \left\langle \mathfrak{e}%
_{\mathbf{x}_{1}},F_{u_{1},u_{2}}\left( H_{\infty },P_{\infty }\right) 
\mathfrak{e}_{\mathbf{x}_{2}}\right\rangle _{\mathcal{H}_{\infty
}}\right\vert \mathrm{d}u_{2}\right\} \ .
\end{multline*}%
Finally, by using approximate maximizers of the suprema over $u_{1}\in
\lbrack 0,\beta +1)$ and $\mathbf{x}_{1}\in \mathbb{X}_{\infty }$, and the
monotone convergence, the lemma follows.
\end{proof}

Now we are in a position to prove the summability of the covariance:

\begin{theorem}[Summability of the covariance]
\label{Therem summabilityI}\mbox{ }\newline
Fix a (finite) spin set $\mathrm{S}$, $d\in \mathbb{N}$ and let $H_{\infty
}\in \mathcal{B}(\mathcal{H}_{\infty })$ be any self-dual Hamiltonian on $(%
\mathcal{H}_{\infty },\mathfrak{A}_{\infty })$. Then, under Condition \ref%
{Condition a la con}, for any $\beta \in \mathbb{R}^{+}$, $\epsilon \in
(0,1] $ and $\upsilon \in \mathbb{R}_{0}^{+}$,%
\begin{equation*}
\mathbf{\omega }_{H_{\infty },\mathcal{P},\upsilon ,\epsilon ,0}\leq 2\left( 
\mathbf{D}_{\mathcal{P},\upsilon ,\epsilon }+1\right) ^{2}\mathbf{D}%
_{H,\beta ,\upsilon ,\epsilon }\left( \beta +1\right) .
\end{equation*}
\end{theorem}

\begin{proof}
The theorem is a consequence of Condition \ref{Condition a la con}, Lemmata %
\ref{Lemma fermi1} and \ref{Lemma fermi1 copy(1)}, and Equation (\ref%
{function a la con encore}) together with straightforward computations.
\end{proof}

By Lemma \ref{Lemma fermi1}, we observe from this theorem that%
\begin{equation}
\mathbf{\omega }_{H_{\infty },\mathcal{P},\upsilon ,\epsilon ,0}=\mathcal{O}%
\left( \left( \beta +1\right) ^{d\epsilon ^{-1}+1}\right) \ ,
\label{estimate}
\end{equation}%
for \emph{all} self-dual Hamiltonians $H_{\infty }\in \mathcal{B}(\mathcal{H}%
_{\infty })$ as soon as $\mathbf{D}_{\mathcal{P},\upsilon ,\epsilon }<\infty 
$ and $4\beta \mathbf{S}(H,\mu )\upsilon <\pi \mu $ for some $\mu >\upsilon $
and $\epsilon \in (0,1]$. Recall that if $H_{\infty }$ is gauge-invariant
then one can take $\mathbf{D}_{\mathcal{P},\upsilon ,\epsilon }=1$. The
scaling (\ref{estimate}) with respect to the inverse temperature $\beta \in 
\mathbb{R}^{+}$ is exactly the same one obtained from the Fourier analysis
of the two-point Green function of translation-invariant, free fermi systems
at positive density. Our result shows that this estimate does \emph{not}
depend on translation invariance.

\subsubsection{Summability of Covariances -- Gapped Case}

Estimate (\ref{estimate}) does not allow for the control of the decay
parameter $\mathbf{\omega }_{H_{\infty },\mathcal{P},\upsilon ,\epsilon ,0}$
(\ref{decay parameter}) in the zero-temperature limit, i.e., when $\beta
\rightarrow \infty $. However, in the special case of \emph{gapped}
self-dual Hamiltonians, that is, self-dual Hamiltonians $H_{\infty }$
satisfying%
\begin{equation}
\mathfrak{g}_{H_{\infty }}\doteq \inf \left\{ \varepsilon >0:\left[
-\varepsilon ,\varepsilon \right] \cap \mathrm{spec}(H_{\infty })\neq
\emptyset \right\} >0,  \label{gap assumption}
\end{equation}%
we can uniformly bound $\mathbf{\omega }_{H_{\infty },\mathcal{P},\upsilon
,\epsilon ,0}$ at arbitrarily large $\beta \gg 1$, for a natural choice of
sequence $\mathcal{P=}\left\{ P_{L}\right\} _{L\in \mathbb{R}_{0}^{+}}$ of
basis projections associated with $(\mathcal{H}_{L},\mathfrak{A}_{L})$ and
diagonalizing $H_{L}$, $L\in \mathbb{R}_{0}^{+}$. To demonstrate this, we
need to adapt Lemma \ref{Lemma fermi1} to this gapped situation:

\begin{lemma}[Summability of the Fermi distribution -- gapped case]
\label{Lemma fermi1 gapped}\mbox{ }\newline
Fix a (finite) spin set $\mathrm{S}$, $d\in \mathbb{N}$, $\beta \in \mathbb{R%
}^{+}$, $\epsilon \in (0,1]$, $\upsilon ,\gimel \in \mathbb{R}_{0}^{+}$ and $%
H=H^{\ast }\in \mathcal{B}(\mathcal{H}_{\infty })$ such that $\mathfrak{g}%
_{H}>0$. Then, 
\begin{eqnarray*}
&&\sup_{u_{1}\in \lbrack 0,\beta +1)}\sup_{\mathbf{x}_{1}\mathbf{=}(x_{1},%
\mathrm{s}_{1},v_{1})\in \mathbb{X}_{\infty }}\left\{ \sum\limits_{\mathbf{x}%
_{2}\mathbf{=}(x_{2},\mathrm{s}_{2},v_{2})\in \mathbb{X}_{\infty }}\mathrm{e}%
^{\upsilon |x_{1}-x_{2}|^{\epsilon }}\right.  \\
&&\left. \times \int_{\lbrack 0,\beta +1)}\mathrm{e}^{\gimel \tilde{\alpha}%
\left( u_{1},u_{2}\right) }\left\vert \left\langle \mathfrak{e}_{\mathbf{x}%
_{1}},F_{u_{1},u_{2}}\left( H,\mathbf{1}\left[ H>0\right] \right) \mathfrak{e%
}_{\mathbf{x}_{2}}\right\rangle _{\mathcal{H}_{\infty }}\right\vert \mathrm{d%
}u_{2}\right\}  \\
&\leq &152\left\vert \mathrm{S}\right\vert \sup_{u_{1}\in \lbrack 0,\beta
+1)}\inf_{\mu \in \mathbb{R}_{0}^{+}}\inf_{\epsilon \in (0,1]}\sum_{x\in 
\mathbb{Z}^{d}}\mathrm{e}^{\left( \upsilon -\mu \min \left\{ 1,\frac{%
\mathfrak{g}_{H}}{4\mathbf{S}(H,\mu )}\right\} \right) |x|^{\epsilon
}}\int_{0}^{\beta +1}\frac{\mathrm{e}^{\left( \gimel -\frac{\mathfrak{g}_{H}%
}{2}\right) \tilde{\alpha}\left( u_{1},u_{2}\right) }}{1-\mathrm{e}^{-\frac{%
\beta \mathfrak{g}_{H}}{2}}}\mathrm{d}u_{2}\text{ },
\end{eqnarray*}%
where we recall that $\tilde{\alpha}$ is the pseudometric on $[0,\beta +1]$
defined by (\ref{sdsdsdsdsdsd}).
\end{lemma}

\begin{proof}
Fix all parameters of the lemma. By Inequality (\ref{inequality combes easy}%
), Theorem \ref{Combes-Thomas} implies the bound 
\begin{equation}
\left\vert \left\langle \mathfrak{e}_{\mathbf{x}},(z-H)^{-1}\mathfrak{e}_{%
\mathbf{y}}\right\rangle \right\vert \leq 4\mathfrak{g}_{H}^{-1}\exp \left(
-\mu \min \left\{ 1,\frac{\mathfrak{g}_{H}}{4\mathbf{S}(H,\mu )}\right\}
|x-y|^{\epsilon }\right)  \label{combes 0}
\end{equation}%
for any $\mathbf{x=}(x,\mathrm{s},v),\mathbf{y=}(y,\mathrm{t},w)\in \mathbb{X%
}_{\infty }$ and $z\in \mathbb{C}$ such that $\Delta (H,z)\geq \mathfrak{g}%
_{H}/2>0$. On the other hand, for every $\eta \in (0,\mathfrak{g}_{H}/2]$,
the function defined by 
\begin{equation*}
G\left( z\right) \doteq \frac{\mathrm{e}^{-\alpha z}}{1+\mathrm{e}^{-\beta z}%
}\ ,\qquad z\in \left( \mathbb{R}_{0}^{+}+\eta +i\eta \left[ -1,1\right]
\right) \ ,
\end{equation*}%
is analytic and uniformly bounded by $\mathrm{e}^{-\alpha \eta }(1-\mathrm{e}%
^{-\beta \eta })^{-1}$. Similar to (\ref{FE}), we again use Cauchy's
integral formula to write, for all real $E\in \mathbb{R}\backslash \{\eta \}$%
, 
\begin{equation*}
\mathbf{1}\left[ E>\eta \right] G\left( E\right) =\frac{1}{2\pi i}\int_{\eta
}^{\infty }\left( \frac{G\left( u-i\eta \right) }{u-E-i\eta }-\frac{G\left(
u+i\eta \right) }{u-E+i\eta }\right) \mathrm{d}u-\frac{1}{2\pi }\int_{-\eta
}^{\eta }\frac{G\left( \eta +iu\right) }{\eta -E+iu}\mathrm{d}u\ ,
\end{equation*}%
which yields%
\begin{eqnarray*}
\mathbf{1}\left[ E>\eta \right] G\left( E\right) &=&\frac{\eta }{\pi }%
\int_{\eta }^{\infty }\frac{G\left( u-i\eta \right) +G\left( u+i\eta \right) 
}{\left( u-E\right) ^{2}+\eta ^{2}}\mathrm{d}u-\frac{2\eta }{\pi }\int_{\eta
}^{\infty }\frac{G\left( u\right) }{\left( u-E\right) ^{2}+4\eta ^{2}}%
\mathrm{d}u \\
&&+\frac{1}{2\pi }\int_{0}^{\eta }\frac{G\left( \eta -iu\right) }{\eta
-iu-E+2i\eta }\mathrm{d}u+\frac{1}{2\pi }\int_{0}^{\eta }\frac{G\left( \eta
+iu\right) }{\eta +iu-E-2i\eta }\mathrm{d}u \\
&&-\frac{1}{2\pi }\int_{-\eta }^{\eta }\frac{G\left( \eta +iu\right) }{\eta
-E+iu}\mathrm{d}u\ .
\end{eqnarray*}%
$E\mapsto \mathbf{1}\left[ E>\eta \right] $ is the characteristic function
of the set $(\eta ,\infty )$. By spectral calculus, together with the last
equality, Inequalities (\ref{Lemma AG982}) and (\ref{combes 0}) and the
Cauchy-Schwarz inequality, it follows that%
\begin{equation*}
\left\vert \left\langle \mathfrak{e}_{\mathbf{x}},\mathbf{1}\left[ H>0\right]
\frac{\mathrm{e}^{-\alpha H}}{1+\mathrm{e}^{-\beta H}}\mathbf{1}\left[ H>0%
\right] \mathfrak{e}_{\mathbf{y}}\right\rangle _{\mathcal{H}_{\infty
}}\right\vert \leq \frac{38\mathrm{e}^{-\frac{\alpha \mathfrak{g}_{H}}{2}}}{%
1-\mathrm{e}^{-\frac{\beta \mathfrak{g}_{H}}{2}}}\mathrm{e}^{-\mu \min
\left\{ 1,\frac{\mathfrak{g}_{H}}{4\mathbf{S}(H,\mu )}\right\}
|x-y|^{\epsilon }}
\end{equation*}%
for all $\mathbf{x=}(x,\mathrm{s},v),\mathbf{y=}(y,\mathrm{t},w)\in \mathbb{X%
}_{\infty }$, $\mu \in \mathbb{R}_{0}^{+}$, $\epsilon \in (0,1]$ and $\alpha
\in \lbrack 0,\beta )$. In the same way,%
\begin{equation*}
\left\vert \left\langle \mathfrak{e}_{\mathbf{x}},\mathbf{1}\left[ H>0\right]
\frac{\mathrm{e}^{\alpha H}}{1+\mathrm{e}^{\beta H}}\mathbf{1}\left[ H>0%
\right] \mathfrak{e}_{\mathbf{y}}\right\rangle _{\mathcal{H}_{\infty
}}\right\vert \leq \frac{38\mathrm{e}^{\frac{\alpha \mathfrak{g}_{H}}{2}}}{%
\mathrm{e}^{\frac{\beta \mathfrak{g}_{H}}{2}}-1}\mathrm{e}^{-\mu \min
\left\{ 1,\frac{\mathfrak{g}_{H}}{4\mathbf{S}(H,\mu )}\right\}
|x-y|^{\epsilon }}
\end{equation*}%
for all $\mathbf{x=}(x,\mathrm{s},v),\mathbf{y=}(y,\mathrm{t},w)\in \mathbb{X%
}_{\infty }$, $\mu \in \mathbb{R}_{0}^{+}$, $\epsilon \in (0,1]$ and $\alpha
\in \lbrack 0,\beta )$. By (\ref{function a la con encore}), this in turn
implies the assertion.
\end{proof}

We are now in a position to prove the uniform summability of the covariance
with respect to the inverse temperature $\beta \in \mathbb{R}^{+}$, in the
special case of gapped Hamiltonians:

\begin{theorem}[Summability of the covariance -- gapped case]
\label{Therem summabilityII}\mbox{ }\newline
Fix a (finite) spin set $\mathrm{S}$, $d\in \mathbb{N}$, $\epsilon \in (0,1]$%
, $\upsilon ,\gimel \in \mathbb{R}_{0}^{+}$ and let $H_{\infty }\in \mathcal{%
B}(\mathcal{H}_{\infty })$ be any self-dual Hamiltonian on $(\mathcal{H}%
_{\infty },\mathfrak{A}_{\infty })$ such that $\mathfrak{g}_{H_{\infty
}}>2\gimel $. Then, there is a sequence $\mathcal{P}\doteq \left\{
P_{L}\right\} _{L\in \mathbb{R}^{+}}$ of basis projections associated with $(%
\mathcal{H}_{L},\mathfrak{A}_{L})$, diagonalizing $H_{L}$, such that, for
any $\beta \in \mathbb{R}^{+}$,%
\begin{equation*}
\mathbf{\omega }_{H_{\infty },\mathcal{P},\upsilon ,\epsilon ,\gimel }\leq
152\left\vert \mathrm{S}\right\vert \sup_{u_{1}\in \lbrack 0,\beta
+1)}\inf_{\mu \in \mathbb{R}_{0}^{+}}\inf_{\epsilon \in (0,1]}\sum_{x\in 
\mathbb{Z}^{d}}\mathrm{e}^{\left( \upsilon -\mu \min \left\{ 1,\frac{%
\mathfrak{g}_{H_{\infty }}}{4\mathbf{S}(H_{\infty },\mu )}\right\} \right)
|x|^{\epsilon }}\int_{0}^{\beta +1}\frac{\mathrm{e}^{(\gimel -\frac{%
\mathfrak{g}_{H_{\infty }}}{2})\tilde{\alpha}\left( u_{1},u_{2}\right) }}{1-%
\mathrm{e}^{-\frac{\beta \mathfrak{g}_{H_{\infty }}}{2}}}\mathrm{d}u_{2}\ .
\end{equation*}%
In particular, if $4\mathbf{S}(H_{\infty },\mu )\upsilon <\mathfrak{g}%
_{H_{\infty }}\mu $ for some $\mu >\upsilon $ and $\mathfrak{g}_{H_{\infty
}}>2\gimel $ then 
\begin{equation*}
\mathbf{\omega }_{H_{\infty },\mathcal{P},\upsilon ,\epsilon ,\gimel }=%
\mathcal{O}\left( (\left( \mathfrak{g}_{H_{\infty }}/2-\gimel \right)
^{-1}+1)\left( \mathfrak{g}_{H_{\infty }}^{-1}+1\right) ^{d}\right) \ .
\end{equation*}
\end{theorem}

\begin{proof}
Fix all parameters of the theorem. Since $H_{\infty }\in \mathcal{B}(%
\mathcal{H}_{\infty })$ is a self-dual\ Hamiltonian on $(\mathcal{H}_{\infty
},\mathfrak{A}_{\infty })$, note again that $H_{L}$, as defined by (\ref%
{definition finite volume hamiltonian}) for any $L\in \mathbb{R}_{0}^{+}$,
is a self-dual\ Hamiltonian on $(\mathcal{H}_{L},\mathfrak{A}_{L})$.
Therefore, for all\ $L\in \mathbb{R}_{0}^{+}$, 
\begin{equation}
\mathfrak{A}_{L}\mathbf{1}\left[ \pm H_{L}>0\right] \mathfrak{A}_{L}=\mathbf{%
1}\left[ \mp H_{L}>0\right] \qquad \text{and}\qquad \mathfrak{A}_{L}\mathbf{1%
}\left[ H_{L}=0\right] \mathfrak{A}_{L}=\mathbf{1}\left[ H_{L}=0\right] \ .
\label{iodiot kljsdlkfj}
\end{equation}%
Because the dimension of $\mathcal{H}_{L}$ is even, for any $L\in \mathbb{R}%
_{0}^{+}$, by (\ref{iodiot kljsdlkfj}), the kernel $\mathrm{ker}H_{L}$ also
has even dimension. Hence, there is a basis projection $P_{L}^{(0)}$
associated with $(\mathrm{ker}H_{L},\mathfrak{A}_{L}|_{\mathrm{ker}H_{L}})$
for any $L\in \mathbb{R}_{0}^{+}$, by \cite[Lemma 3.3]{A68}. In particular,
from (\ref{iodiot kljsdlkfj}), 
\begin{equation*}
P_{L}\doteq P_{L}^{(0)}+\mathbf{1}\left[ H_{L}>0\right] \ ,\qquad L\in 
\mathbb{R}_{0}^{+}\ ,
\end{equation*}%
is a basis projection diagonalizing $H_{L}$. Let $\chi :\mathbb{R}%
\rightarrow \mathbb{R}_{0}^{+}$ be any continuous function, whose support
lies on the interval $(-\mathfrak{g}_{H_{\infty }},\mathfrak{g}_{H_{\infty
}})$, such that $\chi (0)=1$. Then, 
\begin{equation*}
0\leq P_{L}^{(0)}\leq \chi \left( H_{L}\right) \ ,\qquad L\in \mathbb{R}%
_{0}^{+}\ .
\end{equation*}%
Noting that, as $L\rightarrow \infty $, $H_{L}$ strongly converges to $%
H_{\infty }$, we infer from the gap assumption (\ref{gap assumption}) on the
spectrum of $H_{\infty }$ that $\chi \left( H_{L}\right) $ and, hence, $%
P_{L}^{(0)}$ strongly tend to $0$ when $L\rightarrow \infty $. Therefore, $%
P_{L}$ strongly converges to $\mathbf{1}\left[ H_{\infty }>0\right] $, as $%
L\rightarrow \infty $, and Condition \ref{Condition a la con} (a) is
satisfied. Applying now Lemmata \ref{Lemma fermi1 copy(1)} and \ref{Lemma
fermi1 gapped} together with the asymptotics 
\begin{equation*}
\int_{0}^{\beta +1}\frac{\mathrm{e}^{\left( \gimel -\frac{\mathfrak{g}_{H}}{2%
}\right) \tilde{\alpha}\left( u_{1},u_{2}\right) }}{1-\mathrm{e}^{-\frac{%
\beta \mathfrak{g}_{H}}{2}}}\mathrm{d}u_{2}=\mathcal{O}(\left( \mathfrak{g}%
_{H_{\infty }}/2-\gimel \right) ^{-1}+1)
\end{equation*}%
uniformly for $\beta \geq D>0$ and $u_{1}\in \lbrack 0,\beta +1)$ (when $%
\mathfrak{g}_{H_{\infty }}>2\gimel $), the assertion follows.
\end{proof}

\section{CAR Algebra and Second Quantization\label{notation hilbert copy(1)}}

\subsection{CAR Algebra\label{Section Gen func as Grassmann int copy(1)}}

In Section \ref{notation hilbert copy(1)}, for simplicity, $\mathfrak{h}$,
the so-called one-particle Hilbert space, is finite-dimensional. Then, given
such an $\mathfrak{h}$, the associated \emph{CAR algebra} is defined as
follows:

\begin{definition}[CAR algebra]
\label{def Self--dual CAR Algebras copy(1)}\mbox{
}\newline
The CAR algebra $\mathrm{CAR}(\mathfrak{h})\equiv (\mathrm{CAR}(\mathfrak{h}%
),+,\cdot ,\ast )$ is the $C^{\ast }$-algebra generated by a unit $\mathfrak{%
1}$ and a family $\{a(\varphi )\}_{\varphi \in \mathfrak{h}}$ of elements
satisfying Conditions \emph{(a)-(b)}: \newline
\emph{(a)} The map $\varphi \mapsto a(\varphi )^{\ast }$ is (complex) linear.%
\newline
\emph{(b)} The family $\{a(\varphi )\}_{\varphi \in \mathfrak{h}}$ satisfies
the CAR: For all $\varphi _{1},\varphi _{2}\in \mathfrak{h}$,%
\begin{equation}
a(\varphi _{1})a(\varphi _{2})+a(\varphi _{2})a(\varphi _{1})=0,\quad
a(\varphi _{1})a(\varphi _{2})^{\ast }+a(\varphi _{2})^{\ast }a(\varphi
_{1})=\langle \varphi _{1},\varphi _{2}\rangle _{\mathfrak{h}}\mathfrak{1}\ .
\label{CAR}
\end{equation}
\end{definition}

\begin{remark}
\label{bound norm a CAR}\mbox{
}\newline
By the CAR\ (\ref{CAR}), the antilinear map $\varphi \mapsto a\left( \varphi
\right) $ is injective and isometric. In particular, $\Vert a(\varphi )\Vert
_{\mathrm{CAR}(\mathfrak{h})}=\Vert \varphi \Vert _{\mathfrak{h}}$ for any $%
\varphi \in \mathfrak{h}$.
\end{remark}

Strictly speaking, the above conditions only define $\mathrm{CAR}(\mathfrak{h%
})$ up to an isomorphism of $C^{\ast }$-algebra \cite[Theorem 5.2.5]%
{BratteliRobinson}. In particular, if $\{\tilde{a}(\varphi )\}_{\varphi \in 
\mathfrak{h}}$ is a second family of generators of $\mathrm{CAR}(\mathfrak{h}%
)$ satisfying (\ref{CAR}), then there is a $\ast $-automorphism $\tau $ of $%
\mathrm{CAR}(\mathfrak{h})$ such that 
\begin{equation*}
a(\varphi )=\tau (\tilde{a}(\varphi ))\text{ },\text{\qquad }\varphi \in 
\mathfrak{h}\ .
\end{equation*}%
See, e.g., \cite[Section 5.2.2]{BratteliRobinson}.

An important example of a $\ast $-automorphism $\tau =\sigma _{\theta }$ of $%
\mathrm{CAR}(\mathfrak{h})$ is defined, for any fixed $\theta \in \mathbb{R}%
/(2\pi \mathbb{Z)}$, by the condition%
\begin{equation*}
\sigma _{\theta }(a(\varphi ))=\mathrm{e}^{-i\theta }a(\varphi )\text{ },%
\text{\qquad }\varphi \in \mathfrak{h}\ .
\end{equation*}%
A special role is played by $\sigma _{\pi }$. Elements $A,B\in \mathrm{CAR}(%
\mathfrak{h})$ satisfying $\sigma _{\pi }(A)=A$ and $\sigma _{\pi }(B)=-B$
are respectively called \emph{even} and \emph{odd}, whereas elements $A\in 
\mathrm{CAR}(\mathfrak{h})$ satisfying $\sigma _{\theta }(A)=A$ for any $%
\theta \in \lbrack 0,2\pi )$ are called \emph{gauge invariant}. The space of
all even elements forms a $C^{\ast }$-subalgebra, called the even subalgebra
of $\mathrm{CAR}(\mathfrak{h})$.

\subsection{From CAR Algebra to Self-Dual CAR Algebra\label{notation hilbert}%
}

The\ (finite-dimensional) Hilbert space to be used to construct a self-dual
CAR algebra is the direct sum 
\begin{equation}
\mathcal{H}\doteq \mathfrak{h}\oplus \mathfrak{h}^{\ast }\ .
\label{Hilbert space sum}
\end{equation}%
Compare with Equation (\ref{definition H bar}). The scalar product on $%
\mathcal{H}$ is 
\begin{equation}
\left\langle \varphi ,\tilde{\varphi}\right\rangle _{\mathcal{H}}\doteq
\left\langle \mathrm{\varphi }_{1},\mathrm{\tilde{\varphi}}_{1}\right\rangle
_{\mathfrak{h}}+\left\langle \mathrm{\tilde{\varphi}}_{2},\mathrm{\varphi }%
_{2}\right\rangle _{\mathfrak{h}}\ ,\qquad \varphi =(\mathrm{\varphi }_{1},%
\mathrm{\varphi }_{2}^{\ast }),\ \tilde{\varphi}=(\mathrm{\tilde{\varphi}}%
_{1},\mathrm{\tilde{\varphi}}_{2}^{\ast })\in \mathcal{H}\ .
\label{scalar product in H bar}
\end{equation}%
Here, $\varphi ^{\ast }$ denotes the element of the dual $\mathfrak{h}^{\ast
}$ of the Hilbert space $\mathfrak{h}$ which is related to $\varphi $ via
the Riesz representation. See also Notation \ref{remark constant} (i). We
define the canonical antiunitary involution $\mathfrak{A}$ of $\mathcal{H}$
by%
\begin{equation}
\mathfrak{A}\left( \mathrm{\varphi }_{1},\mathrm{\varphi }_{2}^{\ast
}\right) \doteq \left( \mathrm{\varphi }_{2},\mathrm{\varphi }_{1}^{\ast
}\right) \ ,\qquad \varphi =(\mathrm{\varphi }_{1},\mathrm{\varphi }%
_{2}^{\ast })\in \mathcal{H}\ .  \label{antiunitary simple}
\end{equation}%
Note that $\varphi ^{\ast }=\mathfrak{A}\varphi $ for any $\varphi \in 
\mathfrak{h}\subset \mathcal{\mathcal{H}}$. Compare with Definition \ref%
{definition involution}.

Then, the CAR algebra $\mathrm{CAR}(\mathfrak{h})$ and the self-dual CAR
algebra $\mathrm{sCAR}(\mathcal{H},\mathfrak{A})$ are the same $C^{\ast }$%
-algebra, by defining 
\begin{equation}
\mathrm{B}\left( \varphi \right) \equiv \mathrm{B}_{P_{\mathfrak{h}}}\left(
\varphi \right) \doteq a(\mathrm{\varphi }_{1})+a(\mathrm{\varphi }%
_{2})^{\ast }\ ,\qquad \varphi =(\mathrm{\varphi }_{1},\mathrm{\varphi }%
_{2}^{\ast })\in \mathcal{\mathcal{H}}\ ,  \label{map iodiote2}
\end{equation}%
with $P_{\mathfrak{h}}\in \mathcal{B}(\mathcal{\mathcal{H}})$ being the
basis projection of $(\mathcal{H},\mathfrak{A})$ with range $\mathfrak{h}$.
Compare (\ref{CAR}) with (\ref{CAR Grassmann III}), and (\ref{map iodiote2})
with (\ref{map iodiote}). The elements $\mathrm{B}(\varphi +\mathfrak{A}%
\varphi )$, $\varphi \in \mathfrak{h}$, can thus be seen as \emph{field
operators} in the context of CAR algebra.

\subsection{Quadratic Fermionic Hamiltonians as Bilinear Elements\label%
{notation hilbert copy(2)}}

An important class of even elements of $\mathrm{CAR}(\mathfrak{h})$ in
quantum field theory is given by quadratic fermionic Hamiltonians $\mathrm{d}%
\Gamma (h)+\mathrm{d}\Upsilon (g)$ for $h=h^{\ast }\in \mathcal{B}(\mathfrak{%
h})$ and \emph{antilinear} operators $g=-g^{\ast }$ on $\mathfrak{h}$, as
defined by (\ref{form1})-(\ref{form2}). Below, we describe the relationship
between $\mathrm{d}\Gamma (h),\mathrm{d}\Upsilon (g)$ and the bilinear
Hamiltonians from Definition \ref{def trace state copy(1)}, starting with
the second quantization $\mathrm{d}\Gamma (h)$ of one-particle Hamiltonians $%
h$:\medskip

\noindent \underline{Gauge-invariant case:} We lift any operator acting on $%
\mathfrak{h}$ to an operator acting on $\mathcal{H}$ by using the linear map 
$\kappa $ from $\mathcal{B}(\mathfrak{h})$ to $\mathcal{B}(\mathcal{H})$
defined by 
\begin{equation}
\kappa \left( h\right) \doteq \frac{1}{2}\left( P_{\mathfrak{h}}hP_{%
\mathfrak{h}}-\mathfrak{A}P_{\mathfrak{h}}h^{\ast }P_{\mathfrak{h}}\mathfrak{%
A}\right) \text{ },\qquad h\in \mathcal{B}(\mathfrak{h})\text{ }.
\label{kappa}
\end{equation}%
Since $P_{\mathfrak{h}}$ is a basis projection, i.e., $\mathfrak{A}P_{%
\mathfrak{h}}\mathfrak{A}$ is the orthogonal projector on $\mathfrak{h}%
^{\ast }$,%
\begin{equation}
\kappa \left( h\right) ^{\ast }=\kappa \left( h^{\ast }\right) =-\mathfrak{A}%
\kappa \left( h\right) \mathfrak{A}\text{ },\qquad h\in \mathcal{B}(%
\mathfrak{h})\text{ }.  \label{kappabis}
\end{equation}%
Therefore, by Definition \ref{def one particle hamiltinian}, $\kappa \left(
h\right) $ is a self-dual operator on $(\mathcal{H},\mathfrak{A})$. Compare
with (\ref{equation idiote 3bis}) and (\ref{kappabisbiskappabisbis}).

Second quantizations of the form (\ref{form1}) are, up to constants,
bilinear Hamiltonians: Using Equation (\ref{map iodiote2}), one verifies that%
\begin{equation}
\mathrm{d}\Gamma (h)=-\langle \mathrm{B},\kappa \left( h\right) \mathrm{B}%
\rangle +\frac{1}{2}\mathrm{Tr}_{\mathfrak{h}}\left( h\right) \mathfrak{1}\
,\qquad h=h^{\ast }\in \mathcal{B}(\mathfrak{h})\ .  \label{second quant II}
\end{equation}%
Compare with Equation (\ref{second quantizzation00}). Additionally, by (\ref%
{second quant II}), for any one-particle Hamiltonian $h\in \mathcal{B}(%
\mathfrak{h})$, the Gibbs state associated with the quadratic fermionic
Hamiltonians $\mathrm{d}\Gamma (h)$ at fixed inverse temperature $\beta \in
(0,\infty )$, as given by Lemma \ref{Lemma quasi free state}, is the
quasi-free state over $\mathrm{sCAR}(\mathcal{H},\mathfrak{A})$ whose symbol
is $(1+\mathrm{e}^{-\beta \kappa (h)})^{-1}$. In particular, in the
one-particle Hilbert space $\mathfrak{h}$, we recover the Fermi-Dirac
distribution, at inverse temperature $\beta $, associated with the
one-particle Hamiltonian $h$.\medskip

\noindent \underline{Non-gauge-invariant case:} Similar to (\ref{equation
idiote 3bis}) and (\ref{kappa}), let%
\begin{equation*}
\tilde{\kappa}\left( g\right) \doteq \frac{1}{4}\left( P_{\mathfrak{h}%
}\left( g-g^{\ast }\right) P_{\mathfrak{h}}\mathfrak{A}+\mathfrak{A}P_{%
\mathfrak{h}}\left( g^{\ast }-g\right) P_{\mathfrak{h}}\right) \in \mathcal{B%
}(\mathcal{H})
\end{equation*}%
for any antilinear operator $g$ on $\mathfrak{h}$. Note that%
\begin{equation*}
\tilde{\kappa}\left( g\right) ^{\ast }=\tilde{\kappa}\left( g\right) \qquad 
\text{and}\qquad \tilde{\kappa}\left( g\right) =-\mathfrak{A}\tilde{\kappa}%
\left( g\right) \mathfrak{A}\text{ }.
\end{equation*}%
Therefore, $\tilde{\kappa}\left( g\right) $ is a self-dual Hamiltonian on $(%
\mathcal{H},\mathfrak{A})$ and, by Definition \ref{def trace state copy(1)}, 
\begin{equation*}
\mathrm{d}\Upsilon (g)=-\langle \mathrm{B},\tilde{\kappa}\left( g\right) 
\mathrm{B}\rangle
\end{equation*}%
provided $g=-g^{\ast }$. By Equation (\ref{second quant II}), it follows
that 
\begin{equation}
\mathrm{d}\Gamma (h)+\mathrm{d}\Upsilon (g)=-\langle \mathrm{B},\left[
\kappa \left( h\right) +\tilde{\kappa}\left( g\right) \right] \mathrm{B}%
\rangle +\frac{1}{2}\mathrm{Tr}_{\mathfrak{h}}\left( h\right) \mathfrak{1}
\label{second quant III}
\end{equation}%
for any $h=h^{\ast }\in \mathcal{B}(\mathfrak{h})$ and antilinear operator $%
g=-g^{\ast }$ on $\mathfrak{h}$.

\begin{remark}
\mbox{
}\newline
Up to a Bogoliubov $\ast $-automorphism (or unitary transformation) and some
multiple of the unit $\mathfrak{1}$, all bilinear Hamiltonians are of the
form (\ref{second quant II}). It is, however, technically advantageous to
avoid this unitary transformation\ while studying the corresponding
covariance and correlation functions. See, for instance, (\ref%
{kappabisbiskappabisbis}) and Definition \ref{definition isomorphism}, as
well as Sections \ref{Fermionic Path Integral} and \ref{sect det bounds}.
\end{remark}

\subsection{Fock Representation of CAR Algebra\label{Fock}}

Note that a CAR ($C^{\ast }$-) algebra can be constructed from any
pre-Hilbert space \cite[Theorem 5.2.5]{BratteliRobinson}. From such a $%
C^{\ast }$-algebra, there is always an injective (and thus isometric)
homomorphism to the space of bounded operators acting on the corresponding
Fock space, by \cite[Theorem 5.2.5]{BratteliRobinson}. This homomorphism is
called the \emph{Fock representation} of the CAR algebra. In the finite
dimension situation, this homomorphism is even a $\ast $-isomorphism of $%
C^{\ast }$-algebras.

Here we construct the antisymmetric Fock space from the Grassmann algebra $%
\wedge ^{\ast }\mathfrak{h}$ (Definition \ref{Definition Grassmann}):
\medskip

\noindent \underline{(i):} Let $\mathcal{X}$ be a topological vector space.
For every $n\in {\mathbb{N}}$ and $y_{1},\ldots ,y_{n}\in \mathcal{X}$, we
define the completely antisymmetric $n$--linear form $y_{1}\wedge \cdots
\wedge y_{n}$ from $(\mathcal{X}^{\ast })^{n}$ to ${\mathbb{C}}$ by 
\begin{equation*}
y_{1}\wedge \cdots \wedge y_{n}(x_{1}^{\ast },\ldots ,x_{n}^{\ast })\doteq 
\mathrm{det}\left( (x_{k}^{\ast }(y_{l}))_{k,l=1}^{n}\right) \ ,\qquad
x_{1}^{\ast },\ldots ,x_{n}^{\ast }\in \mathcal{X}^{\ast }\ .
\end{equation*}%
Compare with (\ref{eq mulit linear}). Then, similar to (\ref{directsum0})-(%
\ref{direct sums}) for $\wedge ^{\ast }\mathcal{X}$, one defines the vector
space $\wedge \mathcal{X}$. \medskip

\noindent \underline{(ii):} For any $n\in \mathbb{N}$, there is a unique
linear map $\xi \mapsto \langle \xi ,\;\cdot \,\rangle _{n}$ from $\wedge
^{\ast n}\mathcal{X}$ to the space $\left( \wedge ^{n}\mathcal{X}\right)
^{\prime }$ of all linear maps from $\wedge ^{n}\mathcal{X}$ to ${\mathbb{C}}
$ such that, for all $x_{1}^{\ast },\ldots ,x_{n}^{\ast }\in \mathcal{X}%
^{\ast }$, 
\begin{equation*}
\langle x_{1}^{\ast }\wedge \cdots \wedge x_{n}^{\ast },y_{1}\wedge \cdots
\wedge y_{n}\rangle _{n}=x_{1}^{\ast }\wedge \cdots \wedge x_{n}^{\ast
}(y_{1},\ldots ,y_{n})\ ,\qquad y_{1},\ldots ,y_{n}\in \mathcal{X}\ .
\end{equation*}%
For $n=0$, $\langle c,d\rangle _{0}\doteq cd$ with $c,d\in \mathbb{C}$. This
map is injective for any $n\in \mathbb{N}_{0}$. Then, we denote by $\xi
\mapsto \langle \xi ,\;\cdot \,\rangle $ the unique linear map from $\wedge
^{\ast }\mathcal{X}$ to $\left( \wedge \mathcal{X}\right) ^{\prime }$ such
that, for any $n,m\in \mathbb{N}$, $\xi \in \wedge ^{\ast n}\mathcal{X}$ and 
$w\in \wedge ^{m}\mathcal{X}$, 
\begin{equation*}
\langle \xi ,w\rangle =\left\{ 
\begin{array}{ccc}
\langle \xi ,w\rangle _{n} & \text{if} & n=m\ . \\ 
0 & \text{else .} & 
\end{array}%
\right.
\end{equation*}%
\medskip \noindent \underline{(iii):} Given two elements $h\in \wedge 
\mathcal{X}$ and $\xi \in \wedge ^{\ast }\mathcal{X}$, their \emph{interior
product} $h\lrcorner \xi \in \wedge ^{\ast }\mathcal{X}$ is uniquely defined
by the condition 
\begin{equation}
\langle h\lrcorner \xi ,w\rangle =\langle \xi ,h\wedge w\rangle \ ,\qquad
w\in \wedge \mathcal{X}\ .  \label{anihilation Grassmann def}
\end{equation}%
(Notice that $\langle \xi ,h\wedge (\cdot )\rangle $ is an element of $%
(\wedge \mathcal{X})^{\ast }$ of the form $\langle h\lrcorner \xi ,\;\cdot
\;\rangle $ for some $h\lrcorner \xi \in \wedge ^{\ast }\mathcal{X}$.) For
any $h\in \wedge \mathcal{X}$, the map%
\begin{equation}
h\lrcorner :\xi \mapsto h\lrcorner \xi  \label{annihilator}
\end{equation}%
from $\wedge ^{\ast }\mathcal{X}$ to itself is linear. If $h\in \mathcal{X}%
=\wedge ^{1}\mathcal{X}$ then, by (\ref{grassmana anticommute}), $h\lrcorner 
$ is an \emph{antiderivation} of degree $1$ on the graded algebra $\wedge
^{\ast }\mathcal{X}$, i.e., for all $n\in \mathbb{N}_{0}$, $\xi \in \wedge
^{\ast n}\mathcal{X}$ and $\zeta \in \wedge ^{\ast }\mathcal{X}$, 
\begin{equation}
h\lrcorner \left( \xi \wedge \zeta \right) =\left( h\lrcorner \xi \right)
\wedge \zeta +\left( -1\right) ^{n}\xi \wedge \left( h\lrcorner \zeta
\right) \ .  \label{derivative degree 1}
\end{equation}

Similar to the interior product, for any $\xi \in \wedge ^{\ast }\mathcal{X}$%
, the exterior product induces a linear map%
\begin{equation}
\xi \wedge :\zeta \mapsto \xi \wedge \zeta  \label{creator}
\end{equation}%
from $\wedge ^{\ast }\mathcal{X}$ to itself. Then, for all $\xi ,\zeta \in 
\mathcal{X}^{\ast }=\wedge ^{\ast 1}\mathcal{X}$ and $h,w\in \mathcal{X}$,%
\begin{equation}
(\xi \wedge )(\zeta \wedge )=-(\zeta \wedge )(\xi \wedge )\ ,\quad
h\lrcorner w\lrcorner =-w\lrcorner h\lrcorner \ ,\quad h\lrcorner (\xi
\wedge )+(\xi \wedge )h\lrcorner =\xi (h)\mathbf{1}_{\wedge ^{\ast }\mathcal{%
X}}\ ,  \label{grassmana anticommute2}
\end{equation}%
because of (\ref{grassmana anticommute}). These equalities refer to the
canonical anticommutation relations (CAR). Compare with (\ref{CAR Grassmann
III}) and (\ref{CAR}). \medskip

\noindent \underline{(iv):} Assume, from now on, that $\mathcal{X}$ is a 
\emph{finite-dimensional} (complex) Hilbert space. Then we denote by $%
\mathfrak{j}$ the unique antilinear map $\wedge \mathcal{X}\rightarrow
\wedge ^{\ast }\mathcal{X}$ such that, for any $n\in \mathbb{N}$, $z\in
\wedge ^{0}\mathcal{X}\doteq \mathbb{C}$ and $x_{1},\ldots ,x_{n}\in 
\mathcal{X}$, 
\begin{equation*}
\mathfrak{j}\left( x_{1}\wedge \cdots \wedge x_{n}\right) =x_{1}^{\ast
}\wedge \cdots \wedge x_{n}^{\ast }\qquad \text{and}\qquad \mathfrak{j}%
\left( z\right) =\bar{z}\in \wedge ^{\ast 0}\mathcal{X}\ .
\end{equation*}%
Similarly, $\mathfrak{j}^{\ast }$ is the unique antilinear map $\wedge
^{\ast }\mathcal{X}\rightarrow \wedge \mathcal{X}$ such that, for any $n\in 
\mathbb{N}$, $z\in \wedge ^{\ast 0}\mathcal{X}$ and $x_{1},\ldots ,x_{n}\in 
\mathcal{X}$,%
\begin{equation*}
\mathfrak{j}^{\ast }\left( x_{1}^{\ast }\wedge \cdots \wedge x_{n}^{\ast
}\right) =x_{1}\wedge \cdots \wedge x_{n}\qquad \text{and}\qquad \mathfrak{j}%
^{\ast }\left( z\right) =\bar{z}\in \wedge ^{0}\mathcal{X}\ .
\end{equation*}%
Because of the Fr{\'{e}}chet-Riesz\ representation theorem, $\mathfrak{j}$
and $\mathfrak{j}^{\ast }$ are bijective maps satisfying 
\begin{equation*}
\mathfrak{j}\circ \mathfrak{j}^{\ast }=\mathbf{1}_{\wedge ^{\ast }\mathcal{X}%
}\qquad \text{and}\qquad \mathfrak{j}^{\ast }\circ \mathfrak{j}=\mathbf{1}%
_{\wedge \mathcal{X}}\ .
\end{equation*}%
The condition 
\begin{equation*}
(\zeta ,\xi )\doteq \langle \xi ,\mathfrak{j}^{\ast }(\zeta )\rangle \
,\qquad \xi ,\zeta \in \wedge ^{\ast }\mathcal{X}\ ,
\end{equation*}%
defines a scalar product $(\cdot ,\cdot )$ in $\wedge ^{\ast }\mathcal{X}$.

$\wedge ^{\ast }\mathcal{X}$ equipped with this scalar product is a Hilbert
space $\mathcal{F}_{\mathcal{X}^{\ast }}$, the so-called \emph{fermionic
Fock space} associated with the (one-particle) Hilbert space $\mathcal{X}%
^{\ast }$. Moreover, for $\xi \in \mathcal{X}^{\ast }$ and $h\in \mathcal{X}$%
, the linear operators $h\lrcorner \in \mathcal{B}(\mathcal{F}_{\mathcal{X}%
^{\ast }})$ and $\xi \wedge \in \mathcal{B}(\mathcal{F}_{\mathcal{X}^{\ast
}})$, respectively defined by (\ref{annihilator}) and (\ref{creator}), are
related to each other by taking the adjoint: 
\begin{equation}
\left( \xi \wedge \right) ^{\ast }=\mathfrak{j}^{\ast }\left( \xi \right)
\lrcorner \qquad \text{and}\qquad \left( h\lrcorner \right) ^{\ast }=%
\mathfrak{j}\left( h\right) \wedge \ .  \label{toto creations anni}
\end{equation}%
Their operator norms are respectively equal to 
\begin{equation}
\left\Vert h\lrcorner \right\Vert _{\mathcal{B}(\mathcal{F}_{\mathcal{X}%
^{\ast }})}=\left\Vert h\right\Vert _{\mathcal{X}}\qquad \text{and}\qquad
\left\Vert \xi \wedge \right\Vert _{\mathcal{B}(\mathcal{F}_{\mathcal{X}%
^{\ast }})}=\left\Vert \xi \right\Vert _{\mathcal{X}^{\ast }}\ ,
\label{bounded1}
\end{equation}%
by the CAR (\ref{grassmana anticommute2}). \medskip

\noindent \underline{(v):} Again because of (\ref{grassmana anticommute2}),
if $\mathcal{X}=\mathfrak{h}^{\ast }$, then there is an obvious $\ast $%
-isomorphism $\mathfrak{r}$ from $\mathrm{CAR}(\mathfrak{h})$ to $\mathcal{B}%
(\mathcal{F}_{\mathfrak{h}})$ that is uniquely defined by 
\begin{equation}
\mathfrak{r}\left( a(\varphi )\right) =\varphi ^{\ast }\lrcorner \ ,\qquad
\varphi \in \mathfrak{h}\ .  \label{*--isomorphism r}
\end{equation}%
The norm on $\mathrm{CAR}(\mathfrak{h})$ corresponds to the usual operator
norm on $\mathcal{B}(\mathcal{F}_{\mathfrak{h}})$ and, since the
antisymmetric Fock space $\mathcal{F}_{\mathfrak{h}}$ has dimension equal to 
$2^{\mathrm{dim}\mathfrak{h}}$, the CAR algebra is also finite-dimensional: 
\begin{equation*}
\mathrm{dim}\left( \mathrm{CAR}(\mathfrak{h})\right) =2^{2\mathrm{dim}%
\mathfrak{h}}\ .
\end{equation*}

\begin{remark}[Tracial state]
\label{bound norm a CAR copy(2)}\mbox{
}\newline
For any $\ast $-isomorphism $\mathfrak{r}$ from $\mathrm{CAR}(\mathfrak{h})$
to $\mathcal{B}(\mathcal{F}_{\mathfrak{h}})$, the tracial state is written
as a normalized trace $\mathrm{Tr}_{\mathcal{F}_{\mathfrak{h}}}$ on linear
operators acting on the finite-dimensional fermionic Fock space $\mathcal{F}%
_{\mathfrak{h}}$: 
\begin{equation*}
\mathrm{tr}(A)=2^{-\mathrm{dim}\mathfrak{h}}\mathrm{Tr}_{\mathcal{F}_{%
\mathfrak{h}}}(\mathfrak{r}(A))\ ,\qquad A\in \mathrm{CAR}(\mathfrak{h})\ .
\end{equation*}
\end{remark}

\section{Quantum Large Deviations\label{Section Historical Overview}}

\subsection{Large Deviations (LD) in Quantum Statistical Mechanics\label{LD
histiry}}

In probability theory, the law of large numbers refers to the convergence
(at least in probability), as $n\rightarrow \infty $, of the average or
empirical mean of $n$ independent identically distributed (i.i.d.) random
variables towards their expected value (assuming it exists). The central
limit theorem states that, as $n\rightarrow \infty $, the probability
distribution of fluctuations (at scale $\sqrt{n}$) of this finite average
around the expected value is a normal one (assuming the finiteness of the
variance).\ The large deviation (LD) formalism quantitatively describes, for
large $n\gg 1$, the probability of finding an empirical mean that differs
from the expected value. The latter refers to \emph{rare} events, by the law
of large numbers, and an LD principle (LDP) gives their probability as
exponentially small in the limit $n\rightarrow \infty $.

The concept of LD goes beyond the law of large numbers and complements the
central limit theorem. Note, however, that, by the Bryc theorem \cite[%
Proposition 1]{Bryc1993253}, under some assumptions, the central limit
theorem can result from an LDP. Moreover, the LD formalism does not
necessarily require i.i.d. random variables, i.e., it can be studied for
random variables that differ from empirical means of $n$ i.i.d. random
variables. For a comprehensive account of the LD method, see \cite%
{DS89,dembo1998large}.

The LD formalism is applied in physics since probability theory is the
natural mathematical framework of (classical) statistical mechanics. See 
\cite{E85,L86,touchette2009large}. When an LDP holds true, there is a
well-defined exponential rate of convergence of local quantities towards a
thermodynamic one. In physics, this rate of convergence is frequently
related to some notion of entropy. Moreover, in this case, a powerful
extension of Laplace's method for infinite-dimensional spaces, well-known as 
\emph{Varadhan's lemma} \cite{V66}, becomes available. See \cite%
{DS89,dembo1998large}, in particular \cite[Theorem 2.1.10]{DS89} or \cite[%
Theorem 4.3.1]{dembo1998large}.

Since the eighties, the LD formalism has been successfully used in \emph{%
quantum} statistical mechanics, initially at a time when the method of LD
was not well-known in theoretical physics \cite[p. 63]{BLP88}. This was
further developed by the \textquotedblleft Dublin school\textquotedblright\
in 1988-1993. In the series of papers \cite%
{BLP88,DLP89,BDLP90a,BDLP90b,DLP93} the authors studied boson systems in the
continuum with interactions that only depend on occupation-number operators.
This allowed them to map a thermodynamic problem in the bosonic Fock space
to a classical probability problem for occupation numbers seen as a random
variable. It gives a nice application of LDP to the occupation measure in
order to solve a class of diagonal boson models, see \cite[Section 4.1.1]%
{BruZagrebnov8}. In \cite{LZ88} an LDP was proved for the distribution of
the particle density (the so-called Kac distribution) in the Perfect and
Mean-Field Bose gases.

More recently, \emph{non-diagonal} boson models were fully studied using
LDP, see, e.g., \cite{BZ98,B04,BZ08}. For instance, \cite{BZ08} proves an
LDP for both the (generalized) Kac distribution and the Bose condensate
density in order to study the first order phase transition appearing in the
thermodynamics of the superstable Weakly Imperfect Bose Gas (or Superstable
Bogoliubov model). The phenomenon of Bose condensation in quantum
statistical mechanics has inspired new LD studies in probability theory. See 
\cite{ABK06a,ABK06b,A08,AK08,ACK11}. In these works, the random variables
are quite elaborated mathematical objects within infinite-dimensional
spaces. For instance, the authors in \cite{ACK11} derive a variational
formula for the limiting free energy of a general Bose gas with pair
interactions on the continuum by using marked Poisson point processes and an
LD analysis of the so-called stationary empirical field.

The use of the LD formalism in quantum statistical mechanics is not
restricted to Bose systems. In 1987-1989, this method was used to analyze
quantum spin systems \cite{DP87,CLR88,RaggioWerner1}. These analyses
included certain types of translation-invariant fermionic models with
long-range components on the lattice (e.g., reduced BCS models) via the
quantum spin representation of fermions. Much more recently, as already
mentioned, \cite{LLS00,GLM02,B04} focused on LD for the particle density in
a sub-domain both for the perfect and for rarefied quantum gases (Fermi,
Bose or Boltzmann statistics). Finally, in recent years the LD formalism was
used to study quantum lattice systems at thermal equilibrium, as summarized
in the works \cite%
{LLS00,GLM02,netovcny2004large,lenci2005large,Ogata2010,OR11,RoMaNeSc}, to
name a few examples. In most cases, equilibrium states are, by definition,
infinite-volume KMS states, which are supposed to be unique (i.e., there is
no phase transition).

Note that our list of results related to LD methods in quantum statistical
mechanics is not exhaustive. Our aim is to show its usefulness in solving
important models of quantum\ statistical physics. We also observe at this
point that LD methods do not form a real \textquotedblleft
theory\textquotedblright , as explained in \cite[Preface]{DS89}, and they
have been used in many different ways in thermodynamic studies of quantum
(many-body) models. For instance, in \cite{BLP88}, etc., LD methods are used
by mapping the many-body problem to a relatively simple probabilistic one,
for very specific models. In \cite{ACK11}, the partition function for a
general interacting boson model in the canonical ensemble is reformulated
via the Feynman-Kac formula in terms of interacting Brownian bridges in
order to later apply LD methods to some stochastic process.

In the recent works \cite%
{LLS00,GLM02,netovcny2004large,lenci2005large,HiaiMosoOga,Ogata2010,OR11,RoMaNeSc}
on quantum lattice systems, the LD method is applied to (KMS) states on
quantum spin or CAR $C^{\ast }$-algebra, via the algebraic formulation of
Quantum Mechanics. See, e.g., \cite[Section 2]{brupedraLR}. This approach is
also used in the present work to study LD properties associated with any
observable for weakly interacting fermionic gases at thermal equilibrium. We
thus briefly present this approach in Section \ref{LD Formalism within
Algebraic Quantum Mechanics}.

\subsection{LD Formalism for Quantum Observables\label{LD Formalism within
Algebraic Quantum Mechanics}\label{subsec:LDFormalism}}

Let $\mathcal{X}$ be a unital $C^{\ast }$-algebra. In the algebraic
formulation of Quantum Mechanics, each self-adjoint element $A^{\ast }=A$ of 
$\mathcal{X}$, also called an \emph{observable}, represents some apparatus
(or measuring device), and its spectrum, denoted by $\mathrm{spec}(A)$,
corresponds to all values that can arise by measuring the corresponding
physical quantity. The state of the physical system is represented by \emph{%
states} on the $C^{\ast }$-algebra $\mathcal{X}$, that is, by definition,
continuous linear functionals $\rho $ which are normalized and positive,
i.e., $\rho (\mathfrak{1})=1$ and $\rho (A^{\ast }A)\geq 0$ for all $A\in 
\mathcal{X}$.

The commutative $C^{\ast }$-subalgebra of $\mathcal{X}$ generated by any
self-adjoint element $A^{\ast }=A\in \mathcal{X}$ is isomorphic to the
algebra of continuous functions on the compact set $\mathrm{spec}(A)\subset 
\mathbb{R}$. Hence, by the Riesz-Markov theorem (or Riesz representation
theorem), for any observable $A\in \mathcal{X}$ and state $\rho \in \mathcal{%
X}^{\ast }$, there is a unique probability measure $\mu _{\rho ,A}$ on $%
\mathbb{R}$ such that%
\begin{equation}
\mu _{\rho ,A}(\mathrm{spec}(A))=1\qquad \text{and}\qquad \rho \left(
f(A)\right) =\int_{\mathbb{R}}f(x)\mu _{\rho ,A}(\mathrm{d}x)
\label{fluctuation measure}
\end{equation}%
for all complex-valued continuous functions $f\in C(\mathbb{R};\mathbb{C})$.
Note that, for any observable $A\in \mathcal{X}$, $y\in \mathbb{R}$ and any
Borel subset $\mathcal{O}\subset \mathbb{R}$, 
\begin{equation*}
\mu _{\rho ,A-y\mathfrak{1}}\left( \mathcal{O}\right) =\mu _{\rho ,A}\left( 
\mathcal{O}+y\right) \ ,
\end{equation*}%
by the uniqueness of the probability measure in the Riesz-Markov theorem. $%
\mu _{\rho ,A}$ is called the \emph{distribution} of the observable $A$ in
the state $\rho $.

Since distributions are, by construction, probability measures on the
complete, separable space $\mathbb{R}$, the LD formalism naturally arises: A 
\emph{rate} function is a lower semi-continuous function $\mathrm{I}:\mathbb{%
R}\rightarrow \lbrack 0,\infty ]$. If $\mathrm{I}$ is not the $\infty $%
--constant function and has compact level sets, i.e., if $\mathrm{I}%
^{-1}([0,m])=\{x\in \mathbb{R}:\mathrm{I}(x)\leq m\}$ is compact for any $%
m\geq 0$, then one says that $\mathrm{I}$ is a \emph{good }rate function. A
sequence $\{A_{l}\}_{l\in \mathbb{R}^{+}}\subset \mathcal{X}$ of observables
satisfies the LD \emph{upper} bound in a state $\rho \in \mathcal{X}^{\ast }$
with speed $\mathfrak{n}_{l}\in \mathbb{R}$, $l\in \mathbb{R}^{+}$, and rate
function $\mathrm{I}$ if, for any closed subset $\mathcal{C}$ of $\mathbb{R}$%
, 
\begin{equation}
\limsup_{l\rightarrow \infty }\frac{1}{\mathfrak{n}_{l}}\ln \mu _{\rho
,A_{l}}\left( \mathcal{C}\right) \leq -\inf_{\mathcal{C}}\mathrm{I}\left(
x\right) \ ,  \label{LDPupper}
\end{equation}%
and it satisfies the LD \emph{lower} bound in (a state $\rho $ and) the
closed interval $\mathcal{I}\subseteq \mathbb{R}$ if, for any open subset $%
\mathcal{G}\subset \mathcal{I}$ of $\mathbb{R}$, 
\begin{equation}
\liminf_{l\rightarrow \infty }\frac{1}{\mathfrak{n}_{l}}\ln \mu _{\rho
,A_{l}}\left( \mathcal{G}\right) \geq -\inf_{\mathcal{G}}\mathrm{I}\left(
x\right) \ .  \label{LDPlower}
\end{equation}%
If both upper and lower bound are satisfied, we say that $\{A_{l}\}_{l\in 
\mathbb{R}^{+}}\subset \mathcal{X}$ satisfies a \emph{large deviation
principle} (LDP) in a state $\rho $ and the closed interval $\mathcal{I}%
\subseteq \mathbb{R}$. The LDP is called \emph{weak} if the upper bound in (%
\ref{LDPupper}) holds only for compact\textit{\ }sets $\mathcal{C}$.

One of the most important consequences of an LDP satisfied by $%
\{A_{l}\}_{l\in \mathbb{R}^{+}}\subset \mathcal{X}$ in the state $\rho \in 
\mathcal{X}^{\ast }$ and in $\mathcal{I}=\mathbb{R}$ is \emph{Varadhan's
lemma}, which says, in this context, that the limiting generating function 
\begin{equation*}
\mathrm{G}\left( f\right) \doteq \lim_{l\rightarrow \infty }\frac{1}{%
\mathfrak{n}_{l}}\ln \int_{\mathbb{R}}\exp \left( \mathfrak{n}_{l}f\left(
x\right) \right) \mu _{\rho ,A_{l}}(\mathrm{d}x)=\lim_{l\rightarrow \infty }%
\frac{1}{\mathfrak{n}_{l}}\ln \rho \left( \mathrm{e}^{\mathfrak{n}%
_{l}f\left( A_{l}\right) }\right)
\end{equation*}%
is equal, for any bounded and continuous real-valued function $f\in C_{b}(%
\mathbb{R})$, to 
\begin{equation*}
\mathrm{G}\left( f\right) =\sup_{x\in \mathbb{R}}\left\{ f\left( x\right) -%
\mathrm{I}\left( x\right) \right\} \ .
\end{equation*}%
See \cite[Theorem 2.1.10]{DS89} or \cite[Theorem 4.3.1]{dembo1998large}.

The converse holds true when one can ensure that no loss of mass appears in
the limit $l\rightarrow \infty $: We say that a family $\{A_{l}\}_{l\in 
\mathbb{R}^{+}}\subset \mathcal{X}$ of observables is \emph{exponentially
tight} in the state $\rho \in \mathcal{X}^{\ast }$ if the sequence $\{\mu
_{\rho ,A_{l}}\}_{l\in \mathbb{R}^{+}}$ is exponentially tight. This means
that, for every $D<\infty $, there exists a compact set $\mathcal{C}%
_{D}\subset \mathbb{R}$ such that 
\begin{equation*}
\limsup_{l\rightarrow \infty }\frac{1}{\mathfrak{n}_{l}}\ln \mu _{\rho
,A_{l}}\left( \mathbb{R}\backslash \mathcal{C}_{D}\right) <-D\ .
\end{equation*}%
Then, by \cite{B90} (or \cite[Theorem 4.4.2]{dembo1998large}), the existence
of $\mathrm{G}\left( f\right) $ for all $f\in C_{b}(\mathbb{R})$, together
with the exponential tightness of $\{A_{l}\}_{l\in \mathbb{R}^{+}}\subset 
\mathcal{X}$ in the state $\rho \in \mathcal{X}^{\ast }$, implies that $%
\{A_{l}\}_{l\in \mathbb{R}^{+}}$ satisfies an LDP with the good rate
function $\mathrm{I}$ defined, for any $x\in \mathbb{R}$, by 
\begin{equation*}
\mathrm{I}\left( x\right) \doteq \sup_{f\in C_{b}(\mathbb{R})}\left\{
f\left( x\right) -\mathrm{G}\left( f\right) \right\} \ .
\end{equation*}

To prove the existence of an LDP, the knowledge of $\mathrm{G}\left(
f\right) $ for \emph{all} $f\in C_{b}(\mathbb{R})$ is not necessary.
Instead, by the G\"{a}rtner-Ellis theorem, one can consider the (limiting)%
\emph{\ logarithmic moment generating function}. See, e.g., \cite[Section 2.2%
]{DS89} or \cite[Section 2.3 or 4.5]{dembo1998large}. For any sequence $%
\{A_{l}\}_{l\in \mathbb{R}^{+}}\subset \mathcal{X}$ of observables and
state\ $\rho \in \mathcal{X}^{\ast }$, this function is defined by 
\begin{equation}
\mathrm{J}\left( s\right) \doteq \lim_{l\rightarrow \infty }\frac{1}{%
\mathfrak{n}_{l}}\ln \left( \int_{\mathbb{R}}\mathrm{e}^{s\mathfrak{n}%
_{l}x}\mu _{\rho ,A_{l}}(\mathrm{d}x)\right) =\lim_{l\rightarrow \infty }%
\frac{1}{\mathfrak{n}_{l}}\ln \rho \left( \mathrm{e}^{s\mathfrak{n}%
_{l}A_{l}}\right) \ ,\qquad s\in \mathbb{R}\ .  \label{eq:state_LDP}
\end{equation}%
It equals $\mathrm{G}\left( h_{s}\right) $ with $h_{s}(x)\doteq sx$ for $%
s,x\in \mathbb{R}$ and it is the standard limiting logarithmic moment
generating function of the sequence $\{\mu _{\rho ,A_{l}}\}_{l\in \mathbb{R}%
^{+}}$ of measures on $\mathbb{R}$.

Then, a sufficient condition to ensure that a sequence of observables
satisfies an LDP is given by the G\"{a}rtner-Ellis theorem (see, e.g., \cite[%
Corollary 4.5.27]{dembo1998large}), which is better adapted to our methods
than Bryc's inverse Varadhan lemma \cite[Theorem 4.4.2]{dembo1998large}. In
our context, this refers to the following assertion:

\begin{theorem}[G\"{a}rtner-Ellis]
\label{prop Gartner--Ellis}\mbox{ }\newline
Let $\rho \in \mathcal{X}^{\ast }$ be a state. Take any sequence $%
\{A_{l}\}_{l\in \mathbb{R}^{+}}\subset \mathcal{X}$ of observables that is
exponentially tight in the state $\rho $. Assume the existence in a closed
interval $\mathcal{J}\subseteq \mathbb{R}$ of the limiting logarithmic
moment generating function $\mathrm{J}$ defined by (\ref{eq:state_LDP}).
Then, one has: \newline
\emph{(LD1)} $\{A_{l}\}_{l\in \mathbb{R}^{+}}$ satisfies the LD upper bound (%
\ref{LDPupper}) in the state $\rho $, with rate function $\mathrm{I}$ given
as the Legendre-Fenchel transform of $\mathrm{J}$, that is, the convex lower
semi-continuous map from $\mathbb{R}$ to $\left( -\infty ,\infty \right] $
defined, for any $x\in \mathbb{R}$, by 
\begin{equation}
\mathrm{I}(x)\doteq \sup\limits_{s\in \mathcal{J}}\left\{ sx-\mathrm{J}%
\left( s\right) \right\} \ .  \label{rate function}
\end{equation}%
\emph{(LD2)} If $\mathrm{J}$ is differentiable for all $s\in \mathcal{J}$
and 
\begin{equation*}
\lim_{l\rightarrow \infty }\rho (A_{l})\in \left( \inf \mathrm{J}^{\prime
}\left( \mathcal{J}\right) ,\sup \mathrm{J}^{\prime }\left( \mathcal{J}%
\right) \right) \ ,
\end{equation*}%
then $\{A_{l}\}_{l\in \mathbb{R}^{+}}$ satisfies the LD lower bound (\ref%
{LDPlower}) in (the state $\rho $ and) the closed interval 
\begin{equation*}
\mathcal{I}\doteq \left[ \inf \mathrm{J}^{\prime }\left( \mathcal{J}\right)
,\sup \mathrm{J}^{\prime }\left( \mathcal{J}\right) \right]
\end{equation*}%
with good rate function $\mathrm{I}$ given by (\ref{rate function}).
\end{theorem}

\begin{proof}
(LD1) and (LD2) for $\mathcal{I}=\mathbb{R}$ are consequences of \cite[%
Theorem 2.2.4]{DS89} and \cite[Corollary 4.5.27]{dembo1998large},
respectively. The case $\mathcal{I}\varsubsetneq \mathbb{R}$ follows
straightforwardly from the previous one. For an explicit proof in the
special case of finite intervals $\mathcal{I}\varsubsetneq \mathbb{R}$, see 
\cite[Appendix A.2]{JOPP12}.
\end{proof}

\cite{Bryc1993253} explains how, in the zero mean case, the central limit
theorem is deduced from the existence of the limiting logarithmic moment
generating function on a neighborhood - in the complex plane - of the
origin. Bryc's result \cite[Proposition 1]{Bryc1993253} can be rewritten in
our context as follows:

\begin{theorem}[Bryc]
\label{Coro estimate super importante}\mbox{ }\newline
Take a sequence $\{A_{l}\}_{l\in \mathbb{R}^{+}}\subset \mathcal{X}$ of
observables and fix a state $\rho \in \mathcal{X}^{\ast }$. If the limiting
logarithmic moment generating function $\mathrm{J}$ defined by (\ref%
{eq:state_LDP}) exists and admits an analytic continuation to a neighborhood
of $s=0\in \mathbb{C}$, then the family $\mu _{\sqrt{\mathfrak{n}_{l}}\left(
A_{l}-\rho (A_{l})\right) }$, $l\in \mathbb{R}^{+}$, of measures converges
in the weak$^{\ast }$-topology to the normal (or Gaussian) distribution $%
\mathcal{N}_{0,\sigma ^{2}}$ with variance $\sigma ^{2}=\mathrm{J}^{\prime
\prime }\left( 0\right) $.
\end{theorem}

\noindent Note that \cite[Proposition 2]{Bryc1993253} gives a condition to
ensure the hypothesis of this proposition by using an LDP. See also \cite[%
Lemma 2.2.9]{DS89}.

\section{Application of LD to Quantum Statistical Mechanics\label{Spin or
CAR Unital Algebras copy(1)}}

\subsection{Spin or CAR Algebra\label{Spin or CAR Unital Algebras}}

The applications we have in mind refer to quantum spins or fermions on a $d$%
-dimensional lattice $\mathbb{Z}^{d}$, $d\in \mathbb{N}$. Denote by $%
\mathcal{P}_{f}(\mathbb{Z}^{d})\subset 2^{\mathbb{Z}^{d}}$ the set of all 
\emph{finite} subsets of $\mathbb{Z}^{d}$. Fix also a finite set $\mathrm{S}$%
, the elements of which represent the pure spin states of a quantum particle
in any arbitrary site of the lattice.

Then, in the algebraic formulation of Quantum Mechanics, it is standard to
associate with the pair $(\mathbb{Z}^{d},\mathrm{S})$\ a unital $C^{\ast }$%
-algebra $\mathcal{X}=\mathcal{U}$ generated by a net $\{\mathcal{U}%
_{\Lambda }\}_{\Lambda \in \mathcal{P}_{f}(\mathbb{Z}^{d})}\subset \mathcal{U%
}$ of finite-dimensional subalgebras. In fact, $\mathcal{U}$ is a so-called 
\emph{quasi-local algebra} with respect to this generating net, see \cite[%
Definition 2.6.3.]{BratteliRobinsonI}. For quantum spin systems, 
\begin{equation}
\mathcal{U}_{\Lambda }=\mathcal{B}(\mathbb{C}^{\left\vert \Lambda
\right\vert \times \left\vert \mathrm{S}\right\vert })\ ,  \label{algebra1}
\end{equation}%
while, for fermions on the lattice, 
\begin{equation}
\mathcal{U}_{\Lambda }=\mathrm{CAR}(\ell ^{2}(\Lambda \times \mathrm{S}%
))\equiv \mathcal{B}(\mathcal{F}_{\ell ^{2}(\Lambda \times \mathrm{S}%
)}))\equiv \mathcal{B}(\mathbb{C}^{2^{2\left\vert \Lambda \right\vert \times
\left\vert \mathrm{S}\right\vert }})  \label{algebra2}
\end{equation}%
is the (local or finite-volume) CAR ($C^{\ast }$-) algebra associated with
the one-particle Hilbert space $\ell ^{2}\left( \Lambda \times \mathrm{S}%
\right) $, as explained in Section \ref{Section Gen func as Grassmann int
copy(1)}. Recall that $\mathcal{B}\left( \mathfrak{h}\right) $ denotes the $%
C^{\ast }$-algebra of bounded linear operators acting on $\mathfrak{h}$
(Notation \ref{remark constant}), while $\mathcal{F}_{\mathfrak{h}}$ is the
fermionic Fock space associated with the one-particle Hilbert space $%
\mathfrak{h}$ (Section \ref{Fock}). (In the bosonic case, a similar
formalism can be developed with the use of CCR (Canonical Commutation
Relations) algebras. See \cite{BratteliRobinson, EK98}.)

In the context of self-dual CAR algebra, described in Section \ref{notation
hilbert}, for any $\Lambda \in \mathcal{P}_{f}(\mathbb{Z}^{d})$, $\mathcal{U}%
_{\Lambda }$ can be seen as the self-dual CAR algebra associated with $\ell
^{2}(\Lambda \times \mathrm{S})\oplus \ell ^{2}(\Lambda \times \mathrm{S}%
)^{\ast }$. Then, the CAR algebra $\mathcal{U}$ generated by the net $\{%
\mathcal{U}_{\Lambda }\}_{\Lambda \in \mathcal{P}_{f}(\mathbb{Z}^{d})}$ is,
in turn, the self-dual CAR\ algebra associated with $\ell ^{2}(\mathbb{Z}%
^{d}\times \mathrm{S})\oplus \ell ^{2}(\mathbb{Z}^{d}\times \mathrm{S}%
)^{\ast }$, similar to the finite-dimensional case. Note that 
\begin{equation}
\ell ^{2}(\Lambda \times \mathrm{S})\oplus \ell ^{2}(\Lambda \times \mathrm{S%
})^{\ast }\equiv \ell ^{2}(\Lambda ;\mathcal{H}_{\mathrm{S}})\ ,\qquad
\Lambda \subset \mathbb{Z}^{d}\ .  \label{Hilbert self dual}
\end{equation}
Compare with Section \ref{Section summability}.

\subsection{Thermodynamic Sequences of Observables\label{Thermodynamic
Sequences of Observables}}

The speed of the LD we study is typically $\mathfrak{n}_{l}=|\Lambda _{l}|$, 
$l\in \mathbb{R}^{+}$, where $\{\Lambda _{l}\}_{l\in \mathbb{R}^{+}}$ is the
sequence of cubic boxes defined by (\ref{eq:boxesl1}). This choice to define
the thermodynamic limit is technically convenient, but not essential. The
arguments can be adapted to more general van Hove nets of boxes. A typical
sequence $\{A_{l}\}_{l\in \mathbb{R}^{+}}$ of observables analyzed in the
context of Section \ref{LD Formalism within Algebraic Quantum Mechanics} is
the sequence of space averages, of some fixed observable $A$, over the boxes 
$\Lambda _{l}$.

Denote by $\{\chi _{x}\}_{x\in \mathbb{Z}^{d}}$ the family of $\ast $-automor%
%TCIMACRO{\TeXButton{\-}{\-}}%
%BeginExpansion
\-%
%EndExpansion
phisms of the $C^{\ast }$-algebra $\mathcal{U}$ that implements the action
of the group of lattice translations. When $\mathcal{U}=\mathrm{CAR}(\ell
^{2}(\mathbb{Z}^{d}\times \mathrm{S}))$, this family is uniquely defined by
the conditions 
\begin{equation*}
\chi _{x}(a(\varphi ))=a(\varphi (x+\cdot ,\cdot ))\ ,\qquad \varphi \in
\ell ^{2}\left( \mathbb{Z}^{d}\times \mathrm{S};\mathbb{C}\right) \ .
\end{equation*}%
The definition of this family is also straightforward for quantum spin
systems and we omit the details.

Then, the \emph{empirical mean }$\{\mathrm{M}_{l}^{A}\}_{l\in \mathbb{R}%
^{+}} $ of\ observables is defined, for any $l\in \mathbb{R}^{+}$ and $A\in 
\mathcal{U}$, by 
\begin{equation}
\mathrm{M}_{l}^{A}\doteq \frac{1}{|\Lambda _{l}|}\sum\limits_{x\in \Lambda
_{l}}\chi _{x}\left( A\right) \ .  \label{emperical mean}
\end{equation}%
This case obviously includes the sequence $\{\mathrm{D}_{l}\}_{l\in \mathbb{R%
}^{+}}\subset \mathcal{X}$ of fermionic density observables, which is
defined, for any $l\in \mathbb{R}^{+}$, by%
\begin{equation}
\mathrm{D}_{l}\doteq \frac{1}{|\Lambda _{l}|}\sum\limits_{x\in \Lambda
_{l},\ \mathrm{s}\in \mathrm{S}}a^{\ast }\left( \mathfrak{e}_{x,\mathrm{s}%
}\right) a\left( \mathfrak{e}_{x,\mathrm{s}}\right) \in \mathcal{U}_{\Lambda
_{l}}\cap \mathcal{U}^{+}\subset \mathcal{U}\ .  \label{density observables}
\end{equation}%
Here, $\left\{ \mathfrak{e}_{x,\mathrm{s}}\right\} _{(x,\mathrm{s})\in 
\mathbb{Z}^{d}\times \mathrm{S}}$ is the canonical orthonormal basis of $%
\ell ^{2}(\mathbb{Z}^{d}\times \mathrm{S})$, defined, like (\ref{canonical
onb1}), by 
\begin{equation}
\mathfrak{e}_{x,\mathrm{s}}(y,\mathrm{t})\doteq \delta _{x,y}\delta _{%
\mathrm{s},\mathrm{t}}\ ,\qquad x,y\in \mathbb{Z}^{d},\quad \mathrm{s},%
\mathrm{t}\in \mathrm{S}\ .  \label{ONB}
\end{equation}%
Observe also that, for any $A\in \mathcal{U}$, $\{\mathrm{M}_{l}^{A}\}_{l\in 
\mathbb{R}^{+}}\subset \mathcal{X}$ is exponentially tight in any state $%
\rho \in \mathcal{U}^{\ast }$ since 
\begin{equation*}
\mathrm{spec}(\mathrm{M}_{l}^{A})\subset \left[ -\left\Vert A\right\Vert _{%
\mathcal{U}},\left\Vert A\right\Vert _{\mathcal{U}}\right] \ ,\qquad l\in 
\mathbb{R}^{+}\ .
\end{equation*}%
See Equation (\ref{fluctuation measure}).

A similar sequence of observables can be defined via a so-called \emph{%
interaction}, a standard notion in the context of quantum spin and lattice
Fermi systems. This refers to a family $\Psi =\{\Psi _{\Lambda }\}_{\Lambda
\in \mathcal{P}_{f}(\mathbb{Z}^{d})}\subset \mathcal{U}^{+}$ such that $\Psi
_{\Lambda }=\Psi _{\Lambda }^{\ast }\in \mathcal{U}_{\Lambda }\cap \mathcal{U%
}^{+}$ for all $\Lambda \in \mathcal{P}_{f}(\mathbb{Z}^{d})$ and $\Psi
_{\emptyset }=0$. Here, for quantum spin systems, $\mathcal{U}^{+}=\mathcal{U%
}$, while in fermion systems on the lattice, $\mathcal{U}^{+}\varsubsetneq 
\mathcal{U}$ is the even subalgebra of $\mathcal{U=}\mathrm{CAR}(\ell ^{2}(%
\mathbb{Z}^{d}\times \mathrm{S}))$. See end of Section \ref{Section Gen func
as Grassmann int copy(1)}. Then, a thermodynamic sequence of observables
that can also be studied is defined, for any interaction $\Psi $ and $l\in 
\mathbb{R}^{+}$, by 
\begin{equation}
\mathrm{E}_{l}^{\Psi }\doteq \frac{1}{|\Lambda _{l}|}\sum\limits_{\Lambda
\subseteq \Lambda _{l}}\Psi _{\Lambda }\in \mathcal{U}_{\Lambda _{l}}\cap 
\mathcal{U}^{+}\subset \mathcal{U}\ .  \label{average K}
\end{equation}%
It can be seen as a finite-volume, energy-density observable\ of the
interaction $\Psi $, or the energy-per-unit-volume observable\ associated
with the interaction $\Psi $ and the box $\Lambda _{l}$.

An important class of interactions are the so-called translation-invariant,
finite-range interactions:

\begin{itemize}
\item A \emph{translation-invariant} interaction is an interaction $\Psi $
satisfying 
\begin{equation*}
\Psi _{x+\Lambda }=\chi _{x}(\Psi _{\Lambda })\ ,\qquad \Lambda \in \mathcal{%
P}_{f}(\mathbb{Z}^{d}),\ x\in \mathbb{Z}^{d}\ .
\end{equation*}

\item An interaction $\Psi $ has \emph{finite-range} iff there is some $%
R<\infty $ such that ${\o }(\Lambda )>R$ implies $\Psi _{\Lambda }=0$,
where, for any $\Lambda \in \mathcal{P}_{f}(\mathbb{Z}^{d})$, 
\begin{equation*}
{\o }(\Lambda )\doteq \max\limits_{\left( x_{1},\ldots ,x_{d}\right) ,\left(
y_{1},\ldots ,y_{d}\right) \in \Lambda }\sqrt{|x_{1}-y_{1}|^{2}+\dots
+|x_{d}-y_{d}|^{2}}\ .
\end{equation*}
\end{itemize}

For instance, let $\Lambda _{0}\in \mathcal{P}_{f}(\mathbb{Z}^{d})$, $%
A=A^{\ast }\in \mathcal{U}_{\Lambda _{0}}\cap \mathcal{U}^{+}$, and $\Psi
^{A}$ be the interaction defined by $\Psi _{\Lambda }^{A}\doteq \chi
_{x}\left( A\right) $ for $\Lambda =\Lambda _{0}+x$ with $x\in \mathbb{Z}%
^{d} $ and $\Psi _{\Lambda }^{A}=0$ otherwise. This is an example of
translation-invariant and finite-range interactions. Additionally, note that 
\begin{equation}
\mathrm{E}_{l}^{\Psi ^{A}}=\mathrm{M}_{l}^{A}+\mathcal{O}\left( l^{-1}\right)
\label{equation mean}
\end{equation}%
in the $C^{\ast }$-algebra $\mathcal{U}$. The finite-volume energy-density
observables can be approximated by empirical means even for more general
interactions: If $\Psi $ is translation-invariant and finite-range, then
straightforward computations (see as an example \cite[Definition 1.24 and
Lemma 4.17]{BruPedra2}) show that%
\begin{equation}
\lim_{l\rightarrow \infty }\Vert \mathrm{E}_{l}^{\Psi }-\mathrm{M}_{l}^{%
\mathfrak{E}^{\Psi }}\Vert _{\mathcal{U}}=0\ ,
\label{limit observable average}
\end{equation}%
where $\mathrm{M}_{l}^{\mathfrak{E}^{\Psi }}$ is the empirical mean (\ref%
{emperical mean}) associated with the energy-density observable%
\begin{equation}
\mathfrak{E}^{\Psi }\doteq \sum\limits_{\Lambda \in \mathcal{P}_{f}(%
\mathfrak{L}),\;\Lambda \ni 0}\frac{\Psi _{\Lambda }}{|\Lambda |}\in 
\mathcal{U}^{+}\ .  \label{energy observable}
\end{equation}%
Moreover, in this case, $\{\mathrm{E}_{l}^{\Psi }\}_{l\in \mathbb{R}%
^{+}}\subset \mathcal{X}$ is exponentially tight in any state $\rho \in 
\mathcal{U}^{\ast }$. Equation (\ref{limit observable average}), however,
does not have to hold when the interaction $\Psi $ is not
translation-invariant or finite-range.

\subsection{Exponentially Ergodic States}

\emph{Ergodic} states are states whose space averages or empirical means of
observables exist and do not fluctuate, in the limit of large volumes:

\begin{definition}[Ergodic states]
\label{ergodic00}\mbox{ }\newline
We say that a state\footnote{%
I.e., $\rho (\mathbf{1})=1$\ and $\rho (A^{\ast }A)\geq 0$ for $A\in 
\mathcal{U}$.} $\rho \in \mathcal{U}^{\ast }$ is ergodic if, for all $A\in 
\mathcal{U}$, $\rho \left( \mathrm{M}_{l}^{A}\right) $ converges, as $%
l\rightarrow \infty $, and 
\begin{equation*}
\lim_{l\rightarrow \infty }\rho \left( \left( \mathrm{M}_{l}^{A}-\rho \left( 
\mathrm{M}_{l}^{A}\right) \mathfrak{1}\right) ^{\ast }\left( \mathrm{M}%
_{l}^{A}-\rho \left( \mathrm{M}_{l}^{A}\right) \mathfrak{1}\right) \right)
=0\ .
\end{equation*}
\end{definition}

For any $M=M^{\ast }\in \mathcal{U}$, the non-negative number 
\begin{equation*}
\rho \left( \left( M-\rho \left( M\right) \mathfrak{1}\right) ^{2}\right)
\geq 0
\end{equation*}%
quantifies the fluctuation, around the expected value $\rho (M)$, of the
physical quantity associated with the observable $M$ in the state $\rho $.
It is the variance of any real-valued, random variable with distribution
given by $\mu _{\rho ,M}$, as defined by (\ref{fluctuation measure}).

Because of (\ref{equation mean}) and (\ref{limit observable average}), the
ergodicity of states can be expressed equivalently in terms of the sequences 
$\{\mathrm{E}_{l}^{\Psi }\}_{l\in \mathbb{R}^{+}}$ (\ref{average K}):

\begin{lemma}[Ergodic states]
\label{ergodic00 copy(1)}\mbox{ }\newline
A state $\rho \in \mathcal{U}^{\ast }$ is ergodic iff, for any
translation-invariant and finite-range interaction $\Psi $, $\rho \left( 
\mathrm{E}_{l}^{\Psi }\right) $ converges, as $l\rightarrow \infty $,\ and%
\begin{equation*}
\lim_{l\rightarrow \infty }\rho \left( \left( \mathrm{E}_{l}^{\Psi }-\rho
\left( \mathrm{E}_{l}^{\Psi }\right) \mathfrak{1}\right) ^{2}\right) =0\ .
\end{equation*}
\end{lemma}

\begin{proof}
By the Cauchy-Schwarz inequality for states \cite[Lemma 2.3.10 (b)]%
{BratteliRobinsonI}, Definition \ref{ergodic00} can be restricted to
observables $A\in \mathcal{U}$: A state $\rho $ is ergodic iff $\rho \left( 
\mathrm{M}_{l}^{A}\right) $ converges, as $l\rightarrow \infty $, and 
\begin{equation}
\lim_{l\rightarrow \infty }\rho \left( \left( \mathrm{M}_{l}^{A}-\rho \left( 
\mathrm{M}_{l}^{A}\right) \mathfrak{1}\right) ^{2}\right) =0\ ,\qquad
A=A^{\ast }\in \mathcal{U}\ .  \label{ergodic condition1}
\end{equation}%
By (\ref{limit observable average}) and the density of the set of
finite-volume elements of $\mathcal{U}$, the assertion then follows.
\end{proof}

The existence of such ergodic states is well-known. For instance, in the
case of translation-invariant states, ergodicity is equivalent to
extremality (see, e.g., \cite[Section 1.2 and Chapter 4]{BruPedra2}): Denote
the set of all \emph{translation-invariant} states by 
\begin{equation}
E_{1}\doteq \bigcap\limits_{x\in \mathbb{Z}^{d},\text{ }A\in \mathcal{U}%
}\{\rho \in \mathcal{U}^{\ast }\;:\;\rho (\mathbf{1})=1,\;\rho (A^{\ast
}A)\geq 0\text{\quad }\mathrm{with\ }\rho =\rho \circ \chi _{x}\}\ .
\label{periodic invariant states}
\end{equation}%
Ergo, for any state $\rho \in E_{1}$, 
\begin{equation}
\rho \left( \mathrm{M}_{l}^{A}\right) =\rho \left( A\right) \ ,\qquad A\in 
\mathcal{U}\ ,  \label{condition 0}
\end{equation}%
where $\mathrm{M}_{l}^{A}\in \mathcal{U}$ is the empirical mean defined by (%
\ref{emperical mean}). For translation-invariant states, one can view $%
\mathrm{M}_{l}^{A}$ as the empirical mean of $|\Lambda _{l}|$ (not
necessarily independent) identically distributed (i.d.) observables for any $%
l\in \mathbb{R}^{+}$. By \cite[Remark 1.4, Theorem 1.16, Corollary 4.6]%
{BruPedra2}, the set of ergodic states is the set $\mathcal{E}_{1}$ of
extreme points of the weak$^{\ast }$-compact convex set $E_{1}$. Moreover,
by \cite[Remark 1.4, Corollary 4.6]{BruPedra2}, $\mathcal{E}_{1}$ is a weak$%
^{\ast }$-\emph{dense} $G_{\delta }$ subset of $E_{1}$. (Up to an affine
homeomorphism, $E_{1}$ is the Poulsen simplex.)

Definition \ref{ergodic00} (or Lemma \ref{ergodic00 copy(1)}) does not
automatically yield an LDP for sequences of finite-volume energy-density
observables. We thus define the following subset of ergodic states:

\begin{definition}[Exponentially ergodic states]
\label{lemma Al space average copy(1)}\mbox{ }\newline
Exponentially ergodic states are ergodic states $\rho $ with the property
that, for any translation-invariant and finite-range interaction $\Psi $,
the sequence $\{\mathrm{E}_{l}^{\Psi }\}_{l\in \mathbb{R}^{+}}$ (\ref%
{average K}) satisfies an LDP in some closed interval $\mathcal{I}\ni
\lim_{l\rightarrow \infty }\rho (\mathrm{E}_{l}^{\Psi })$ with a good rate
function $\mathrm{I}$.
\end{definition}

Using results of this paper, we will prove in \cite{LD2} that translation
invariant (or more generally periodic) \emph{KMS states} of weakly
interacting fermions on the lattice are exponentially ergodic states.

\subsection{Thermal Equilibrium\ States for Fermion Systems\label{Thermal eq
states}}

By definition, an interaction is a family $\Phi =\{\Phi _{\Lambda
}\}_{\Lambda \in \mathcal{P}_{f}(\mathfrak{L})}$ of (even) elements such
that $\Phi _{\Lambda }=\Phi _{\Lambda }^{\ast }\in \mathcal{U}_{\Lambda
}\cap \mathcal{U}^{+}$ for all $\Lambda \in \mathcal{P}_{f}(\mathbb{Z}^{d})$
and $\Phi _{\emptyset }=0$. For quantum spin systems, $\mathcal{U}^{+}=%
\mathcal{U}$, while in fermion systems on the lattice, $\mathcal{U}%
^{+}\varsubsetneq \mathcal{U}$ is the even subalgebra.

Fix two interactions $\Phi _{\mathrm{f}}$ and $\Phi _{\mathrm{i}}$
respectively representing the free (or unperturbed) and interparticle
components of the total interaction $\Phi =\Phi _{\mathrm{f}}+\Phi _{\mathrm{%
i}}$. These interactions define the quantum spin or fermion system via a
sequence of (internal) energy observables or Hamiltonians%
\begin{equation}
H_{L_{\mathrm{f}},L_{\mathrm{i}}}\doteq \sum\limits_{\Lambda _{1}\subseteq
\Lambda _{L_{\mathrm{f}}}}\Phi _{\mathrm{f},\Lambda
_{1}}+\sum\limits_{\Lambda _{2}\subseteq \Lambda _{L_{\mathrm{i}}}}\Phi _{%
\mathrm{i},\Lambda _{2}}=\left\vert \Lambda _{L_{\mathrm{f}}}\right\vert 
\mathrm{E}_{L_{\mathrm{f}}}^{\Phi _{\mathrm{f}}}+\left\vert \Lambda _{L_{%
\mathrm{i}}}\right\vert \mathrm{E}_{L_{\mathrm{i}}}^{\Phi _{\mathrm{i}}}\
,\qquad L_{\mathrm{f}},L_{\mathrm{i}}\in \mathbb{R}_{0}^{+}\ .
\label{Hamiltonians}
\end{equation}%
Then, the Gibbs state\footnote{%
Or finite-volume thermal-equilibrium state, that is, the unique state
minimizing the finite-volume free energy.} $\rho _{L_{\mathrm{f}},L_{\mathrm{%
i}}}\in \mathcal{U}_{\Lambda _{L_{\mathrm{f}}}}^{\ast }$ is defined, for any 
$L_{\mathrm{f}},L_{\mathrm{i}}\in \mathbb{R}_{0}^{+}$ and $\beta \in \mathbb{%
R}^{+}$, by 
\begin{equation}
\rho _{L_{\mathrm{f}},L_{\mathrm{i}}}(A)\doteq \frac{\mathrm{tr}\left( A%
\mathrm{e}^{-\beta H_{L_{\mathrm{f}},L_{\mathrm{i}}}}\right) }{\mathrm{tr}%
\left( \mathrm{e}^{-\beta H_{L_{\mathrm{f}},L_{\mathrm{i}}}}\right) }\
,\qquad A\in \mathcal{U}\ .  \label{def Gibbs Phi}
\end{equation}%
Here, $\mathrm{tr}\in \mathcal{U}^{\ast }$ is the tracial state of the $%
C^{\ast }$-algebra $\mathcal{U}$. Cf. Lemma \ref{Lemma quasi free state}.
For fermion systems, see Definition \ref{def trace state} and Remark \ref%
{bound norm a CAR copy(2)}. For any $L_{\mathrm{f}},L_{\mathrm{i}}\in 
\mathbb{R}_{0}^{+}$, $\rho _{L_{\mathrm{f}},L_{\mathrm{i}}}$ is the unique
KMS state \cite[Section 5.3]{BratteliRobinson} associated with the inverse
temperature $\beta \in \mathbb{R}^{+}$ and the continuous group $\{\tau
_{t}^{(L_{\mathrm{f}},L_{\mathrm{i}})}\}_{t\in {\mathbb{R}}}$ of $\ast $-auto%
%TCIMACRO{\TeXButton{\-}{\-}}%
%BeginExpansion
\-%
%EndExpansion
morphisms of $\mathcal{U}$ defined by%
\begin{equation*}
\tau _{t}^{(L_{\mathrm{f}},L_{\mathrm{i}})}(A)=\mathrm{e}^{itH_{L_{\mathrm{f}%
},L_{\mathrm{i}}}}A\mathrm{e}^{-itH_{L_{\mathrm{f}},L_{\mathrm{i}}}}\
,\qquad A\in \mathcal{U}\ .
\end{equation*}

The limit $(L_{\mathrm{f}},L_{\mathrm{i}})\rightarrow (\infty ,\infty )$
refers to the thermodynamic limit. Similar to the set of probability
measures in the commutative setting, the set of all states on any $C^{\ast }$%
-algebra is a weak$^{\ast }$-compact (convex) set and so, for any
interaction $\Phi _{\mathrm{f}}$, $\Phi _{\mathrm{i}}$, the family $\{\rho
_{L_{\mathrm{f}},L_{\mathrm{i}}}\}_{L_{\mathrm{f}},L_{\mathrm{i}}\in \mathbb{%
R}_{0}^{+}}$ has at least one weak$^{\ast }$-accumulation point, which is a
state. If $\Phi _{\mathrm{f}}$, $\Phi _{\mathrm{i}}$ are finite-range
interactions (Section \ref{Thermodynamic Sequences of Observables}), the
groups $\{\tau _{t}^{(L_{\mathrm{f}},L_{\mathrm{i}})}\}_{t\in {\mathbb{R}}}$%
, $L_{\mathrm{f}},L_{\mathrm{i}}\in \mathbb{R}_{0}^{+}$, converge strongly,
as $(L_{\mathrm{f}},L_{\mathrm{i}})\rightarrow (\infty ,\infty )$, to a
strongly continuous group $\{\tau _{t}\}_{t\in {\mathbb{R}}}$ of $\ast $-auto%
%TCIMACRO{\TeXButton{\-}{\-}}%
%BeginExpansion
\-%
%EndExpansion
morphisms on $\mathcal{U}$. See for instance \cite[Theorem 4.8]{brupedraLR}.
By \cite[Proposition 5.3.25]{BratteliRobinson}, in this case, any weak$%
^{\ast }$-accumulation point $\rho $ of $\{\rho _{L_{\mathrm{f}},L_{\mathrm{i%
}}}\}_{L_{\mathrm{f}},L_{\mathrm{i}}\in \mathbb{R}_{0}^{+}}$ is a KMS state
associated with $\{\tau _{t}\}_{t\in {\mathbb{R}}}$ and the inverse
temperature $\beta \in \mathbb{R}^{+}$. Note that, in our study, we take the
limit $L_{\mathrm{i}}\rightarrow \infty $ after $L_{\mathrm{f}}\rightarrow
\infty $. See, for instance, Equation (\ref{decay parameter}).

By \cite[Theorem 6.2.42 and discussions on page 294]{BratteliRobinson}
(quantum spin systems) or \cite[Theorems 6.4, 11.7, 12.11]{Araki-Moriya}
(fermion systems), for any translation-invariant and finite-range
interaction $\Phi _{\mathrm{f}}$, $\Phi _{\mathrm{i}}$, there is at least
one translation-invariant KMS state $\rho \in E_{1}$. In this case, it is
well-known that the set of all translation-invariant KMS states associated
with $\{\tau _{t}\}_{t\in {\mathbb{R}}}$ and $\beta \in \mathbb{R}^{+}$ is a
weak$^{\ast }$-closed face of $E_{1}$ and thus, there is at least one such
KMS state which is ergodic. In particular, the uniqueness of the KMS state
automatically yields its ergodicity. This is the typical situation studied
in previous results \cite{LLS00,RoMaNeSc,lenci2005large,Ogata2010}.

\begin{remark}[LDP and uniqueness of KMS states]
\label{LDP and KMS states}\mbox{ }\newline
\cite{netovcny2004large} is a study of LD in quantum spin systems at inverse
temperatures $\beta <\beta _{0}$ done by \textquotedblleft polymer-cluster
expansions\textquotedblright , without explicitly using the uniqueness of
KMS\ states. For sufficiently small $\beta >0$, however, the KMS state is
unique. See discussions of \cite[Remark 2.14]{netovcny2004large}. In \cite%
{HiaiMosoOga} LD associated with particular ergodic states (finitely
correlated ergodic states) are studied in dimension 1. See also \cite%
{Ogata2010}. In \cite{OR11} the authors impose strong conditions on the
state, which are known to be satisfied in concrete models only when the KMS
state is unique. In \cite{LD2}, similar to \cite{GLM02}, we will consider
weakly interacting fermions on the lattice at any inverse temperature $\beta
\in \mathbb{R}^{+}$. In this situation, the uniqueness of KMS\ is generally
unknown.
\end{remark}

\noindent \textit{Acknowledgments:} This work is supported by FAPESP
(2017/22340-9) and CNPq (309723/2020-5), as well as by the Basque Government
through the grant IT641-13 and the BERC 2018-2021 program, and by the
Spanish Ministry of Science, Innovation and Universities: BCAM Severo Ochoa
accreditation SEV-2017-0718, MTM2017-82160-C2-2-P. We are very grateful to
Y. Ogata and L. Rey-Bellet for hints and discussions. Finally, we thank
Zosza Lefevre for linguistic corrections.

\end{document}